\documentclass[a4paper,11pt]{article}
\pdfoutput=1
\usepackage{jcappub}
\usepackage{bm}
\usepackage{color}
\usepackage{array}
\usepackage{graphicx}
\usepackage{multirow}
\usepackage{afterpage}
\newcolumntype{P}[1]{>{\centering\arraybackslash}p{#1}}
\newcolumntype{M}[1]{>{\centering\arraybackslash}m{#1}}

\newcommand{\be}{\begin{equation}}
\newcommand{\ee}{\end{equation}}
\newcommand{\een}{\end{subequations}}
\newcommand{\ben}{\begin{subequations}}
\newcommand{\beq}{\begin{eqalignno}}
\newcommand{\eeq}{\end{eqalignno}}

\newcommand{\lsim}{\mathrel{\mathop{\kern 0pt \rlap
      {\raise.2ex\hbox{$<$}}}\lower.9ex\hbox{\kern-.190em $ \sim$}}}
\newcommand{\gsim}{\mathrel{\mathop{\kern 0pt
      \rlap{\raise.2ex\hbox{$>$}}}\lower.9ex\hbox{\kern-.190em $\sim$}}}

\newcommand{\CO}{\mathcal{O}}

\newcommand{\erf}{\mbox{erf}}

\title{COSINE--100 and DAMA/LIBRA-phase2 in WIMP effective models}

\author{The COSINE-100 Collaboration\hspace{15em}}
\author[a]{G.~Adhikari}
\author[a,1]{P.~Adhikari}
\author[b]{E.~Barbosa~de~Souza}
\author[c]{N.~Carlin}
\author[d]{S.~Choi}
\author[e]{M.~Djamal}
\author[f]{A.~C.~Ezeribe}
\author[g]{C.~Ha}
\author[h]{I.~S.~Hahn}
\author[g]{E.~J.~Jeon}
\author[b]{J.~H.~Jo}
\author[d]{H.~W.~Joo}
\author[g]{W.~G.~Kang}
\author[i]{W.~Kang}
\author[j]{M.~Kauer}
\author[k]{G.~S.~Kim}
\author[g]{H.~Kim}
\author[k]{H.~J.~Kim}
\author[g]{K.~W.~Kim}
\author[g]{N.~Y.~Kim}
\author[d]{S.~K.~Kim}
\author[g,a,m]{Y.~D.~Kim}
\author[g,l,m]{Y.~H.~Kim}
\author[g]{Y.~J.~Ko}
\author[f]{V.~A.~Kudryavtsev}
\author[g,m]{H.~S.~Lee}
\author[g]{J.~Lee}
\author[k]{J.~Y.~Lee}
\author[g,m]{M.~H.~Lee}
\author[g]{D.~S.~Leonard}
\author[f]{W.~A.~Lynch}
\author[b]{R.~H.~Maruyama}
\author[f]{F.~Mouton}
\author[g]{S.~L.~Olsen}
\author[m]{B.~J.~Park}
\author[n]{H.~K.~Park}
\author[l]{H.~S.~Park}
\author[g]{K.~S.~Park}
\author[c]{R.~L.~C.~Pitta}
\author[e]{H.~Prihtiadi}
\author[g]{S.~J.~Ra}
\author[i]{C.~Rott}
\author[g]{K.~A.~Shin}
\author[f]{A.~Scarff}
\author[f]{N.~J.~C.~Spooner}
\author[b]{W.~G.~Thompson}
\author[o]{L.~Yang}
\author[i]{and G.~H.~Yu\note{Present address : Department of Physics, Carleton University, Ottawa, ON K1S 5B6, Canada}}
\author{\hspace{15em}}
\author{\hspace{15em}}

\affiliation[a]{Department of Physics, Sejong University, Seoul 05006, Republic of Korea}
\affiliation[b]{Department of Physics and Wright Laboratory, Yale University, New Haven, CT 06520, USA}
\affiliation[c]{Physics Institute, University of S\~{a}o Paulo, 05508-090, S\~{a}o Paulo, Brazil}
\affiliation[d]{Department of Physics and Astronomy, Seoul National University, Seoul 08826, Republic of Korea}
\affiliation[e]{Department of Physics, Bandung Institute of Technology, Bandung 40132, Indonesia}
\affiliation[f]{Department of Physics and Astronomy, University of Sheffield, Sheffield S3 7RH, United Kingdom}
\affiliation[g]{Center for Underground Physics, Institute for Basic Science (IBS), Daejeon 34126, Republic of Korea}
\affiliation[h]{Department of Science Education, Ewha Womans University, Seoul 03760, Republic of Korea} 
\affiliation[i]{Department of Physics, Sungkyunkwan University, Suwon 16419, Republic of Korea}
\affiliation[j]{Department of Physics and Wisconsin IceCube Particle Astrophysics Center, University of Wisconsin-Madison, Madison, WI 53706, USA}
\affiliation[k]{Department of Physics, Kyungpook National University, Daegu 41566, Republic of Korea}
\affiliation[l]{Korea Research Institute of Standards and Science, Daejeon 34113, Republic of Korea}
\affiliation[m]{IBS School, University of Science and Technology (UST), Daejeon 34113, Republic of Korea}
\affiliation[n]{Department of Accelerator Science, Korea University, Sejong 30019, Republic of Korea}
\affiliation[o]{Department of Physics, University of Illinois at Urbana-Champaign, Urbana, IL 61801, USA}

\author{The Sogang Phenomenology Group\hspace{15em}}
\author[aa]{Sunghyun Kang}
\author[aa]{Stefano Scopel}
\author[aa]{Gaurav Tomar}
\author[aa,bb]{and Jong--Hyun Yoon}
\affiliation[aa]{Department of Physics, Sogang University, Seoul 04107, Republic of Korea}
\affiliation[bb]{Department of Physics, University of Helsinki, FI-00014 Helsinki, Finland}
\emailAdd{scopel@sogang.ac.kr}
\emailAdd{changhyon.ha@gmail.com}

\abstract{ Assuming a standard Maxwellian for the WIMP velocity
  distribution, we obtain the bounds from null WIMP search results
  of 59.5 days of COSINE--100 data on the DAMA/LIBRA--phase2 modulation
  effect within the context of the non--relativistic
  effective theory of WIMP--nucleus scattering.  Here, we
  systematically assume that one of the effective operators allowed by
  Galilean invariance dominates in the effective Hamiltonian of a
  spin--1/2 dark matter (DM) particle. We find that, although
  DAMA/LIBRA and COSINE--100 use the same sodium--iodide target, the
  comparison of the two results still depends on the particle--physics
  model. This is mainly due to two reasons: i) the WIMP signal
  spectral shape; ii)
  the expected modulation fractions, when the upper bound on the
  time--averaged rate in COSINE--100 is converted into a constraint on
  the annual modulation component in DAMA/LIBRA. We find that the
  latter effect is the dominant one.  For several effective operators
  the expected modulation fractions are larger than in
  the standard spin--independent or spin--dependent interaction
  cases. As a consequence, compatibility between the modulation effect
  observed in DAMA/LIBRA and the null result from COSINE--100 is still
  possible for several non--relativistic operators.
  At low WIMP masses
  such relatively high values of the modulation fractions arise
  because COSINE--100 is mainly sensitive to WIMP--sodium scattering
  events, due to the higher threshold compared to DAMA/LIBRA.
  A next COSINE analysis is expected to have a full sensitivity
  for the 5\,$\sigma$ region of DAMA/LIBRA.
}

\begin{document}

\maketitle

\section{Introduction}
\label{sec:introduction}

Since its first presentation, the DAMA/NaI, DAMA/LIBRA--phase1 and
DAMA/LIBRA--phase2's (DAMA for short)
observation~\cite{Bernabei:2008yi,Bernabei:2010mq,Bernabei:2013xsa,Bernabei:2018yyw}
of an annual modulation signal~\cite{Savage:2008er,Freese:2012xd}
detected in an array of low--background sodium--iodide crystals has
produced continuous speculation about whether or not it is caused by
dark matter.  Given the substantial statistical significance
($>10\,\sigma$) of the signal, an independent verification of the
result with the same sodium--iodide, or NaI(Tl) target crystals is
required.  Experimental efforts by several groups using the same
target medium are on--going and in a few years will confirm the
presence of the signal in a model--independent
way~\cite{Adhikari:2018ljm,Amare:2018sxx,Montini:2018ubd,Angloher:2016ooq}.
On the other hand, a Weakly Interacting Massive
Particle~(WIMP)~\cite{Lee:1977ua,Goodman:1984dc} dark-matter
interpretations for the positive signal have been ruled out by other
experiments~\cite{sckim12,xenon17,Tan:2016zwf,Aprile:2017iyp} using
time-averaged rates and shapes in the specific context of the Standard
Galactic Halo Model~\cite{Lewin:1995rx,freese1987}, assuming a
standard spin--independent (SI) or spin--dependent (SD) interaction.

Although preliminary modulation analyses from ANAIS~\cite{Amare:2019jul} and COSINE-100~\cite{Adhikari:2019off}
have been released, these experiments need several more years of data-taking~\cite{Thompson:2017yvq}
before they can reach the modulation sensitivity required to probe the DAMA signal.
In the meantime, the yearly modulation amplitude and the time--averaged
rate can be compared in the context of specific models of
WIMP--nucleus interactions.
The first such analysis using 59.5 days of COSINE-100 data has excluded the SI WIMP
case as a potential model for the DAMA signal interpretation~\cite{Adhikari:2018ljm}.

Since COSINE-100 uses the same target material, fewer assumptions
about how the WIMP elastic scattering rates scale with the target nucleus
are needed in order to interpret the result.  The dominance of
WIMP--sodium or WIMP--iodine scattering events for different WIMP
masses can then be determined entirely by kinematics.  As a
consequence, expected rates can be directly expressed in terms of
WIMP-nucleus cross sections.  For example, this enabled the Korea
Invisible Mass Search (KIMS) experiment, using a CsI(Tl) target, to
constrain in a model--independent way WIMP--iodine scattering events
in DAMA for WIMP masses above approximately
20\,GeV/c$^2$~\cite{sckim12}.  On the other hand, for lower WIMP
masses, where only WIMP--sodium events are kinematically accessible,
the DAMA result could only be probed using different target nuclei.
Moreover, the recent DAMA/LIBRA--phase2 result with a lower threshold
at 1 keV electron-equivalent (keVee), is more sensitive to
WIMP--iodine scattering events for WIMP masses below 20 GeV/c$^2$ and
the expected relative numbers of WIMP--iodine and WIMP--sodium
scattering events is model dependent.  To address these aspects, we
use non--relativistic (NR) effective field theory in an interpretation
of the DAMA signal with the COSINE--100 data.

COSINE-100~\cite{Adhikari:2017esn,Prihtiadi:2017inr,Adhikari:2018fpo}
is a joint dark matter search experiment
of KIMS~\cite{kykim15,adhikari16,Adhikari:2017gbj} and
DM-Ice~\cite{dmice,deSouza:2016fxg} with an array of low
radioactive NaI(Tl) crystals at the Yangyang underground
laboratory.  The experiment is composed
of eight encapsulated NaI(Tl) crystals (a total of 106\,kg)
placed in the middle of a copper box
which is filled with 2 tons of liquid scintillator~\cite{Park:2017jvs}
and further shielded by lead and plastic scintillator panels from 
external radiations.
The plastic scintillator panels veto cosmic-ray muons
and liquid scintillator actively reduces background radiations that
originate from crystals or vicinity of the crystal detectors~\cite{Prihtiadi:2017inr}.
The details of the COSINE-100 experimental setup
and analysis methods can be found elsewhere~\cite{Adhikari:2017gbj,Adhikari:2018ljm}.
Here, we extend the interpretation of the data collected
in Ref.~\cite{Adhikari:2018ljm} to the general NR effective theory of
nuclear scattering for a WIMP of spin 1/2.

The paper is organized by the following structure:
Section~\ref{sec:eft} summarizes the theory of WIMP--nucleus scattering in the NR effective
theory of a particle of spin 1/2; Section~\ref{sec:analysis} is
devoted to our quantitative analysis; Section~\ref{sec:conclusions}
contains our conclusions.

\section{WIMP--nucleus scattering rates in non--relativistic effective models}
\label{sec:eft}

The expected rate in a given visible energy bin $E_1^{\prime}\le
E^{\prime}\le E_2^{\prime}$ of a direct detection experiment is given
by:

\begin{eqnarray}
R_{[E_1^{\prime},E_2^{\prime}]}(t)&=&MT_{exp}\int_{E_1^{\prime}}^{E_2^{\prime}}\frac{dR}{d
  E^{\prime}}(t)\, dE^{\prime} \label{eq:start}\\
 \frac{dR}{d E^{\prime}}(t)&=&\sum_T \int_0^{\infty} \frac{dR_{\chi T}(t)}{dE_{ee}}{\cal
   G}_T(E^{\prime},E_{ee})\epsilon(E^{\prime})\label{eq:start2}\,d E_{ee} \\
E_{ee}&=&q(E_R) E_R \label{eq:start3},
\end{eqnarray}

\noindent with $\epsilon(E^{\prime})\le 1$ the experimental
efficiency/acceptance. In the equations above $E_R$ is the recoil
energy deposited in the scattering process (indicated in keVnr), while
$E_{ee}$ (indicated in keVee) is the fraction of $E_R$ that goes into
the experimentally detected process (ionization, scintillation, heat)
and $q(E_R)$ is the quenching factor, ${\cal
  G_T}(E^{\prime},E_{ee}=q(E_R)E_R)$ is the probability that the
visible energy $E^{\prime}$ is detected when a WIMP has scattered off
an isotope $T$ in the detector target with recoil energy $E_R$, $M$ is
the fiducial mass of the detector and $T_{exp}$ the live--time
exposure of the data taking.

For a given recoil energy imparted to the target the differential rate
for the WIMP--nucleus scattering process is given by:

\be
\frac{d R_{\chi T}}{d E_R}(t)=\sum_T N_T\frac{\rho_{\mbox{\tiny WIMP}}}{m_{\mbox{\tiny WIMP}}}\int_{v_{min}}d^3 v_T f(\vec{v}_T,t) v_T \frac{d\sigma_T}{d E_R},
\label{eq:dr_de}
\ee

\noindent where $\rho_{\mbox{\tiny WIMP}}$ is the local WIMP mass
density in the neighborhood of the Sun, $N_T$ the number of the
nuclear targets of species $T$ per unit mass in the detector (the sum
over $T$ applies in the case of more than one target), while

\be
\frac{d\sigma_T}{d E_R}=\frac{2 m_T}{4\pi v_T^2}\left [ \frac{1}{2 j_{\chi}+1} \frac{1}{2 j_{T}+1}|\mathcal{M}_T|^2 \right ],
\label{eq:dsigma_de}
\ee

\noindent where $m_T$ is the nuclear target mass and the squared
amplitude in parenthesis is given explicitly in
Eq.(\ref{eq:squared_amplitude}).

WIMP--nucleus scattering is a non--relativistic process that can be
fully described in a non--relativistic Effective Theory approach. In
the case of a spin--1/2 DM particle the corresponding Hamiltonian
density is given by~\cite{haxton1,haxton2}:

\begin{eqnarray}
{\bf\mathcal{H}}({\bf{r}})&=& \sum_{\tau=0,1} \sum_{j=1}^{15} c_j^{\tau} \mathcal{O}_{j}({\bf{r}}) \, t^{\tau} ,
\label{eq:H}
\end{eqnarray}

\begin{table}[]
\begin{center}
\begin{tabular}{|l|l|l|l|l|l|l|l|}
\hline\hline
\multicolumn{4}{l|}{\multirow{7}{*}{}} & \multicolumn{4}{l}{\multirow{7}{*}{}} \\
\multicolumn{4}{l|}{$ \CO_1 = 1_\chi 1_N$} \\
\multicolumn{4}{l|}{$\CO_2 = (v^\perp)^2$} & \multicolumn{4}{l}{$\CO_9 = i \vec{S}_\chi \cdot (\vec{S}_N \times {\vec{q} \over m_N})$}\\
\multicolumn{4}{l|}{$\CO_3 = i \vec{S}_N \cdot ({\vec{q} \over m_N} \times \vec{v}^\perp)$} & \multicolumn{4}{l}{$\CO_{10} = i \vec{S}_N \cdot {\vec{q} \over m_N}$} \\
\multicolumn{4}{l|}{$\CO_4 = \vec{S}_\chi \cdot \vec{S}_N$} & \multicolumn{4}{l}{$\CO_{11} = i \vec{S}_\chi \cdot {\vec{q} \over m_N}$} \\
\multicolumn{4}{l|}{$\CO_5 = i \vec{S}_\chi \cdot ({\vec{q} \over m_N} \times \vec{v}^\perp)$} & \multicolumn{4}{l}{$\CO_{12} = \vec{S}_\chi \cdot (\vec{S}_N \times \vec{v}^\perp)$} \\
\multicolumn{4}{l|}{$\CO_6=
  (\vec{S}_\chi \cdot {\vec{q} \over m_N}) (\vec{S}_N \cdot {\vec{q} \over m_N})$} & \multicolumn{4}{l}{$\CO_{13} =i (\vec{S}_\chi \cdot \vec{v}^\perp  ) (  \vec{S}_N \cdot {\vec{q} \over m_N})$} \\
\multicolumn{4}{l|}{$\CO_7 = \vec{S}_N \cdot \vec{v}^\perp$} & \multicolumn{4}{l}{$\CO_{14} = i ( \vec{S}_\chi \cdot {\vec{q} \over m_N})(  \vec{S}_N \cdot \vec{v}^\perp )$} \\
\multicolumn{4}{l|}{$\CO_8 = \vec{S}_\chi \cdot \vec{v}^\perp$} & \multicolumn{4}{l}{$\CO_{15} = - ( \vec{S}_\chi \cdot {\vec{q} \over m_N}) ((\vec{S}_N \times \vec{v}^\perp) \cdot {\vec{q} \over m_N})$} \\ \hline
\end{tabular}
\caption{Non-relativistic Galilean invariant operators for dark matter with spin $1/2$.}
\label{tab:operators}
\end{center}
\end{table}
\noindent where, $\mathcal{O}_{j}$'s are non-relativistic Galilean
invariant operators which have been collected in
Table~\ref{tab:operators}. In the same Table $1_{\chi}$ and $1_{N}$
are identity operators, $\vec{q}$ is the transferred momentum,
$\vec{S}_{\chi}$ and $\vec{S}_{N}$ are the WIMP and nucleon spins,
respectively, while $\vec{v}^\perp = \vec{v} + \frac{\vec{q}}{2\mu_T}$
(with $\mu_T$ the WIMP--nucleus reduced mass) is the relative
transverse velocity operator satisfying
$\vec{v}^{\perp}\cdot\vec{q}=0$. In particular, one has
$(v^{\perp}_T)^2=v^2_T-v_{min}^2$, where, for WIMP--nucleus elastic
scattering, $v_{min}^2=\frac{q^2}{4 \mu_{T}^2}=\frac{m_T E_R}{2
  \mu_{T}^2}$ represents the minimal incoming WIMP speed required to
impart the nuclear recoil energy $E_R$, while $v_T\equiv|\vec{v}_T|$
is the WIMP speed in the reference frame of the nuclear center of
mass. Moreover $t^0=1$, $t^1=\tau_3$ denote the $2\times2$ identity
and third Pauli matrix in isospin space, respectively, and the
isoscalar and isovector (dimension -2) coupling constants $c^0_j$ and
$c^{1}_j$, are related to those for protons and neutrons $c^{p}_j$ and
$c^{n}_j$ by $c^{p}_j=(c^{0}_j+c^{1}_j)$ and
$c^{n}_j=(c^{0}_j-c^{1}_j)$.

Operator ${\cal O}_2$ is of higher order in $v$ compared
to all the others, implying a cross section suppression of order
${\cal O}(v/c)^4)\simeq 10^{-12}$ for the non--relativistic WIMPs in
the halo of our Galaxy. Moreover it cannot be obtained from the
leading-order non--relativistic reduction of a manifestly relativistic
operator \cite{haxton1}.  So, following Refs.\cite{haxton1,haxton2},
we will not include it in our analysis.

Assuming that the nuclear interaction is the sum of the interactions
of the WIMPs with the individual nucleons in the nucleus the WIMP
scattering amplitude on the target nucleus $T$ can be written in the
compact form:

\begin{equation}
  \frac{1}{2 j_{\chi}+1} \frac{1}{2 j_{T}+1}|\mathcal{M}|^2=
  \frac{4\pi}{2 j_{T}+1}
  \sum_{\tau=0,1}\sum_{\tau^{\prime}=0,1}\sum_{k}
  R_k^{\tau\tau^{\prime}}\left
  [c^{\tau}_j,c^{\tau^{\prime}}_j,(v^{\perp}_T)^2,\frac{q^2}{m_N^2}\right
  ] W_{T k}^{\tau\tau^{\prime}}(y).
\label{eq:squared_amplitude}
\end{equation}

\noindent In the above expression $j_{\chi}$ and $j_{T}$ are the WIMP
and the target nucleus spins, respectively, $q=|\vec{q}|$ while the
$R_k^{\tau\tau^{\prime}}$'s are WIMP response functions (that can be
found in Ref.~\cite{haxton2}) which depend on the couplings
$c^{\tau}_j$ as well as the transferred momentum $\vec{q}$ and
$(v^{\perp}_T)^2$. In equation (\ref{eq:squared_amplitude}) the
$W^{\tau\tau^{\prime}}_{T k}(y)$'s are nuclear response functions and
the index $k$ represents different effective nuclear operators, which,
crucially, under the assumption that the nuclear ground state is an
approximate eigenstate of $P$ and $CP$, can be at most eight:
following the notation in \cite{haxton1,haxton2}, $k$=$M$,
$\Phi^{\prime\prime}$, $\Phi^{\prime\prime}M$,
$\tilde{\Phi}^{\prime}$, $\Sigma^{\prime\prime}$, $\Sigma^{\prime}$,
$\Delta$, $\Delta\Sigma^{\prime}$. The $W^{\tau\tau^{\prime}}_{T
  k}(y)$'s are function of $y\equiv (qb/2)^2$, where $b$ is the size
of the nucleus. For the target nuclei $T$ used in most direct
detection experiments the functions $W^{\tau\tau^{\prime}}_{T k}(y)$,
calculated using nuclear shell models, have been provided in
Refs.\cite{haxton2,catena}\footnote{Setting $k=M$ and
  $W^{p,n}_{TM}(q)\equiv(W_{TM}^{00}(q)\pm W_{TM}^{01}(q)\pm
  W_{TM}^{10}(q) +W_{TM}^{11}(q))/4$, in the case of a standard
  spin--independent interaction one has $16/(2
  j_T+1)W^p_{TM}(q)$=$Z^2_T F^2(q)$ and $16/(2
  j_T+1)W^n_{TM}(q)$=$(A_T-Z_T)^2 F^2(q)$, with $Z_T$ and $A_T-Z_T$
  the number of protons and neutrons in target $T$, and $F(q)$ the SI
  nuclear form factor, for which the parameterization in~\cite{helm}
  is commonly assumed.}. The correspondence between models and nuclear
response functions can be directly read off from the WIMP response
functions $R^{\tau\tau^{\prime}}_{k}$~\cite{haxton2}. In particular,
using the decomposition: \be
R_k^{\tau\tau^{\prime}}=R_{0k}^{\tau\tau^{\prime}}+R_{1k}^{\tau\tau^{\prime}}
(v^{\perp}_T)^2=R_{0k}^{\tau\tau^{\prime}}+R_{1k}^{\tau\tau^{\prime}}\left
(v_T^2-v_{min}^2\right ),
\label{eq:r_decomposition}
\ee

\noindent such correspondence is summarized for convenience in Table
\ref{table:eft_summary}.

\begin{table}[t]
\begin{center}
{\begin{tabular}{c|c|c|c|c|c}
\hline
\rule[-15.5pt]{0pt}{30pt}\multirow{2}{*}
{$\mathbf{c_j}$}  &  $R^{\tau \tau^{\prime}}_{0k}$  & $R^{\tau \tau^{\prime}}_{1k}$ & {$\mathbf{c_j}$}  &  $R^{\tau \tau^{\prime}}_{0k}$  & $R^{\tau \tau^{\prime}}_{1k}$ \\
\hline\hline
$c_1$  &   $M(q^0)$ & - & $c_3$  &   $\Phi^{\prime\prime}(q^4)$  & $\Sigma^{\prime}(q^2)$\\
$c_4$  & $\Sigma^{\prime\prime}(q^0)$,$\Sigma^{\prime}(q^0)$   & - & $c_5$  &   $\Delta(q^4)$  & $M(q^2)$\\
$c_6$  & $\Sigma^{\prime\prime}(q^4)$ & - & $c_7$  &  -  & $\Sigma^{\prime}(q^0)$\\
$c_8$  & $\Delta(q^2)$ & $M(q^0)$ & $c_9$  &  $\Sigma^{\prime}(q^2)$  & - \\
$c_{10}$  & $\Sigma^{\prime\prime}(q^2)$ & - & $c_{11}$  &  $M(q^2)$  & - \\
$c_{12}$  & $\Phi^{\prime\prime}(q^2)$,$\tilde{\Phi}^{\prime}(q^2)$ & $\Sigma^{\prime\prime}(q^0)$,$\Sigma^{\prime}(q^0)$ & $c_{13}$  & $\tilde{\Phi}^{\prime}(q^4)$  & $\Sigma^{\prime\prime}(q^2)$ \\
$c_{14}$  & - & $\Sigma^{\prime}(q^2)$ & $c_{15}$  & $\Phi^{\prime\prime}(q^6)$  & $\Sigma^{\prime}(q^4)$ \\
\hline
\end{tabular}}
\caption{Nuclear response functions corresponding to each coupling $c_j$ of the effective Hamiltonian (\ref{eq:H}),
  for the velocity--independent and the velocity--dependent components
  parts of the WIMP response function, decomposed as in
  Eq.(\ref{eq:r_decomposition}).  In parenthesis the power of $q$ in
  the WIMP response function.
  \label{table:eft_summary}}
\end{center}
\end{table}

Finally, $f(\vec{v}_T)$ is the WIMP velocity distribution, for which
we assume a standard isotropic Maxwellian at rest in the Galactic rest
frame truncated at the escape velocity $u_{esc}$, and boosted to the
Lab frame by the velocity of the Earth. So, for the former we assume:

\begin{eqnarray}
  f(\vec{v}_T,t)&=&N\left(\frac{3}{ 2\pi v_{rms}^2}\right )^{3/2}
  e^{-\frac{3|\vec{v}_T+\vec{v}_E|^2}{2 v_{rms}^2}}\Theta(u_{esc}-|\vec{v}_T+\vec{v}_E(t)|)\\
  N&=& \left [ \erf(z)-\frac{2}{\sqrt{\pi}}z e^{-z^2}\right ]^{-1},  
  \label{eq:maxwellian}
  \end{eqnarray}

\noindent with $z=3 u_{esc}^2/(2 v_{rms}^2)$. In the isothermal sphere
model hydrothermal equilibrium between the WIMP gas pressure and
gravity is assumed, leading to $v_{rms}$=$\sqrt{3/2}v_0$ with $v_0$
the galactic rotational velocity. The yearly modulation effect is due
to the time dependence of the Earth's speed with respect to the
Galactic frame:

\begin{equation}
|\vec{v}_E(t)|=v_{Sun}+v_{orb}\cos\gamma \cos\left [\frac{2\pi}{T_0}(t-t_0)
  \right ],
\label{eq:modulation}
  \end{equation}

\noindent where $\cos\gamma\simeq$0.49 accounts for the inclination of
the ecliptic plane with respect to the Galactic plane, $T_0$=1 year,
$t_0$=2 June, $v_{orb}$=2$\pi r_{\oplus}/(T_0)\simeq$ 29 km/sec
($r_{\oplus}$=1 AU, neglecting the small eccentricity of the Earth's
orbit around the Sun) while $v_{Sun}$=$v_0$+12 km/sec, accounting for a
peculiar component of the solar system with respect to the galactic
rotation. For the two parameters $v_0$ and $u_{esc}$ we take $v_0$=220
km/sec \cite{v0_koposov} and $u_{esc}$=550 km/sec \cite{vesc_2014}.
In the isothermal model the time dependence of
Eq. (\ref{eq:modulation}) induces an expected rate with the functional
form $S(t)=S_0+S_m \cos(2\pi/T-t_0)$, with $S_m>0$ at large values of
$v_{min}$ and turning negative when $v_{min}\lsim$ 200 km/s.  In such
regime of $v_{min}$ and below the phase is modified by the focusing
effect of the Sun's gravitational potential \cite{gf}, while when
$S_m\ll S_0$ the time dependence differs from a simple cosine due the
contribution of higher harmonics~\cite{Freese:2012xd}.

In particular, in each visible energy bin DAMA is sensitive to the
yearly modulation amplitude $S_m$, defined as the cosine transform of
$R_{[E_1^{\prime},E_2^{\prime}]}(t)$:

\begin{equation}
S_{m,[E_1^{\prime},E_2^{\prime}]}\equiv \frac{2}{T_0}\int_0^{T_0}
\cos\left[\frac{2\pi}{T_0}(t-t_0)\right]R_{[E_1^{\prime},E_2^{\prime}]}(t)dt,
\label{eq:sm}
\end{equation}  

\noindent while other experiments put upper bounds on the time average
$S_0$:

\begin{equation}
S_{0,[E_1^{\prime},E_2^{\prime}]}\equiv \frac{1}{T_0}\int_0^{T_0}
R_{[E_1^{\prime},E_2^{\prime}]}(t)dt.
\label{eq:s0}
\end{equation}  

In the present paper, we will systematically consider the possibility
that one of the couplings $c_{j}$ dominates in the effective
Hamiltonian of Eq. (\ref{eq:H}). In this case it is possible to
factorize a term $|c_j^p|^2$ from the squared amplitude of
Eq.(\ref{eq:squared_amplitude}) and express it in terms of the {\it
  effective} WIMP--proton cross section:

\begin{equation}
\sigma_p=(c_j^p)^2\frac{\mu_{\chi{\cal N}}^2}{\pi},
  \label{eq:conventional_sigma}
\end{equation}

\noindent (with $\mu_{\chi{\cal N}}$ the WIMP--nucleon reduced mass)
and the ratio $r\equiv c_j^n/c_j^p$. It is worth pointing out here
that among the generalized nuclear response functions arising from the
effective Hamiltonian (\ref{eq:H}) only the ones corresponding to $M$
(SI interaction), $\Sigma^{\prime\prime}$ and $\Sigma^{\prime}$ (both
related to the standard spin--dependent interaction) do not vanish for
$q\rightarrow$0, and so allow to interpret $\sigma_p$ in terms of a
long--distance, point--like cross section. In the case of the other
interactions $\Phi^{\prime\prime}$, $\tilde{\Phi}^{\prime}$ and
$\Delta$ the quantity $\sigma_p$ is just a convenient alternative to
directly parameterizing the interaction in terms of the $c_j^p$
coupling. Since we will not consider interferences among different
couplings the response functions $W^{\tau\tau^{\prime}}_{T k}(y)$ for
$k$=$\Phi^{\prime\prime}M$, $\Delta\Sigma^{\prime}$ will not play any
role in our analysis.
\begin{table}[ht!]
\begin{center}
\begin{tabular}{|M{1cm}|M{3cm}|M{2cm}|M{2cm}|M{2cm}|} \hline
\rule[-12.5pt]{0pt}{30pt}  $\mathbf{c_j}$  & $\mathbf{m_{\chi,min}}$ \textbf{(GeV)} & $\mathbf{r_{\chi,min}}$ & $\mathbf{\sigma~(\mbox{\bf cm}^2)}$ & $\mathbf{\chi^2_{min}}$ \\\hline\hline
  \multirow{ 2}{*}{$c_1$}     &  11.17 & -0.76 &  2.67e-38 & 11.38 \\ 
                 &  45.19 & -0.66 &  1.60e-39 & 13.22 \\\hline
   \multirow{ 2}{*}{$c_3$}      &  8.10 & -3.14 &  2.27e-31 & 11.1 \\ 
                 &  35.68 & -1.10 & 9.27e-35 & 14.23 \\\hline
 \multirow{ 2}{*}{$c_4$}      &  11.22 & 1.71 &  2.95e-36 &  11.38\\ 
                 &  44.71 & -8.34 & 5.96e-36 & 27.7 \\\hline
   \multirow{ 2}{*}{$c_5$}     &  8.34 & -0.61 &  1.62e-29 &  10.83\\ 
                 &  96.13 & -5.74 & 3.63e-34 &  11.11\\\hline  
 \multirow{ 2}{*}{$c_6$}      &  8.09 & -7.20 &  5.05e-28 &  11.11\\ 
                 &  32.9 & -6.48 & 5.18e-31  &  12.74\\\hline
  \multirow{ 2}{*}{$c_7$}      &  13.41 & -4.32 & 4.75e-30  &  13.94\\ 
                 &  49.24 & -0.65 & 1.35e-30  & 38.09\\\hline 
  \multirow{ 2}{*}{$c_8$}      &  9.27 & -0.84 & 8.67e-33  &  10.82\\ 
                 &  42.33 & -0.96 & 1.30e-34  &  11.6\\\hline
  \multirow{ 2}{*}{$c_9$}      &  9.3 & 4.36 & 8.29e-33  &  10.69\\ 
                 &  37.51 & -0.94 & 1.07e-33  & 15.23 \\\hline
  \multirow{ 2}{*}{$c_{10}$}   &  9.29 &  3.25 & 4.74e-33   &  10.69\\ 
                 &  36.81 & 0.09 &  2.25e-34 & 12.40 \\\hline
  \multirow{ 2}{*}{$c_{11}$}   &  9.27 &  -0.67 & 1.15e-34   &  10.69\\ 
                 &  38.51 & -0.66 &  9.17e-37 & 13.02 \\\hline
  \multirow{ 2}{*}{$c_{12}$}   &  9.26 &  -2.85 & 3.92e-34   &  10.69\\ 
                 &  35.22 & -1.93 &  2.40e-35 & 12.47 \\\hline
  \multirow{ 2}{*}{$c_{13}$}   &  8.65&  -0.26 & 1.21e-26   & 10.76\\ 
                 &  29.42 & 0.10 &  5.88e-29  & 14.28 \\\hline
  \multirow{ 2}{*}{$c_{14}$}   &  10.28 & -0.59  & 2.61e-26   & 11.21\\ 
                 &  38.88 &  -1.93 &  2.19e-27   & 14.48  \\\hline
  \multirow{ 2}{*}{$c_{15}$}   &  7.32 &  -3.58 &  2.04e-27  & 12.91\\ 
                 &  33.28 & 4.25  &  2.05e-33  &  16.26\\\hline
\end{tabular}
\caption{Absolute and local minima of the DAMA--phase2 modulation
  result $\chi^2$ of Eq.(\ref{eq:chi2}) for each of the couplings
  $c_j$ of the effective Hamiltonian (\ref{eq:H}). From
  Ref.~\cite{dama_eft_2018}.}
\label{tab:best_fit_values}
\end{center}
\end{table}

\begin{table}[t]
\begin{center}
{\begin{tabular}{@{}c|c|c|c|c|c@{}}
\hline
\rule[-12.5pt]{0pt}{30pt}
\multirow{2}{*}{$\mathbf{c_j}$} & \multicolumn{2}{l|}{$\left (\frac{ S^{DAMA}_m}{S^{DAMA}_0}\right)_{E^{\prime}<3.5~\rm keVee}$} & \multirow{2}{*}{$\mathbf{c_j}$} & \multicolumn{2}{l}{$\left (\frac{ S^{DAMA}_m}{S^{DAMA}_0}\right)_{E^{\prime}<3.5~\rm keVee}$} \\ \cline{2-3} 
\cline{5-6} 
& Low-mass & High-mass &  & Low-mass & High-mass \\
 & local minimum & local minimum &  & local minimum & local minimum \\
\hline\hline
$c_1$  &   0.066   & 0.054  & $c_3$  & 0.120   & 0.098 \\
$c_4$  &  0.065  & 0.047 & $c_5$  &  0.122  & 0.059\\
$c_6$  & 0.121 & 0.111  & $c_7$  & 0.097  & 0.080\\
$c_8$  & 0.094 & 0.072 & $c_9$ & 0.093 & 0.079 \\
$c_{10}$  & 0.093 & 0.085 & $c_{11}$  & 0.094  & 0.083 \\
$c_{12}$  &  0.094 & 0.096 & $c_{13}$  & 0.123 & 0.139 \\
$c_{14}$  & 0.126  & 0.122 & $c_{15}$  & 0.146  & 0.113\\
\hline
\end{tabular}}
\caption{Minimum value of the modulation fraction
  $(S^{DAMA}_m/S^{DAMA}_0)_{E^{\prime}<3.5~\rm keVee}$ in the three
  DAMA energy bins for 2 keVee$\le E^{\prime}\le$3.5 keVee, where the
  bulk of the DAMA modulation effect above the COSINE-100 threshold is
  concentrated
  \label{tab:modulation}}
\end{center}
\end{table}
\section{Analysis}
\label{sec:analysis}

This analysis uses the COSINE-100 data
from October 20, 2016 to December 19, 2016.
After application of data quality criteria,
the 59.5 live days of good data are used for the results presented here.
A total of 11 hours of data did not pass these quality requirement where
abrupt high PMT noise triggers and electronic interference triggers were rejected.
Total exposure is 6303.9 kg$\cdot$day.
During this period, light yield, gain, and other environmental data show stable behavior.
The overall crystal PMT gain is changed by less than 1\% relative to the beginning of the physics run.

An event is triggered if a photon is observed in each PMT within 200\,ns in a crystal.
When this happens the data acquisition system reads out
the full veto detectors including liquid scintillator and plastic scintillators
and other crystal signals simultaneously~\cite{Adhikari:2018fpo}.
The detector stability is checked by using crystal internal gamma calibrations
which show consistent results with external source calibrations.
The low energy electron signals produced from $^{60}$Co calibration
of the Compton scattering and tagged by neighboring crystals
are used to separate the PMT noise events from data.

First, muon-induced events are rejected by requiring the time difference between muon veto events in the plastic panels and the crystal to be less than 30\,ms. This efficiently removes most of muon-induced events that directly pass through the crystals.
We, then, require that leading edges of the trigger pulses start later than 2.0\,$\mu$s, each waveform contain more than two pulses, and integrated charge below the baseline should be small enough. These reject muon-induced phosphor events and electronic interference events.
Next, we demand a single-site condition where neighboring crystals should not have more than four photons
and an energy deposit by the surrounding liquid scintillator should be less than 20 keV.

To identify scintillation signals, one must reject two types of backgrounds
which are more than the desired signals especially below 20\,keVee region.
The first class is thin pulses that are originated from PMTs.
These noise events are triggered partly from the PMT individually
and partly from radioactivities inside the PMT circuitry which make
the crystal scintillate. 
The second class, less often than the first, consists of bell-shaped waveforms that occur
sporadically and in a few PMTs only.
These bell-shaped pulses are produced due to occasional PMT discharge
and the shape of a waveform looks more symmetric than a typical scintillation signal.

\begin{figure*}
\begin{tabular}{ccc}
  \includegraphics[width=0.32\columnwidth]{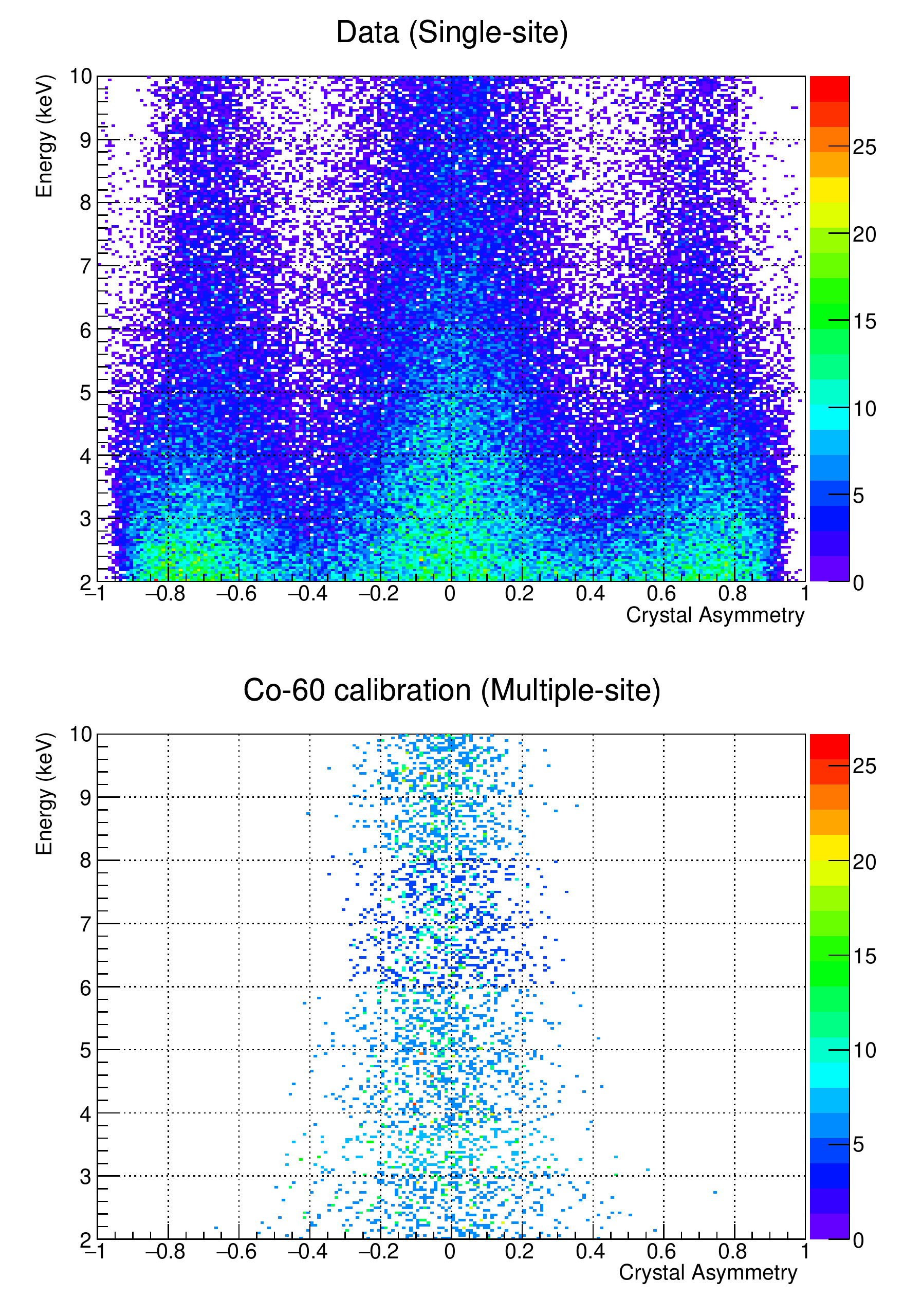} &
  \includegraphics[width=0.32\columnwidth]{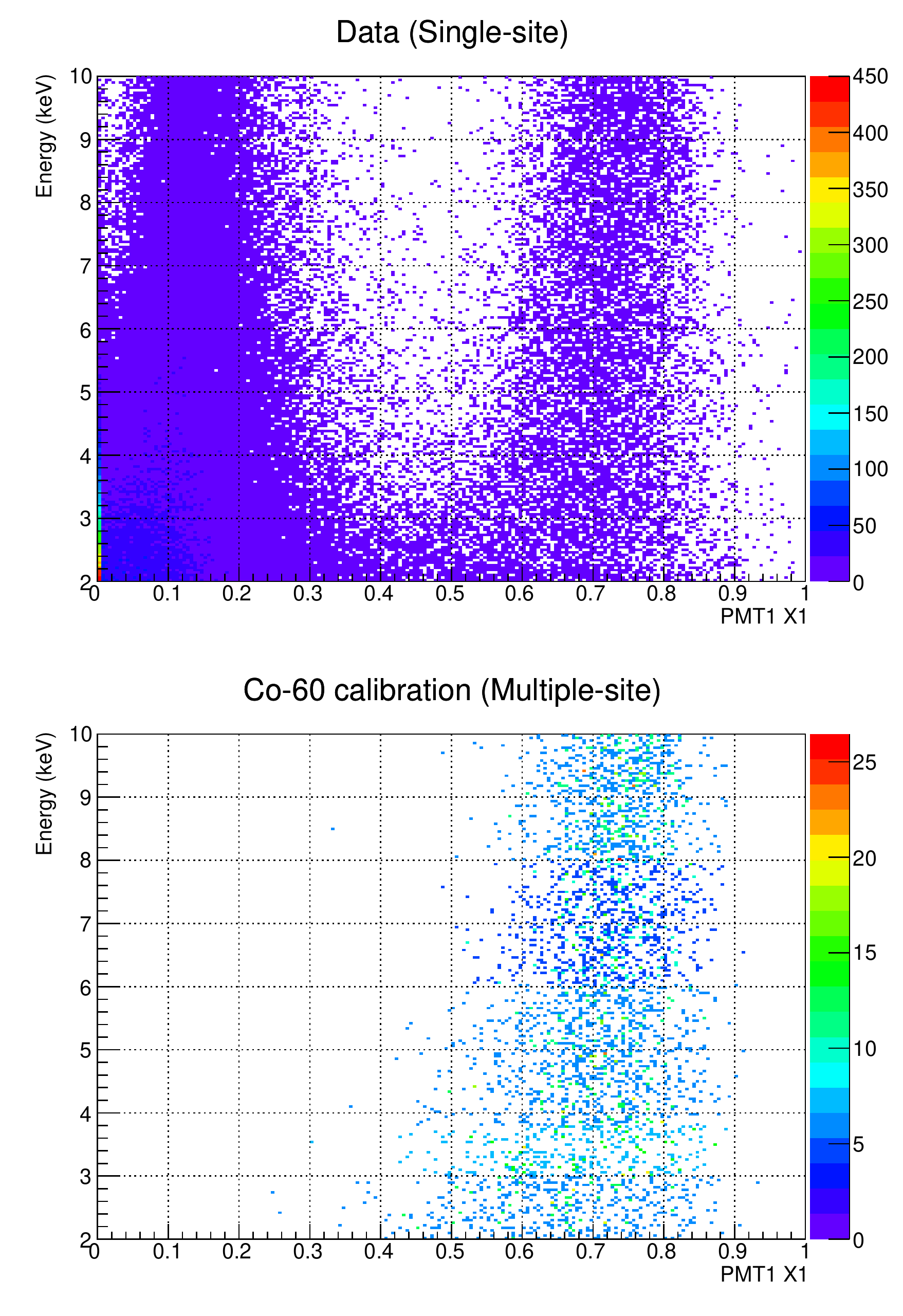} &
  \includegraphics[width=0.32\columnwidth]{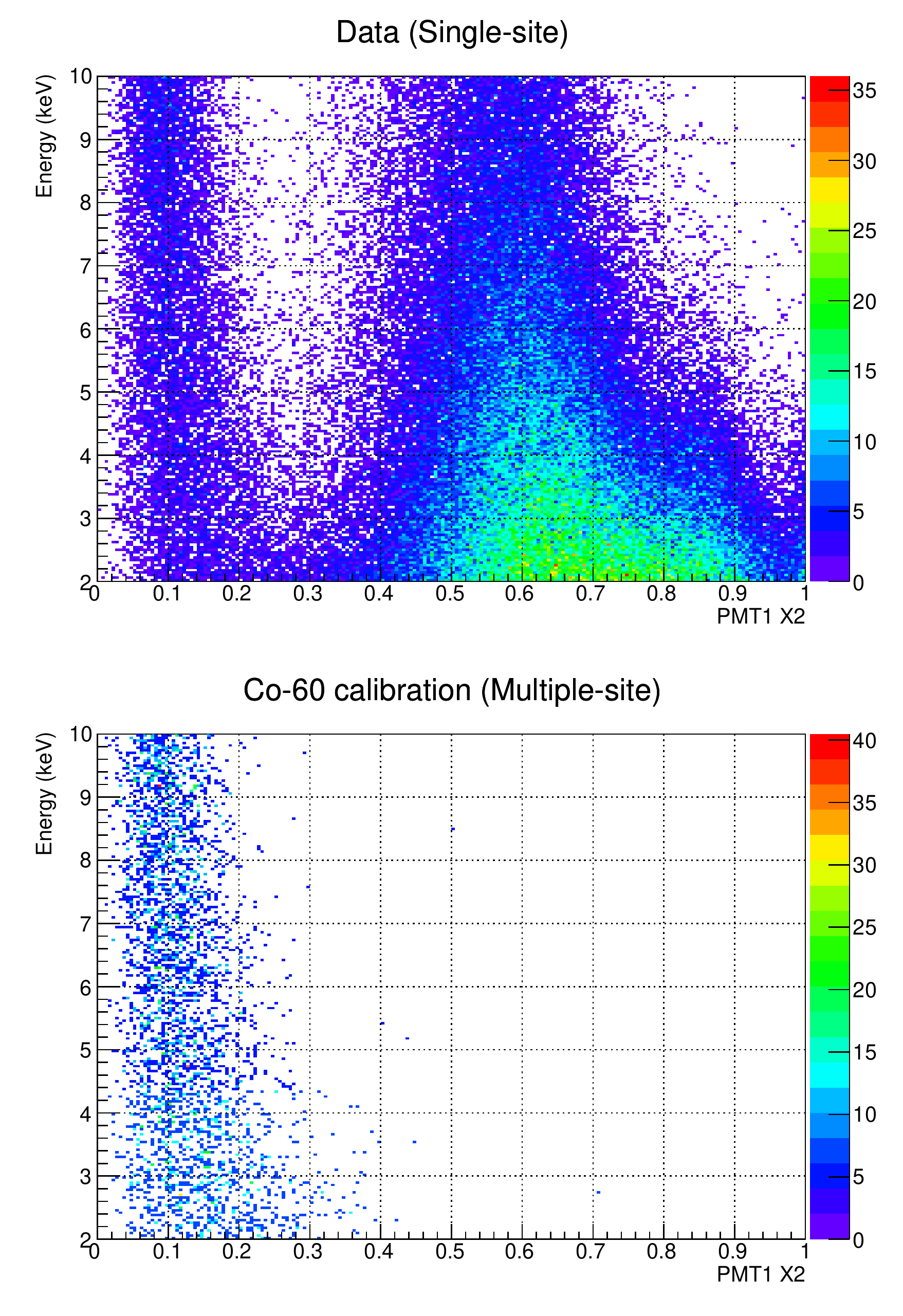} \\
  a) Asymmetry & b) X1 & c) X2  \\
  \includegraphics[width=0.32\columnwidth]{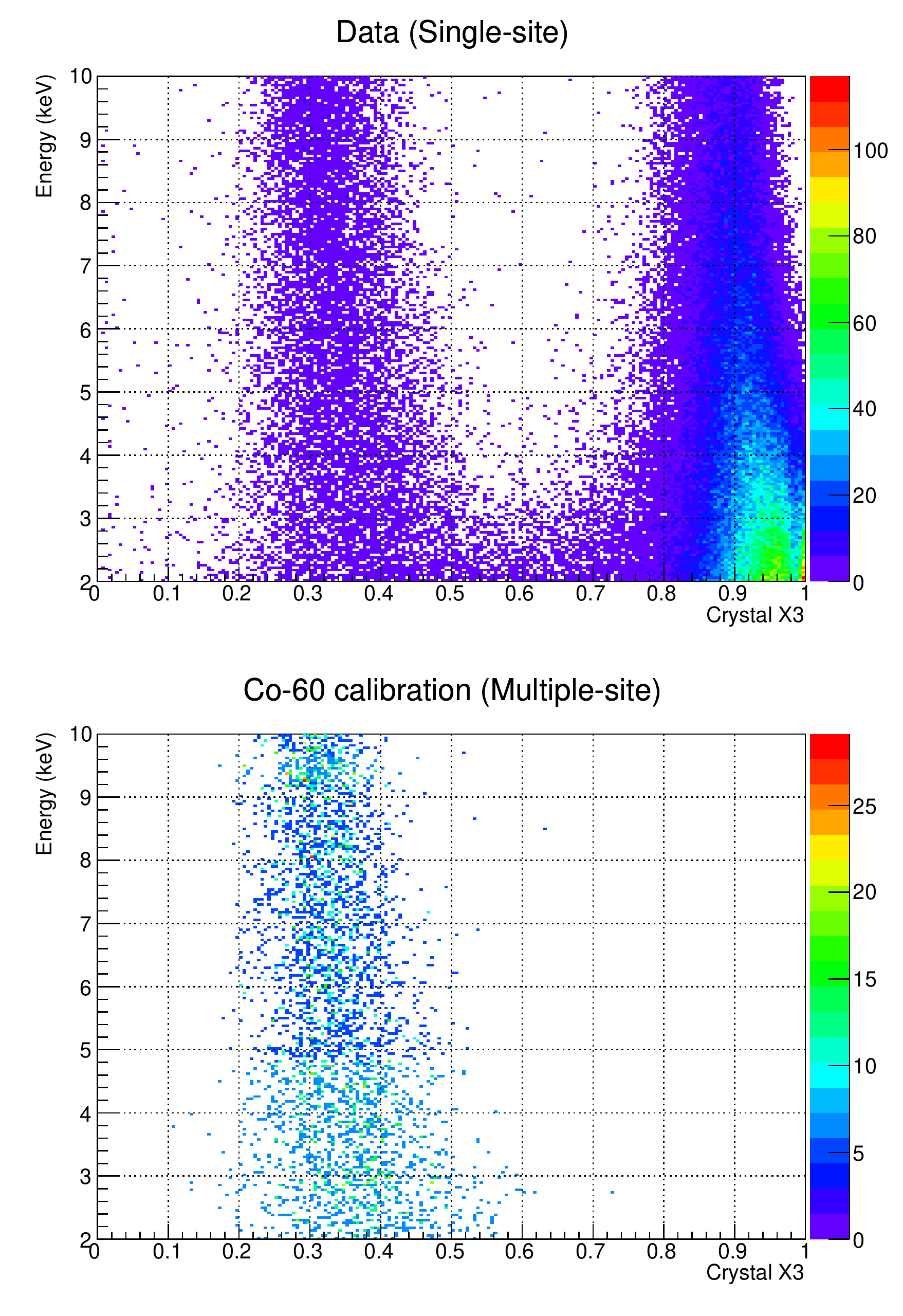} &
  \includegraphics[width=0.32\columnwidth]{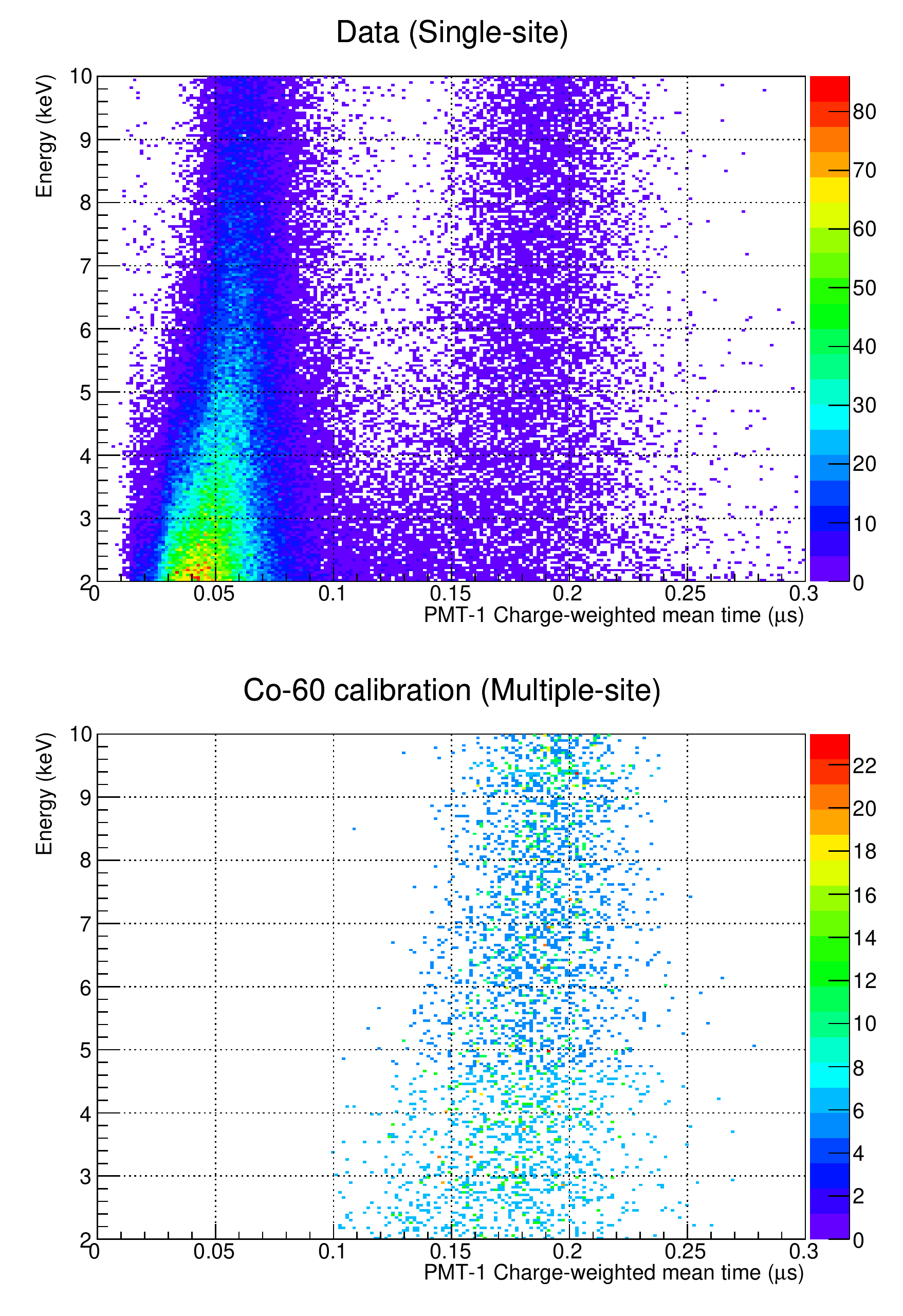}&
  \includegraphics[width=0.32\columnwidth]{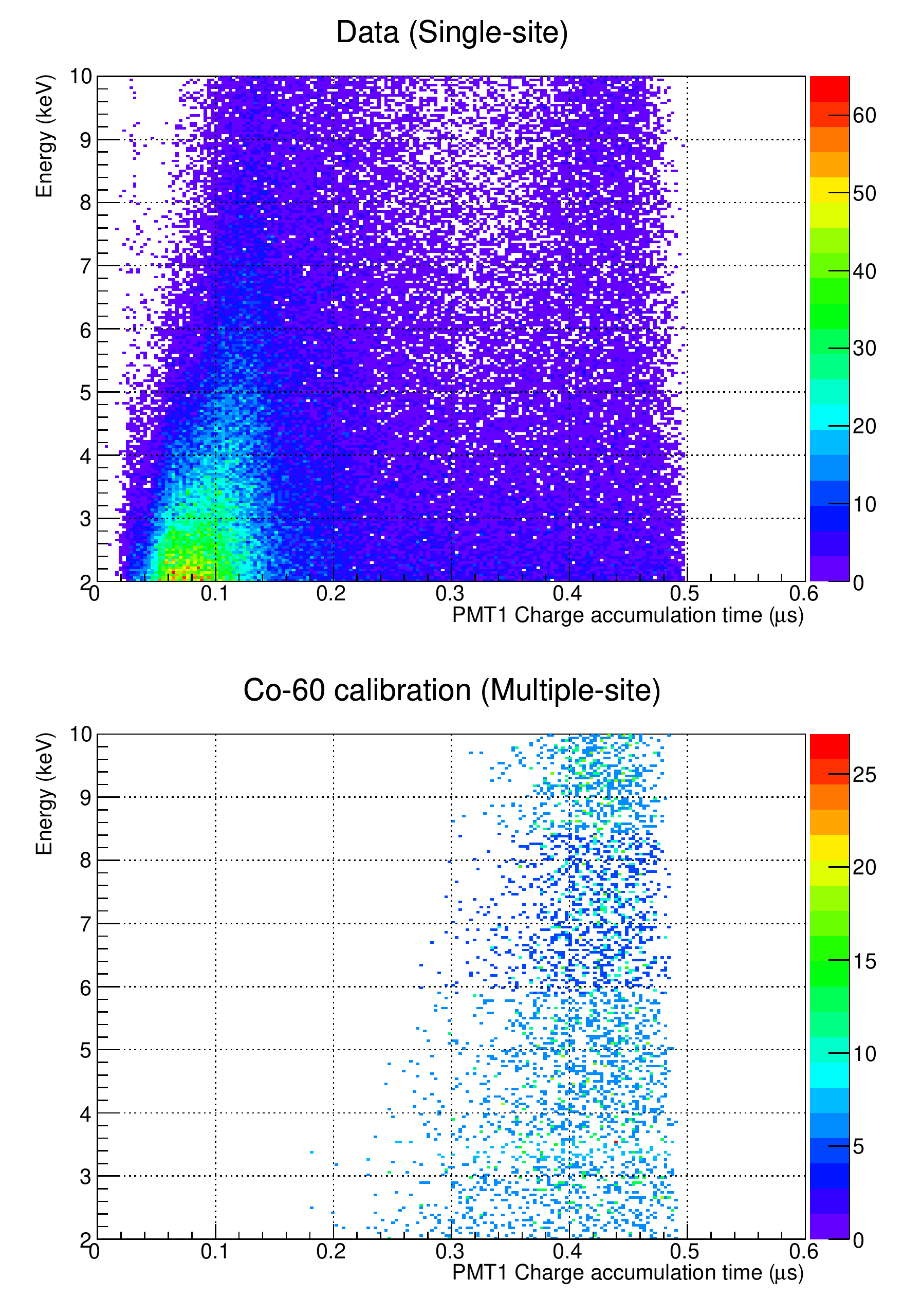} \\
  d) X3 & e) MT & f) CAT \\
\end{tabular}
\caption{Selective input parameters for training BDTs. All parameters are drawn as a function of energy between 2 and 10 keV.The upper panel of each parameter is showing noise-contained data while the lower panel shows calibration signal which contains no noise.}
\label{fig:params}
\end{figure*}

The initial rejection algorithms focus on eliminating the thin
pulses and other pathological events.  We calculate the
balance of the deposited charge from two PMTs (Asymmetry : Eq.~\ref{eq:asym} shown in a) of Fig.~\ref{fig:params}), the charge fraction of 500\,ns to 600\,ns from the first 600\,ns (X1 : Eq.~\ref{eq:x1} shown in b) of Fig.~\ref{fig:params}), the charge fraction of the first 50\,ns to first 600\,ns (X2 : Eq.~\ref{eq:x2} shown in c) of Fig.~\ref{fig:params}),
the charge--weighted mean time of pulses within first 500\,ns (MT : Eq.~\ref{eq:mt} shown in e) of Fig.~\ref{fig:params}),
the total charge (QC : Eq.~\ref{eq:qc}) and the number of pulses (NC : Eq.~\ref{eq:nc}). Boosted Decision Trees (BDTs) were trained
using aforementioned variables.
The electron/gammas signal model is obtained from the energy--weighted
$^{60}$Co multiple-site distributions and data is used for the noise model.
Each crystal is trained for a separate BDT.
Six parameters comparing noise-containing data with $^{60}$Co multiple signals
are shown in Fig.~\ref{fig:params}.

The definitions of each variable used in the rejection algorithms are following, 
\begin{eqnarray}
  Asymmetry = (Q_1 - Q_2)/(Q_1 + Q_2) \label{eq:asym}\\
  X1 = \sum^{600\,ns}_{100\,ns}{q_i}/\sum^{600\,ns}_{0\,ns}{q_i}   \label{eq:x1}\\
  X2 = \sum^{50\,ns}_{0\,ns}{q_i}/\sum^{600\,ns}_{0\,ns}{q_i}   \label{eq:x2} \\
  X3 = \sum^{120\,ns}_{0\,ns}{q_i}/\sum^{600\,ns}_{0\,ns}{q_i}   \label{eq:x3} \\
  X4 = \sum^{150\,ns}_{100\,ns}{q_i}/\sum^{600\,ns}_{0\,ns}{q_i}   \label{eq:x4} \\
  MT = \sum^{500\,ns}_{0\,ns}{q_it_i }/\sum^{500\,ns}_{0\,ns}{q_i}   \label{eq:mt}\\
  MTL= \sum^{30\,ns}_{0\,ns}{q_it_i }/\sum^{30\,ns}_{0\,ns}{q_i}   \label{eq:mtl}\\
  MV = \sum^{1000\,ns}_{0\,ns}{q_it_i^2 }/\sum^{1000\,ns}_{0\,ns}{q_i} - (\sum^{1000\,ns}_{0\,ns}{q_it_i }/\sum^{1000\,ns}_{0\,ns}{q_i})^2   \label{eq:mv} \\
  NC = \text{the number of pulses}   \label{eq:nc} \\
  QC = \sum^{5000\,ns}_{0\,ns}{q_i}  \label{eq:qc} \\
  CAT = \text{time of 95\% charge accumulation} (\sum^{95\%}_{0\%}{q_i}) \label{eq:cat},
\end{eqnarray}
where Q$_{1,2}$ indicates PMT integrated charges within 5~$\mu$s and q$_i$ and t$_i$
are waveform amplitudes and times for each 2\,ns bin, respectively.

For the rejection of the bell-shaped pulse events in the Crystal-1 in the later quarter of the data,
we trained another BDT (BDTA) using the variance of charge--weighted mean time (MV : Eq.~\ref{eq:mv}),
the charge ratios of waveform leading edges (X3 : Eq.~\ref{eq:x3} shown in d) of Fig.~\ref{fig:params} and X4 : Eq.~\ref{eq:x4}), the charge--weighted mean time(MT : Eq.~\ref{eq:mt} and MTL : Eq.~\ref{eq:mtl}), the charge accumulation time (CAT : Eq.~\ref{eq:cat} shown in f) of Fig.~\ref{fig:params}) and the energy of the event.
These effectively identify the shape distortion compared to the regular scintillation signals. Unlike the previous BDT training, early three quarter of the Crystal-1 data as a good data and the later quarter of the data as a noise-contained data are used for signal and background in the training process and later the same training BDT output is applied to all other crystals. 

\begin{figure}
\begin{center}
  \includegraphics[width=0.7\columnwidth]{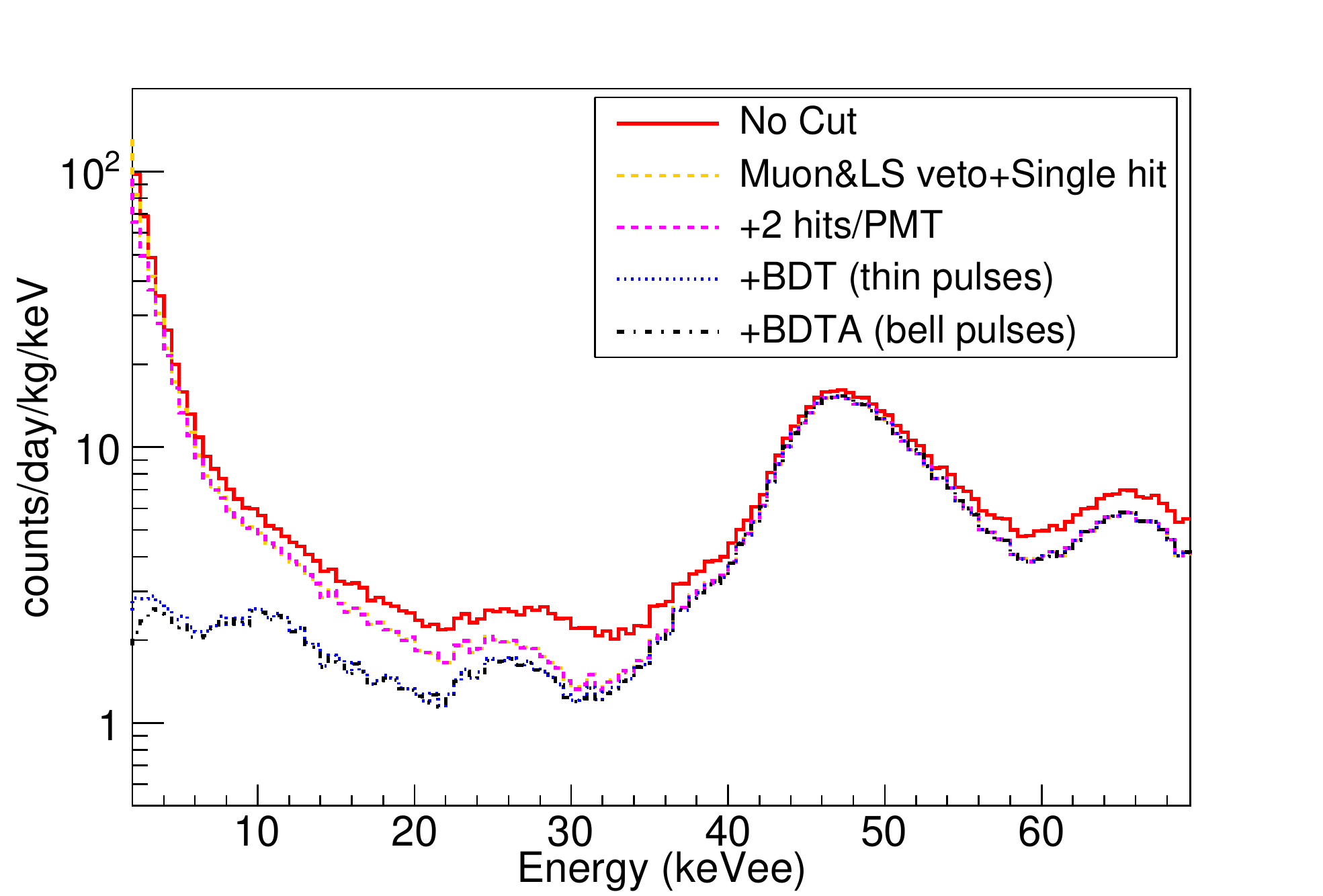}
\end{center}
\caption{Staged event selection versus energy.
  The low-energy spectrum between 2 and 70 keVee is displayed with
  the progression of application of the selection criteria.
  The selection efficiency is uncorrected here.
}
\label{fig:selection}
\end{figure}

The single-site event spectrum is obtained after the application of
the selection criteria and their efficiencies are measured from the
$^{60}$Co multiples. On average, event selection efficiency of 70\% at
2\,keVee is obtained for six crystals except Crystal-5 and Crystal-8
which show higher energy threshold at 4\,keVee and 8\,keVee, respectively,
due to their low light yield.  The progression of event selection for
Crystal-7 is shown in Fig.~\ref{fig:selection}.

The selection efficiency of these event selections are separately checked with neutron-induced nuclear recoils 
obtained with a small rectangular (2~cm$\times$2~cm$\times$1.5~cm) NaI(Tl) crystals from the same ingots of the detector crystals in front of 2.42~MeV mono-energetic neutron beams~\cite{Joo:2018hom}.
The efficiency of nuclear recoil events in energies between 2 and 20 keVee
is consistent with the $^{60}$Co calibration efficiency,
which indicates that the selections do not affect the possible WIMP signal region.

To compare the results of COSINE--100 and DAMA/LIBRA--phase2 we start
from the best--fit analysis of the DAMA/LIBRA--phase2 modulation
effect in terms of NR WIMP effective models in
Ref.~\cite{dama_eft_2018}, as summarized in
Table~\ref{tab:best_fit_values}. In such table each of the NR
couplings is assumed to be the only term in the effective Hamiltonian
of Eq.~(\ref{eq:H}) and the $\chi^2$:

\begin{equation}
\chi^2(m_{\chi},\sigma_p,r)=\sum_{k=1}^{15} \frac{\left [S_{m,k}(m_{\chi},\sigma_p,r)-S^{exp}_{m,k} \right ]^2}{\sigma_k^2}
  \label{eq:chi2}
  \end{equation}

\noindent (where we consider 14 energy bins, of 0.5 keVee width, from
1 keVee to 8 keVee, and one high--energy control bin from 8 keVee to
16 keVee) is minimized in terms of the WIMP mass $m_{\chi}$, the
neutron--over--proton coupling ratio $r\equiv c^n/c^p$ and of the
effective cross section $\sigma_p$ as defined in
Eq.~(\ref{eq:conventional_sigma}).  In Eq.~(\ref{eq:chi2})
$S_{m,k}\equiv S_{m,[E_k^{\prime},E_{k+1}^{\prime}]}$ is given by
Eq.~(\ref{eq:sm}), while $S_{m,k}^{exp}$ and $\sigma_k$ represent the
modulation amplitudes and 1--$\sigma$ uncertainties as measured by
DAMA/LIBRA--phase2~\cite{Bernabei:2018yyw} and reported in Table~\ref{tab:dama_data}.

\begin{table}
  \begin{center}
	\setlength{\tabcolsep}{8pt}
	\begin{tabular}{ccc}
		\hline
		Energy (keVee) & $S_{m,k}$ & $\sigma_k$ \\
		\hline\hline
		$1.0 - 1.5$  & 0.0242      & 0.0056       \\
		$1.5 - 2.0$  & 0.0211      & 0.0043      \\
		$2.0 - 2.5$  & 0.0179      & 0.0023      \\
		$2.5 - 3.0$  & 0.0197      & 0.0030       \\
		$3.0 - 3.5$  & 0.0186      & 0.0027       \\
		$3.5 - 4.0$  & 0.0110      & 0.0026       \\
		$4.0 - 4.5$  & 0.0109      & 0.0021       \\
		$4.5 - 5.0$  & 0.0032      & 0.0019       \\
		$5.0 - 5.5$  & 0.0065      & 0.0019       \\
		$5.5 - 6.0$  & 0.0059      & 0.0019       \\
		$6.0 - 6.5$  & 0.0010      & 0.0016       \\
		$6.5 - 7.0$  & 0.0008      & 0.0017       \\
		$7.0 - 7.5$  & 0.0009      & 0.0016       \\
		$7.5 - 8.0$  & 0.0009      & 0.0016       \\
		$8.0 - 16.0$ & 0.0003      & 0.0004       \\                                        
		\hline
	\end{tabular}
	\caption{Combination of the DAMA/LIBRA--phase1 and the DAMA/LIBRA--phase2
          measurements for the modulation amplitudes $S_{m,k}$ with
          statistical errors $\sigma_k$ used in the present analysis
          (from Ref.~\cite{Bernabei:2018yyw}).}
	\label{tab:dama_data}
        \end{center}
\end{table}

As shown in Table~\ref{tab:best_fit_values} for each NR coupling two
local minima of the $\chi^2$ are found (low and high-- mass)
with the low--mass solution corresponding in all cases to the absolute
minimum. With the exception of the high--mass minima for the ${\cal
  O}_4$ and ${\cal O}_7$ operators, all the $\chi^2$ minima of
Table~\ref{tab:best_fit_values} are acceptable with $15-3$ degrees of
freedom. In particular, the DAMA/LIBRA--phase2 has lowered the
energy threshold to 1 keVee, implying that it is sensitive to
WIMP--iodine scattering events for WIMP masses below $\simeq$ 20 GeV/c$^2$.
In the SI case this requires to highly tune the parameters to suppress
the iodine contribution, in order to avoid an otherwise too steeply
increasing spectrum at low energy of the modulation amplitudes
compared to the data~\cite{freese_2018}. On the other hand, if the
WIMP--nucleus cross section is driven by other operators the fine
tuning required to suppress iodine is reduced and/or the hierarchy
between the WIMP--iodine and the WIMP--sodium cross section is less
pronounced in the first place~\cite{dama_eft_2018}.

\begin{figure*}
\begin{tabular}{ccc}
  \includegraphics[width=0.32\textwidth]{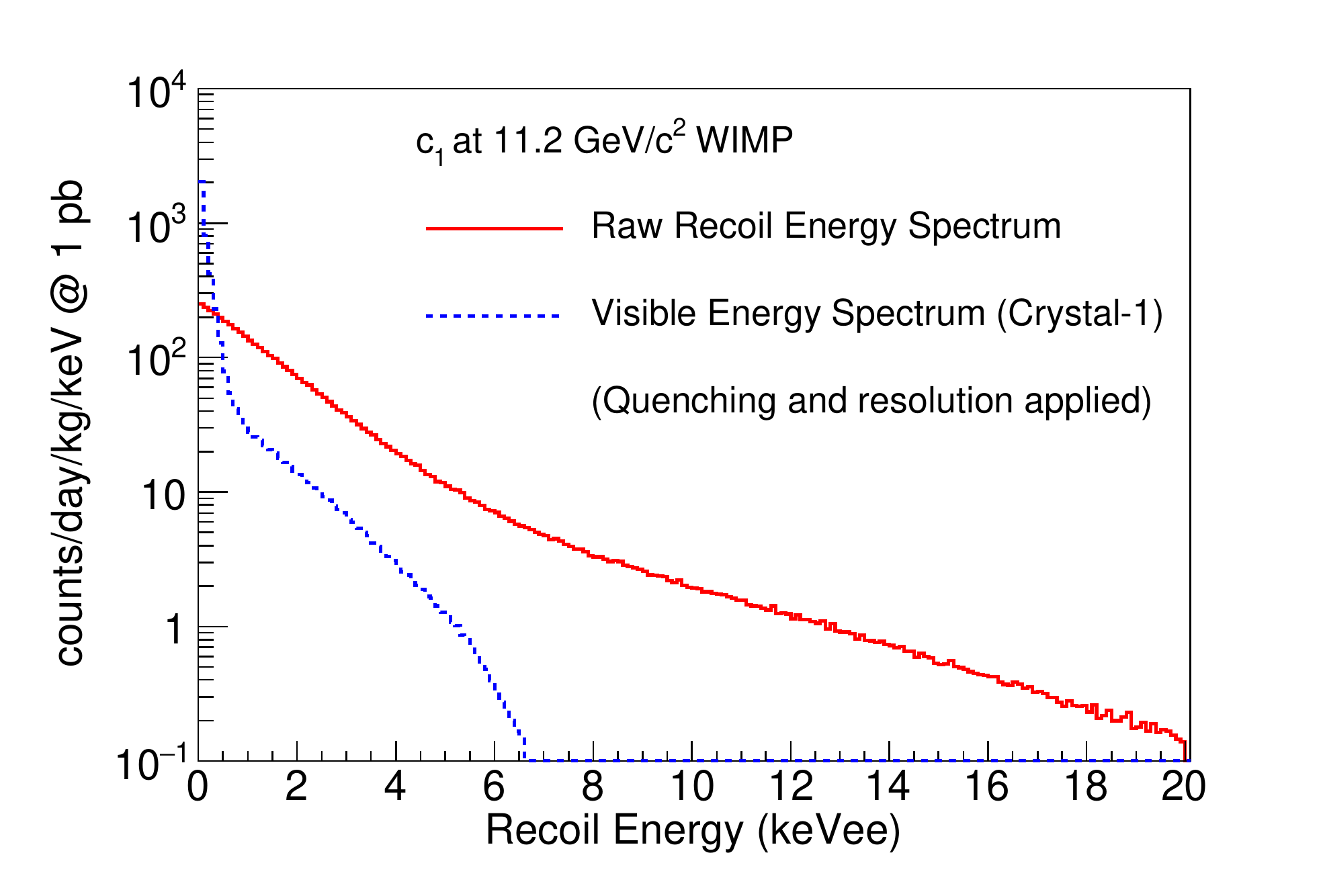} &
  \includegraphics[width=0.32\textwidth]{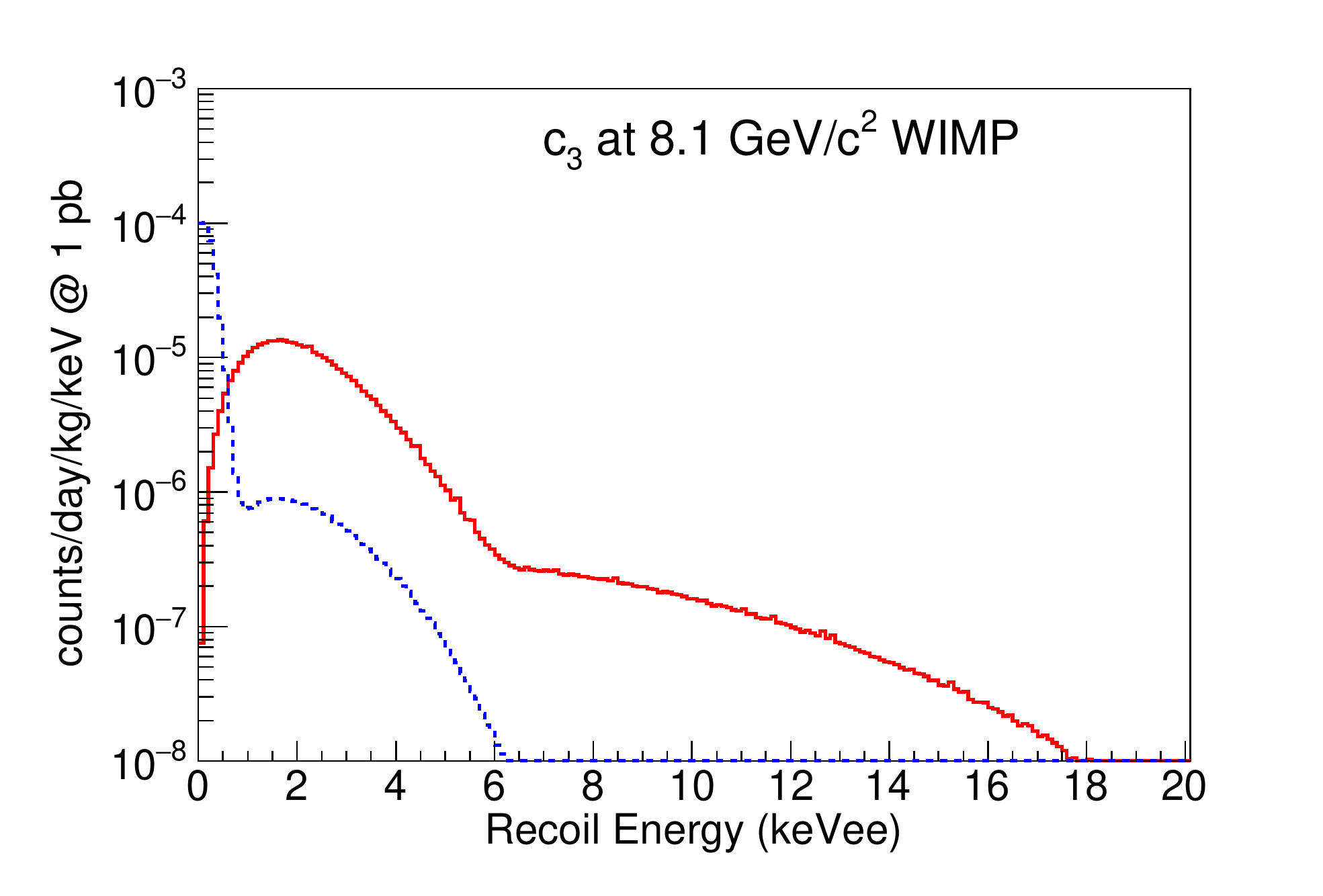} \\
  \includegraphics[width=0.32\textwidth]{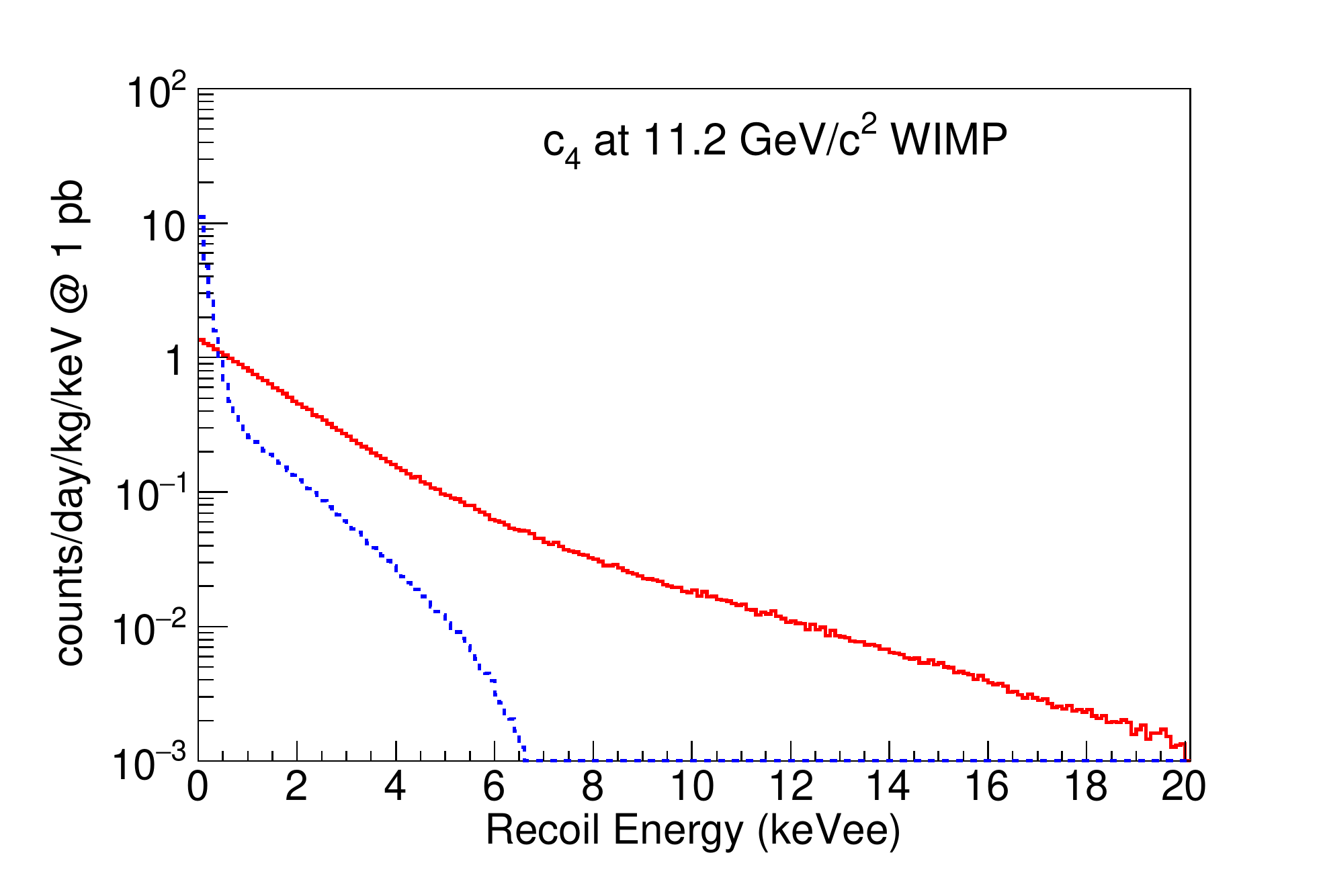} &
  \includegraphics[width=0.32\textwidth]{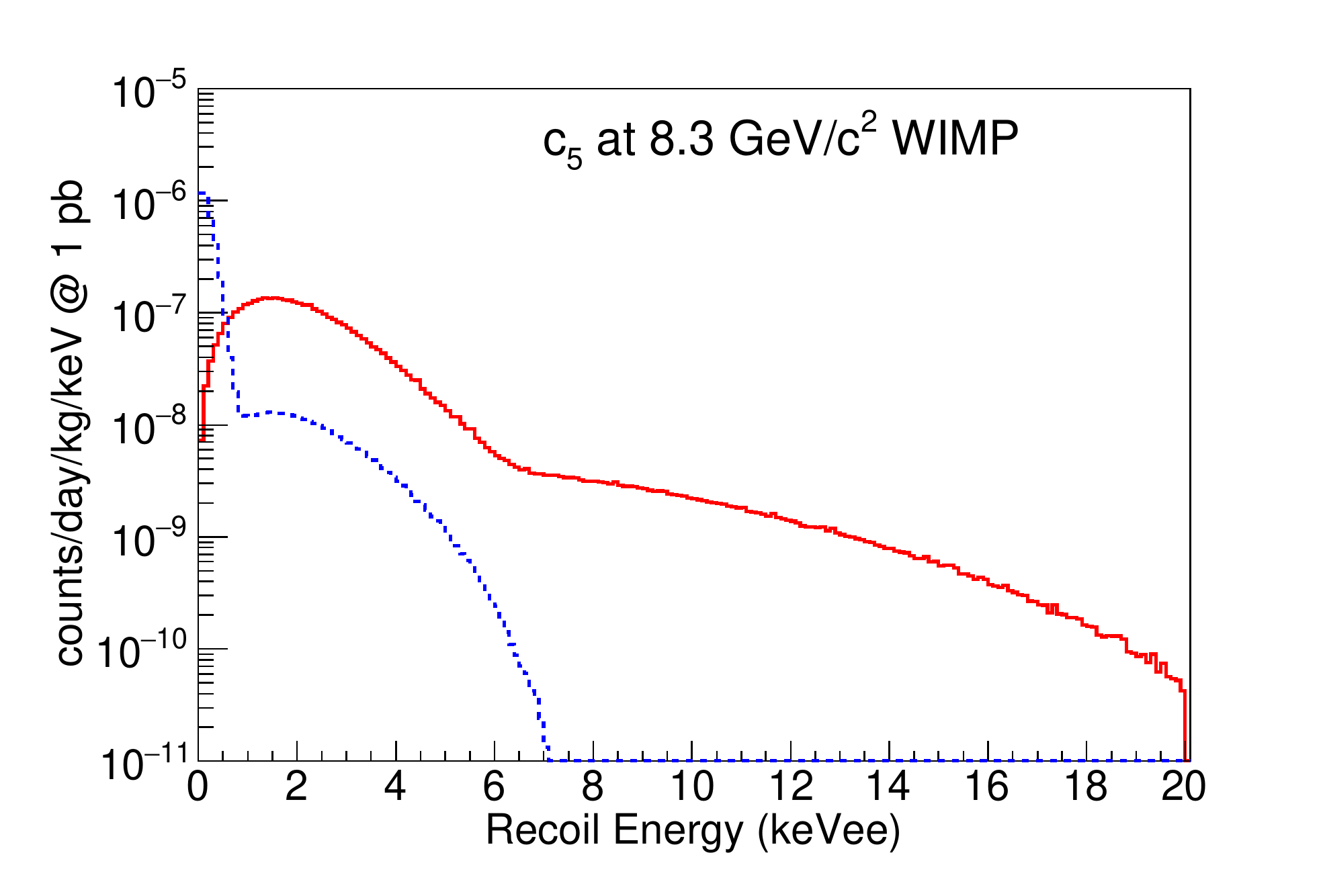} &
  \includegraphics[width=0.32\textwidth]{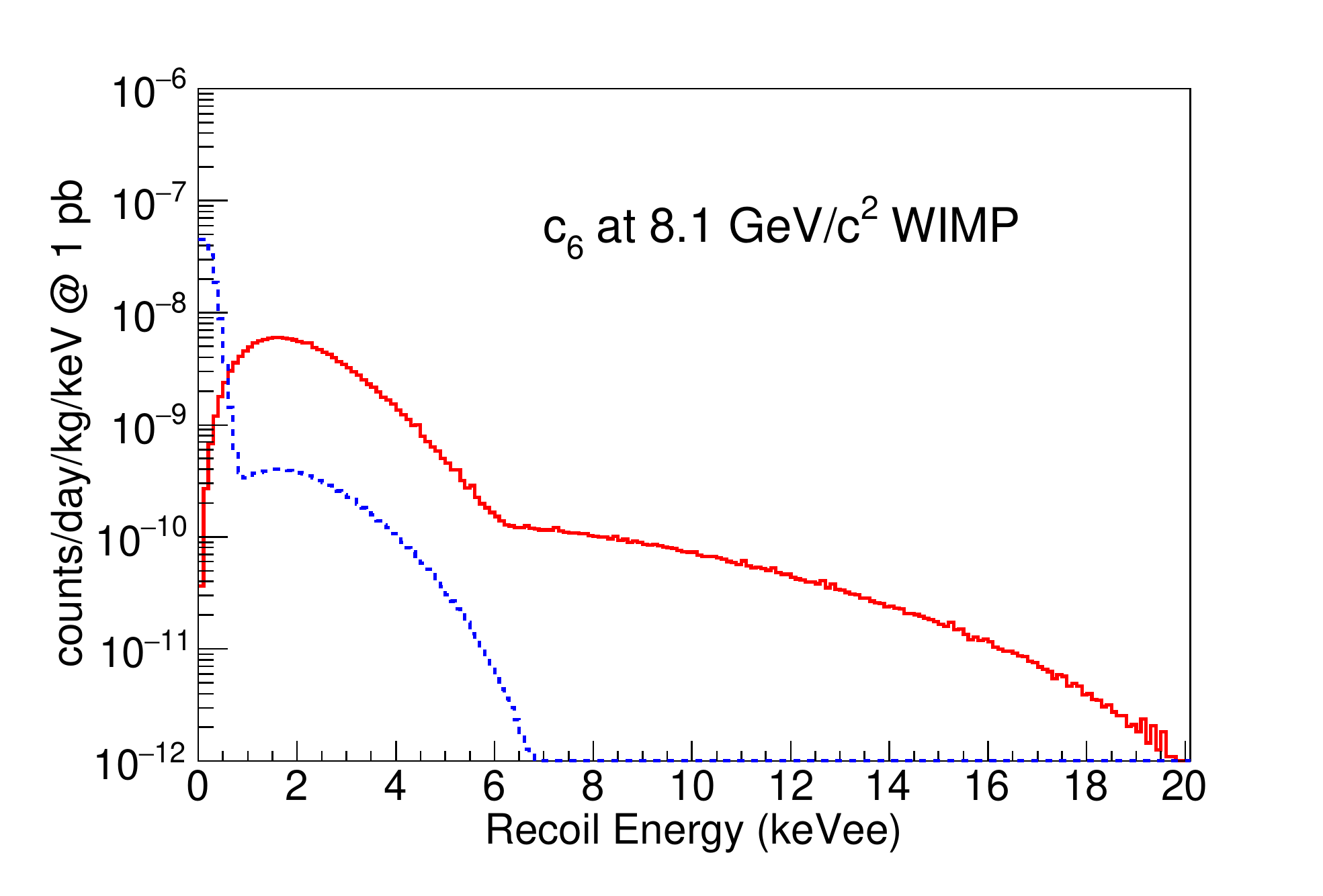} \\
  \includegraphics[width=0.32\textwidth]{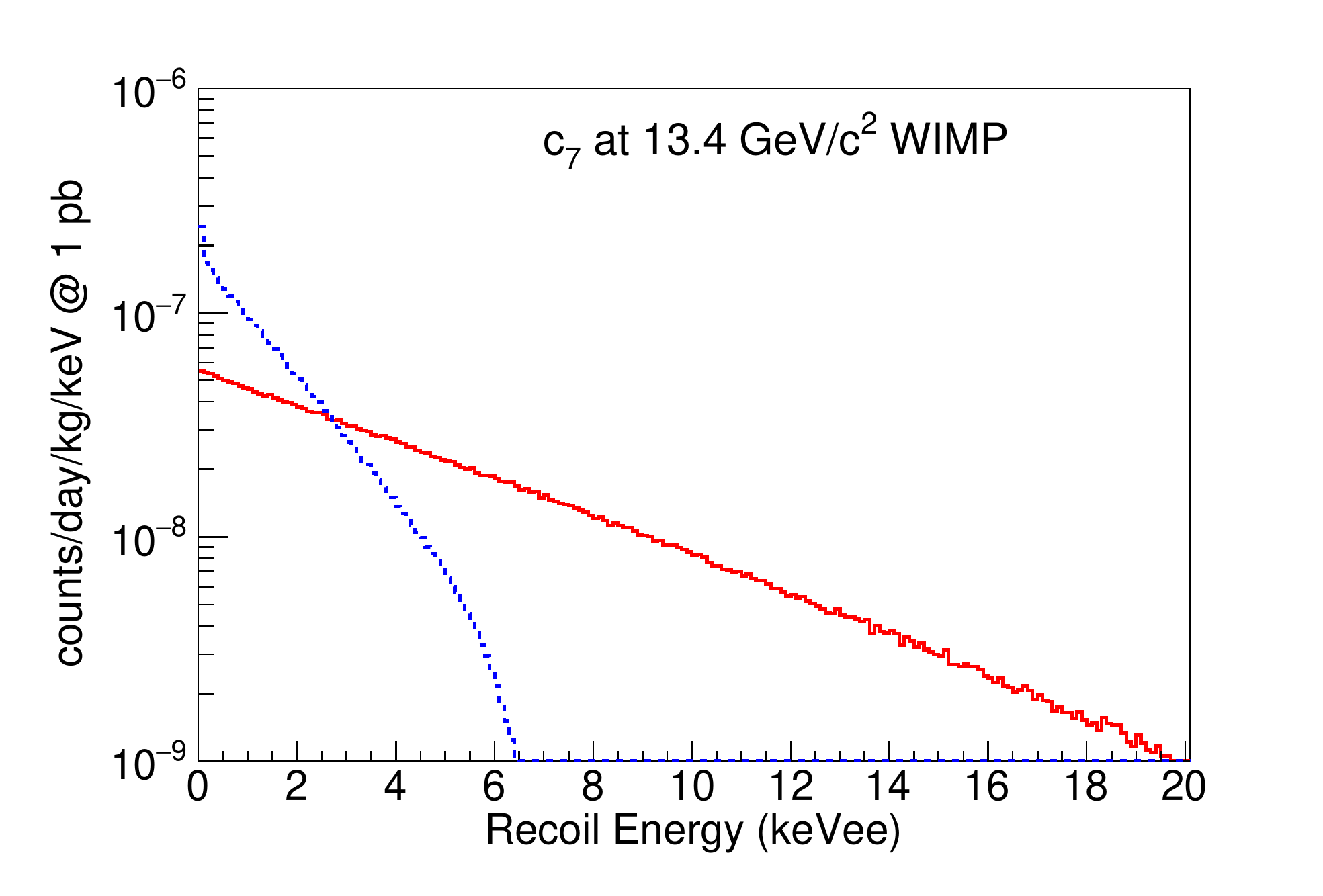} &
  \includegraphics[width=0.32\textwidth]{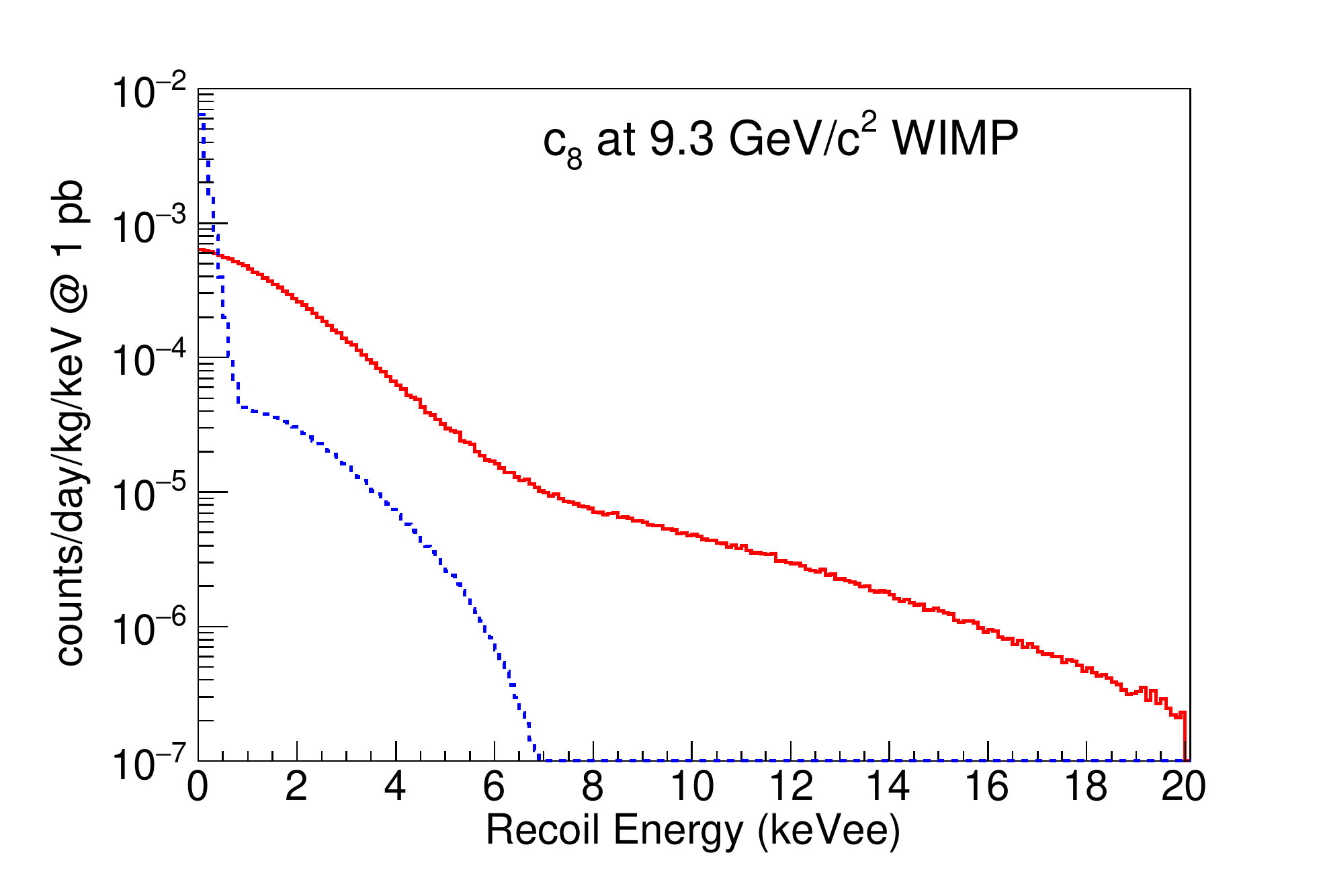} &
  \includegraphics[width=0.32\textwidth]{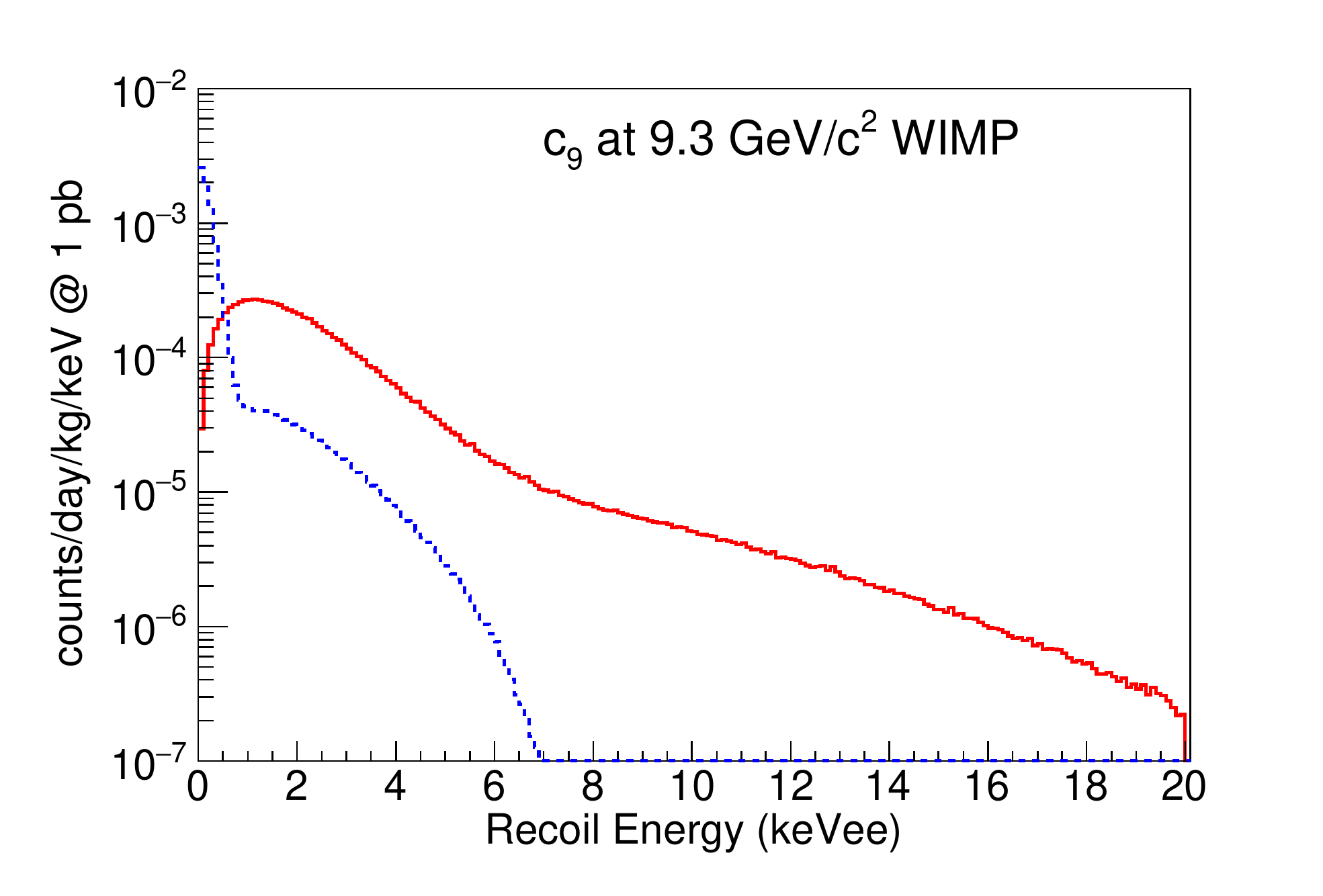} \\
  \includegraphics[width=0.32\textwidth]{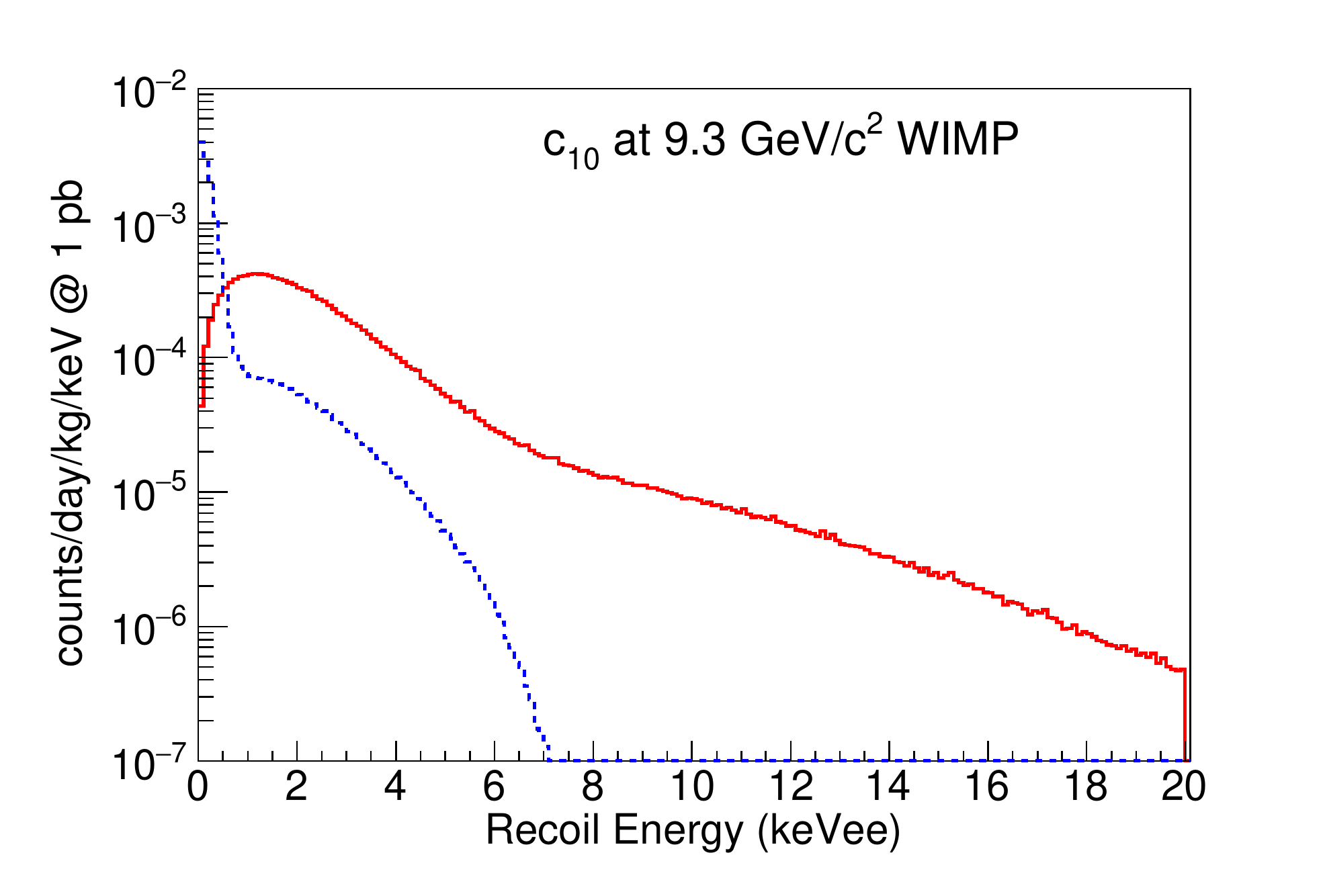} &
  \includegraphics[width=0.32\textwidth]{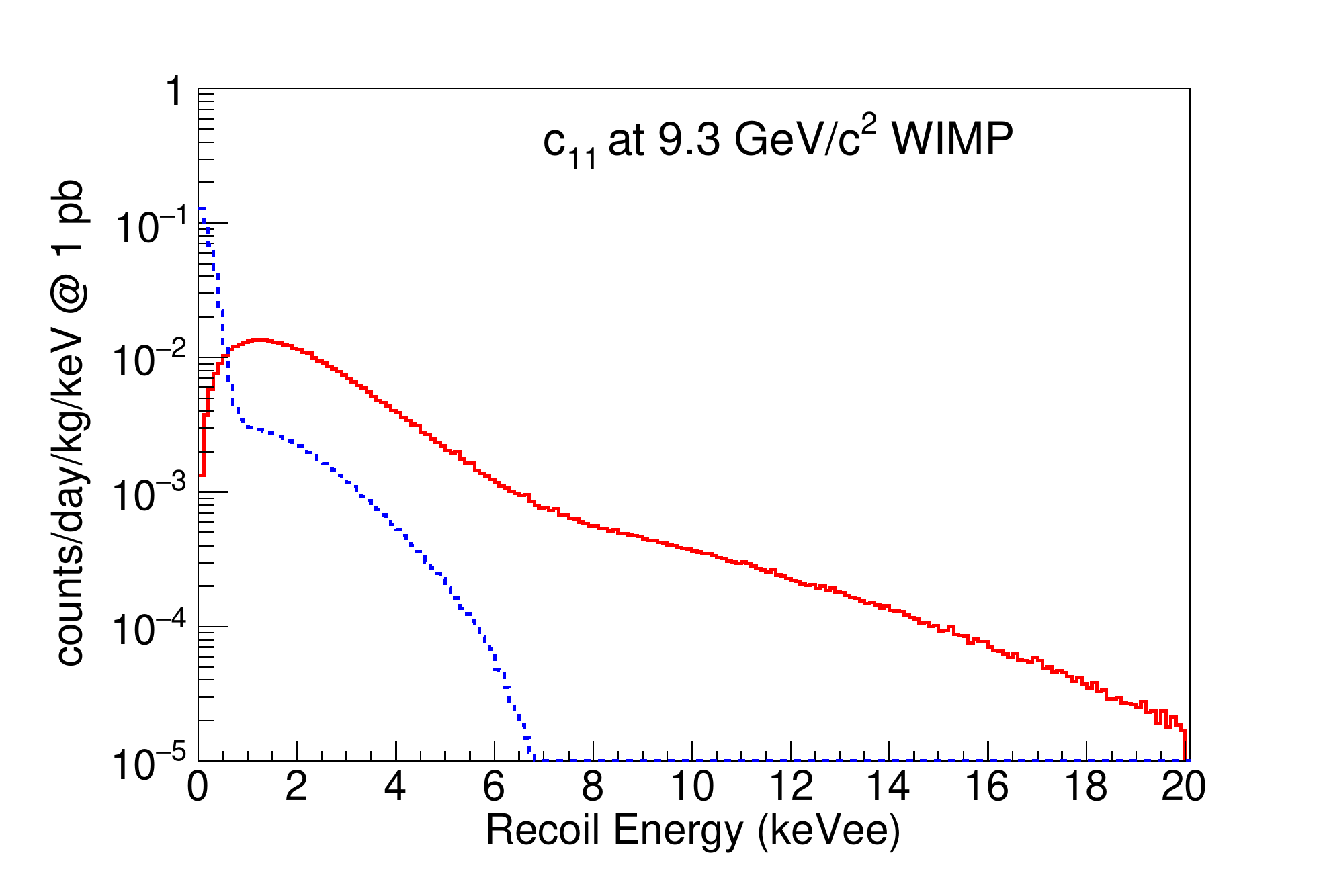} &
  \includegraphics[width=0.32\textwidth]{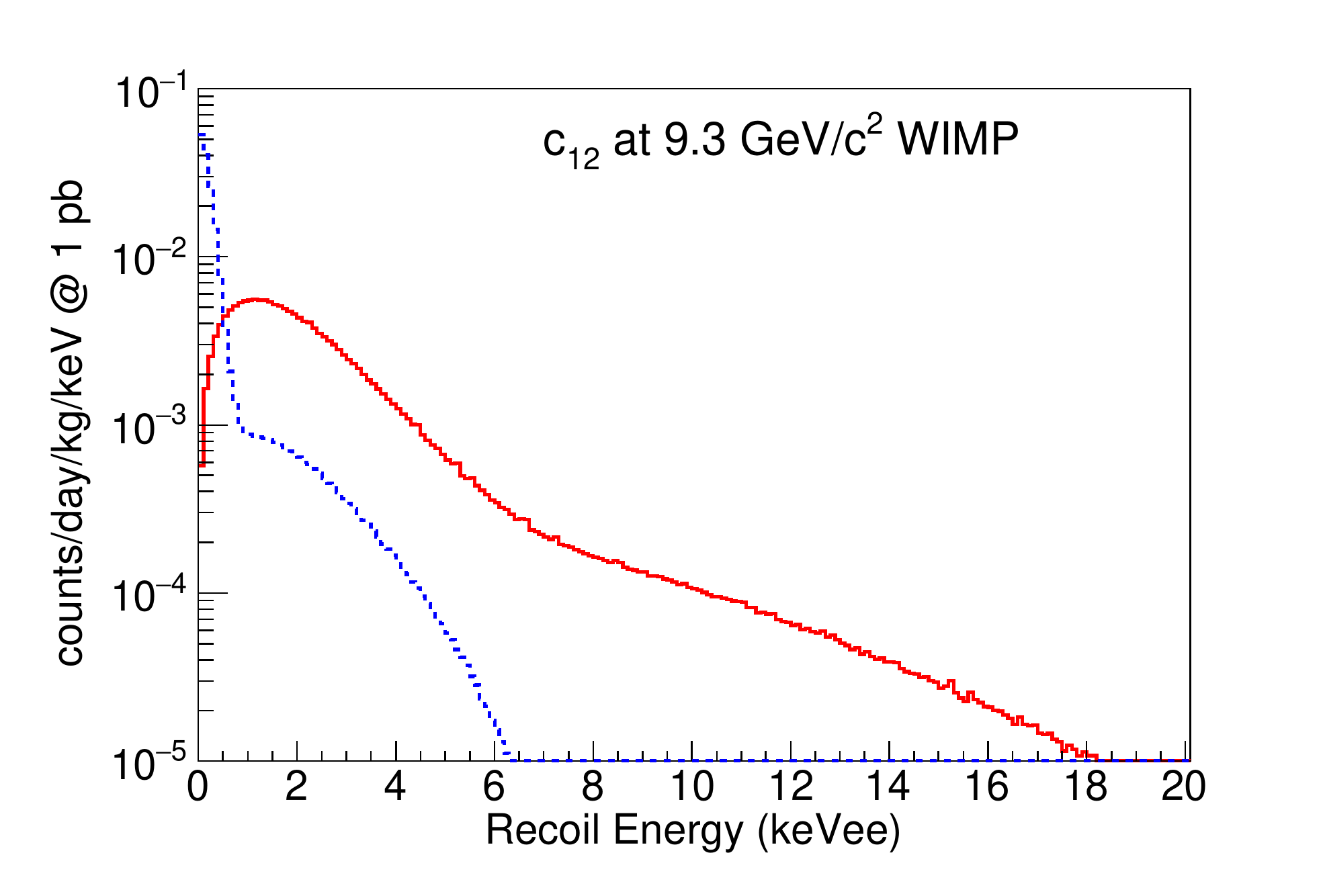} \\
  \includegraphics[width=0.32\textwidth]{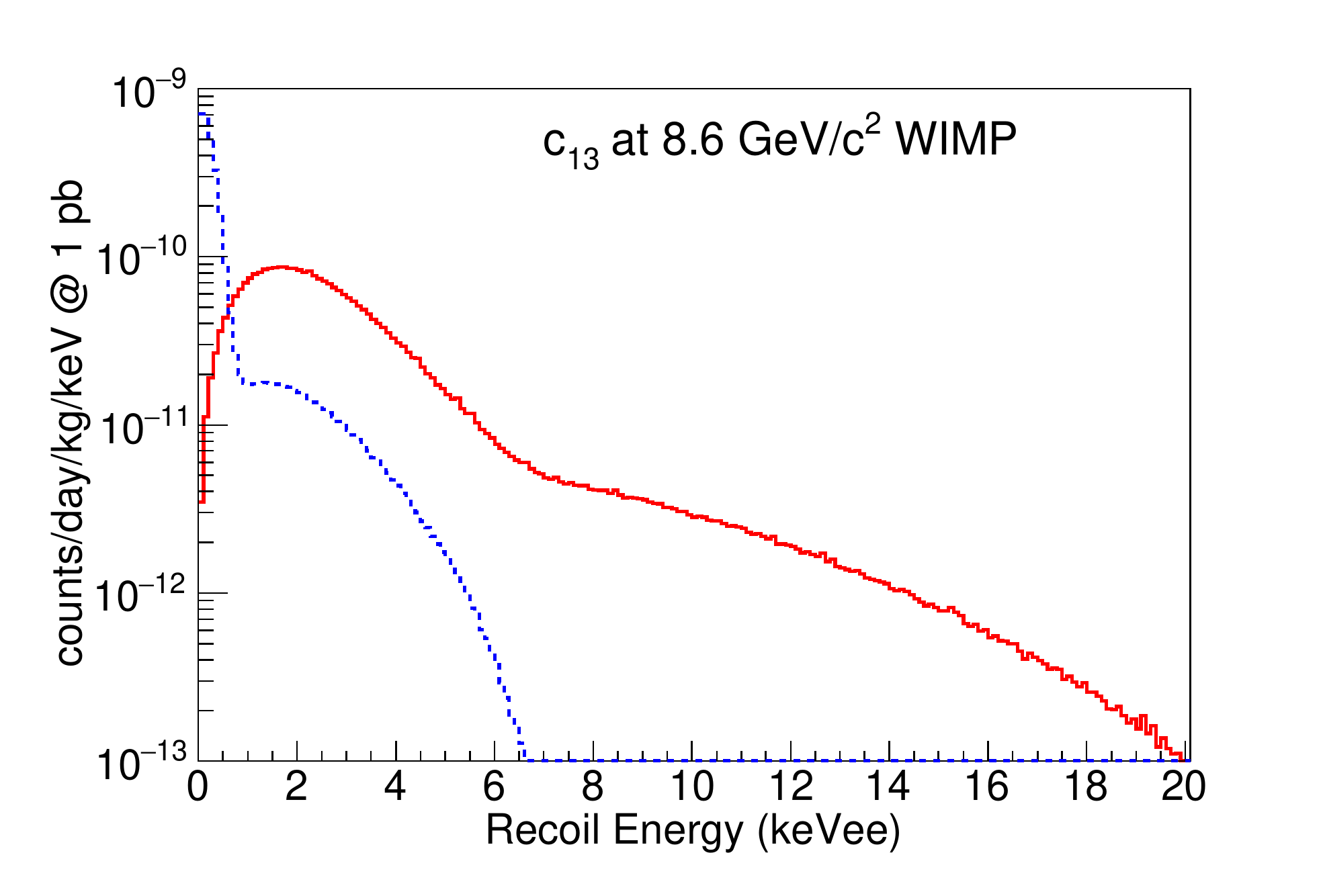} &
  \includegraphics[width=0.32\textwidth]{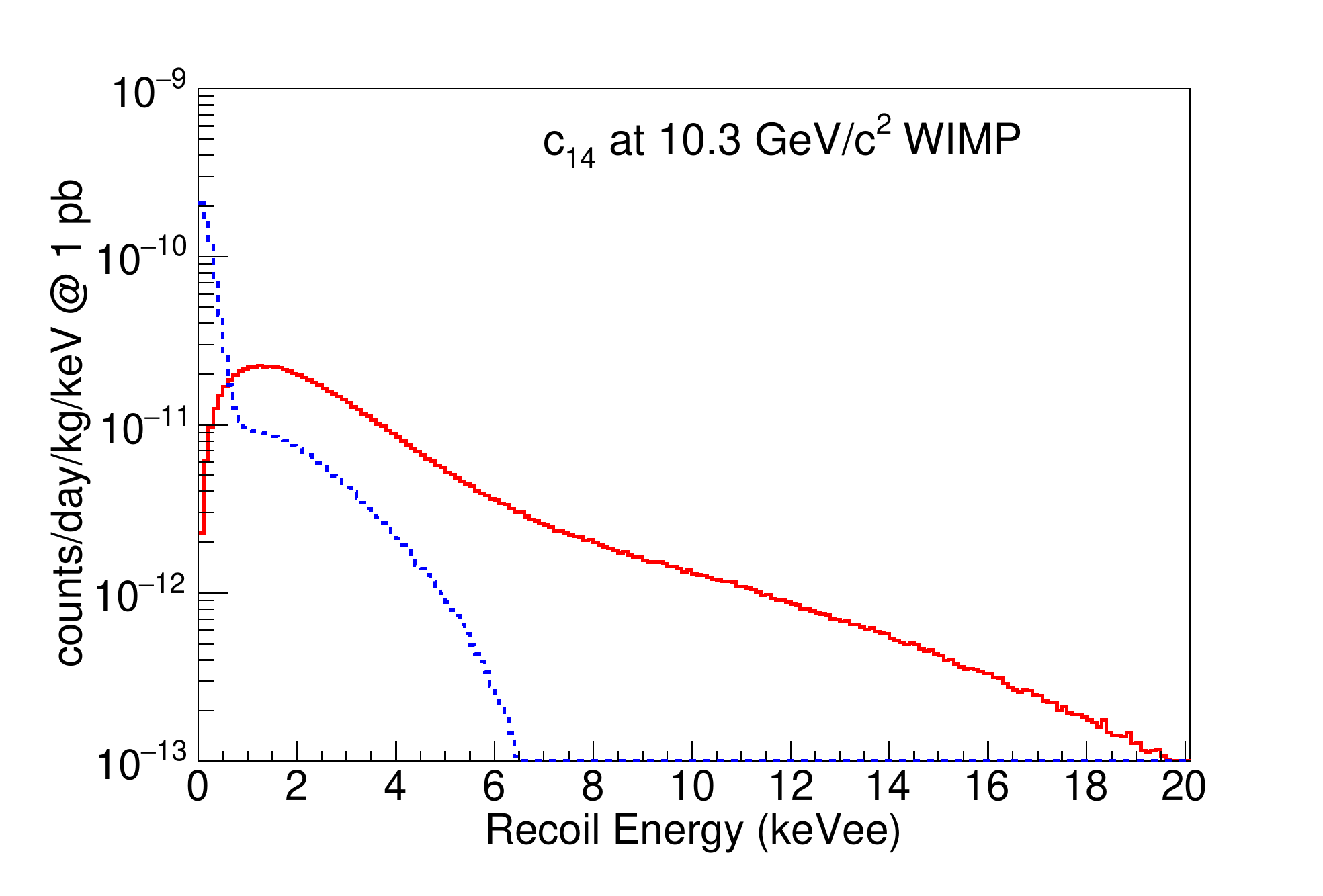} &
  \includegraphics[width=0.32\textwidth]{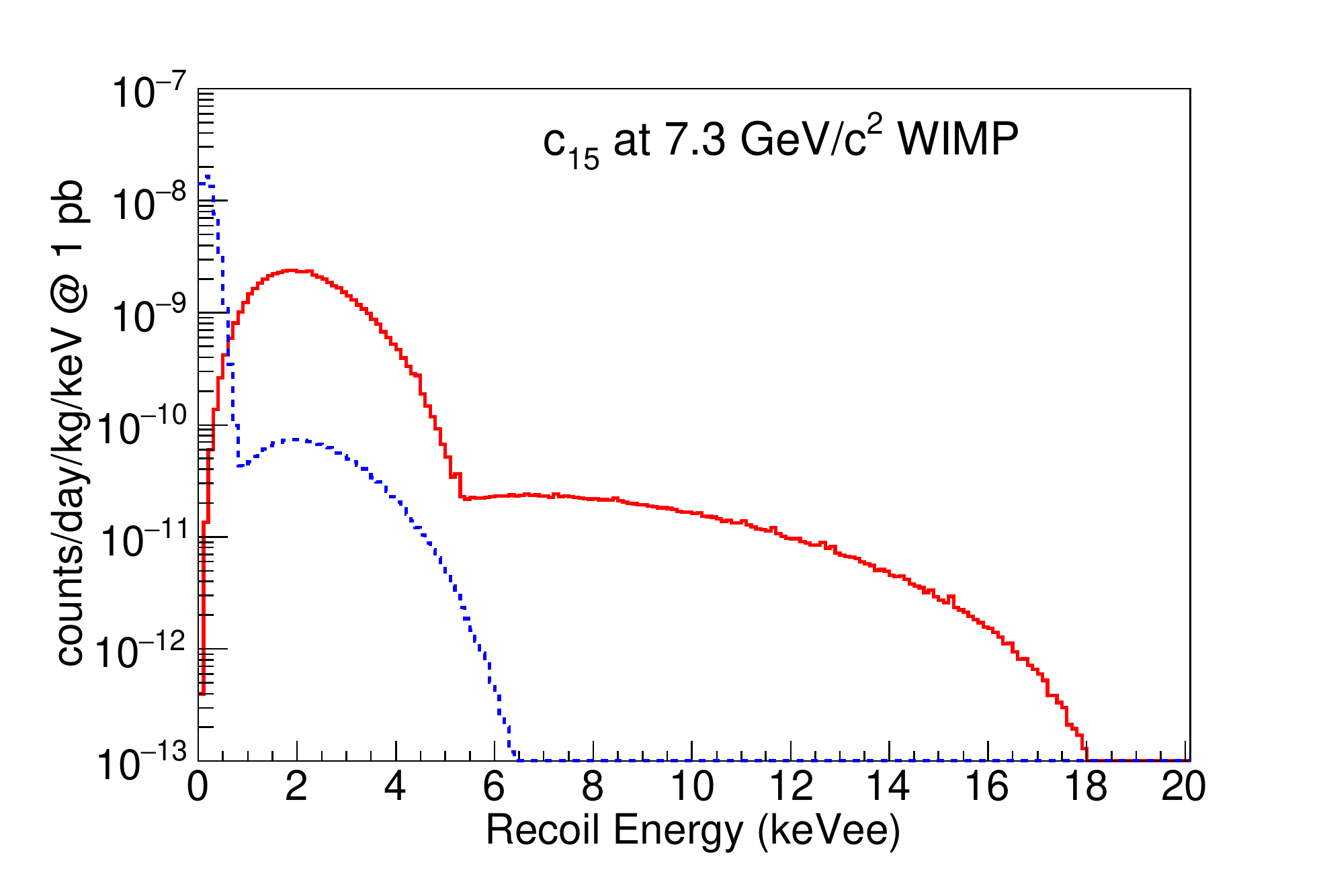} \\
\end{tabular}
\caption{Count rate versus recoil energy spectrum.  Raw energy spectra
  for the case of couplings c$_1$--c$_{15}$ at their best fit low-mass
  positions are compared to the visible energy spectra at Crystal-1
  where the DAMA quenching factor and the crystal resolution are
  applied.  The event rate is normalized to a unit of
  counts/day/kg/keV assuming 1\,picobarn cross-section.  The largest
  impact is from the quenching factor.  After the 2\,keVee threshold
  is applied for the analysis, the effect from the iodine component is
  largely negligible.}
\label{fig:recoil}
\end{figure*}

The raw and observed WIMP spectra corresponding to the
best--fit values of Table~\ref{tab:best_fit_values} for operators
$c_1$--$c_{15}$ are shown in Fig.~\ref{fig:recoil}. The raw spectra
are generated by using Eq.~(\ref{eq:dr_de}), while the energy
resolutions obtained from individual crystal measurements and the DAMA
quenching factors (0.3 for Na and 0.09 for I) are applied to
the raw signal to create the WIMP signal models.

To test the presence of a WIMP signal in the COSINE--100 data
that is consistent to the
modulation effect measured by DAMA/LIBRA, we generated WIMP
spectra for 5 mass values centered around each of the low--mass and
high--mass modulation minima of Table~\ref{tab:best_fit_values}.

To extract the WIMP signal from our data, a Bayesian approach
with a likelihood function based on the Poisson probability is used.
A WIMP is
not expected to have multiple scatterings within our detector volume,
so our WIMP search window is the 2--20 keVee region in the single--hit
spectrum where the average event rate of all crystal is recorded to
3.5 counts/day/kg/keV.
All crystals are fitted
together with crystal-specific background models and a single WIMP
signal at a given mass.  The constraints are applied as 1\,$\sigma$
Gaussian priors for those
obtained from the background understanding.  Similarly, systematic
parameters that change the shape of the background distributions are
added as nuisance parameters.

\begin{figure}
\begin{center}
  \includegraphics[width=0.9\columnwidth]{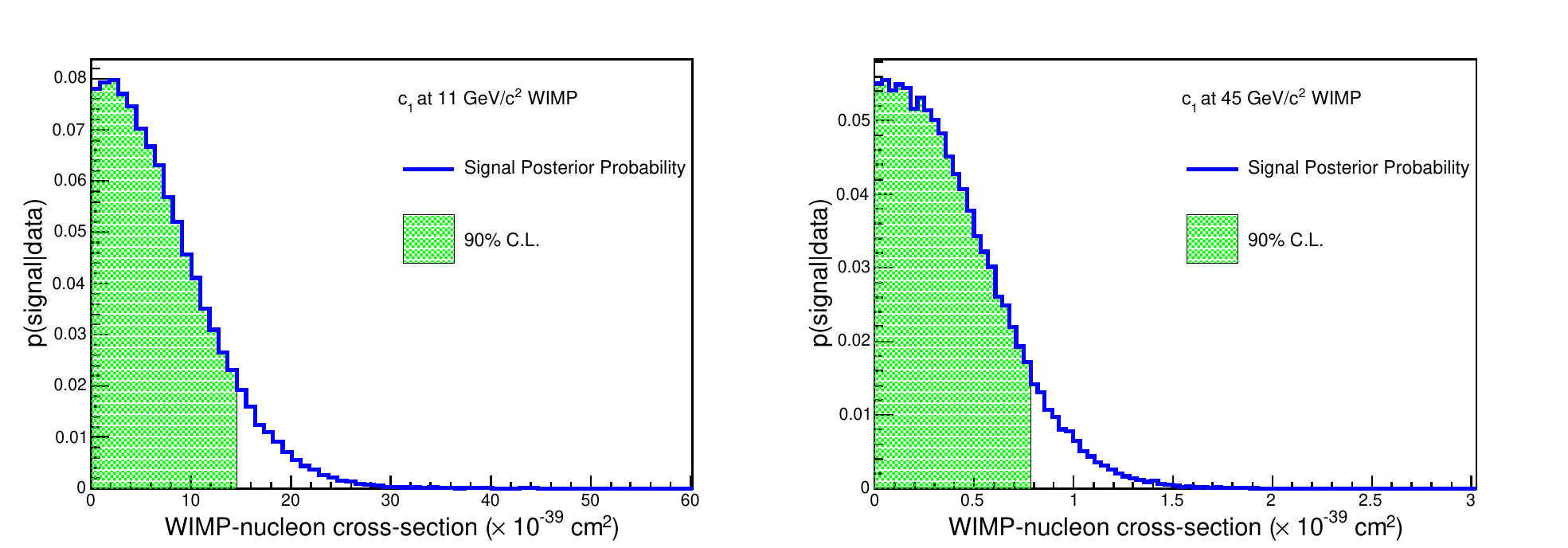}
  \includegraphics[width=0.9\columnwidth]{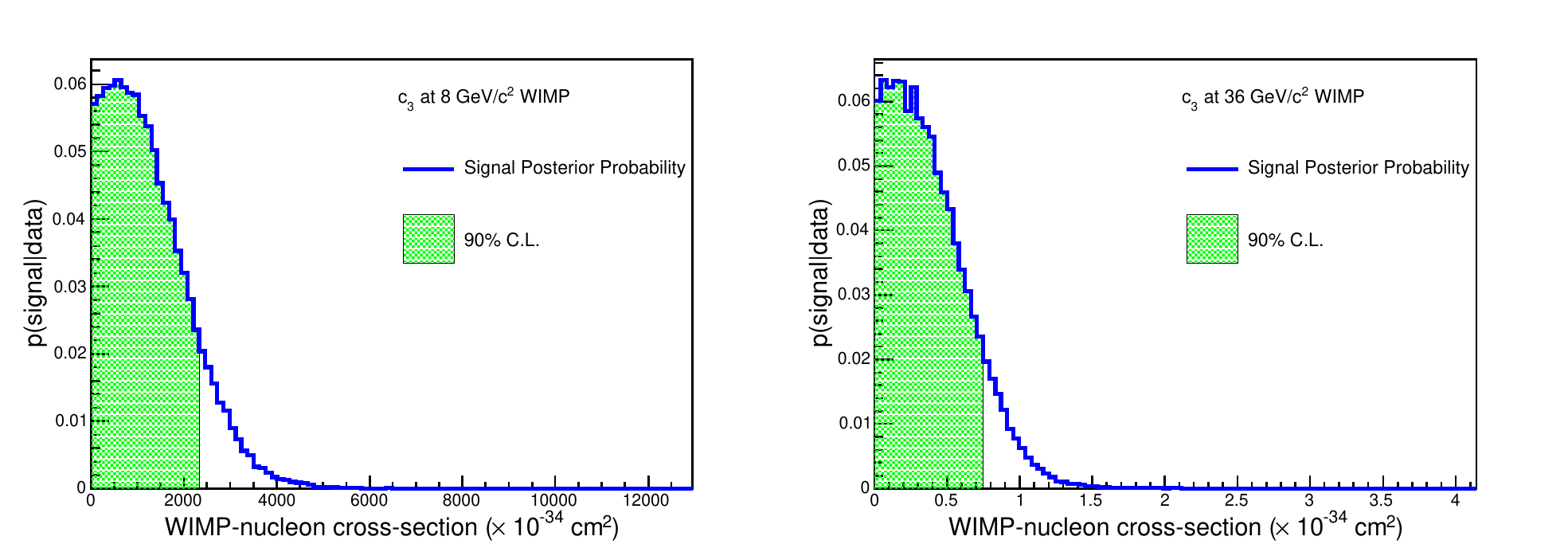}
  \includegraphics[width=0.9\columnwidth]{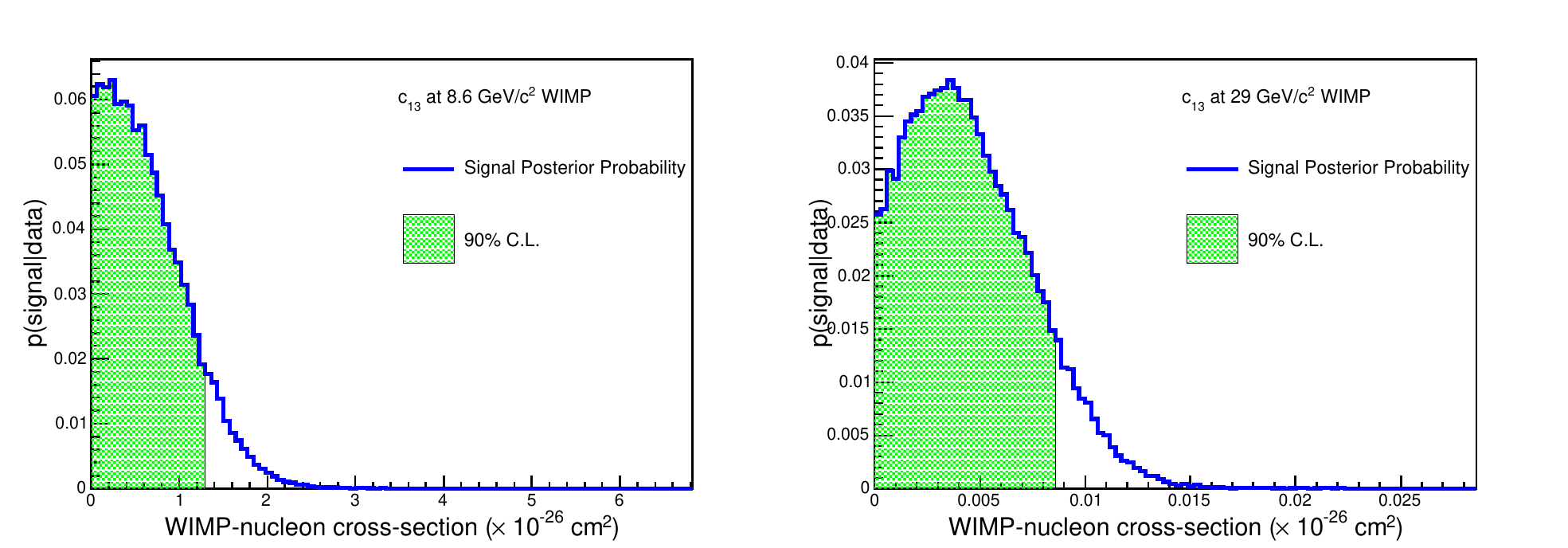}
\end{center}
\caption{Posterior probability versus signal strength for selective operators;
  $c_1$, $c_3$ and $c_{13}$.
  A posterior probability of a signal given data are drawn as
  a function of WIMP-nucleon cross-section scaled at 10$^{-39}$cm$^2$, 10$^{-34}$cm$^2$
  and 10$^{-26}$cm$^2$ respectively for $c_1$, $c_3$ and $c_{13}$.
  No signal is observed for those WIMP mass points.  Therefore, 90\% confidence
  level (C.L.) limit was obtained by integrating the probability from
  zero.
  For the high mass case of $c_{13}$, a positive 0.9\,$\sigma$
  fluctuation in the mean of the probability is observed implying a
  weaker bound.
}
\label{fig:post}
\end{figure}

In this way for each NR coupling and WIMP mass value we produce a
posterior probability of the effective cross section $\sigma_{p}$.
Examples of the posterior probability versus $\sigma_p$ are provided
in Fig.~\ref{fig:post} for the couplings $c_1$, $c_3$ and
$c_{13}$. For all the couplings $c_1$--$c_{15}$ we find no signal, so
a 90\% confidence level (C.L.) upper limit is obtained by integrating
the posterior probability from zero.

\begin{figure}
\begin{center}
  \includegraphics[width=0.32\columnwidth]{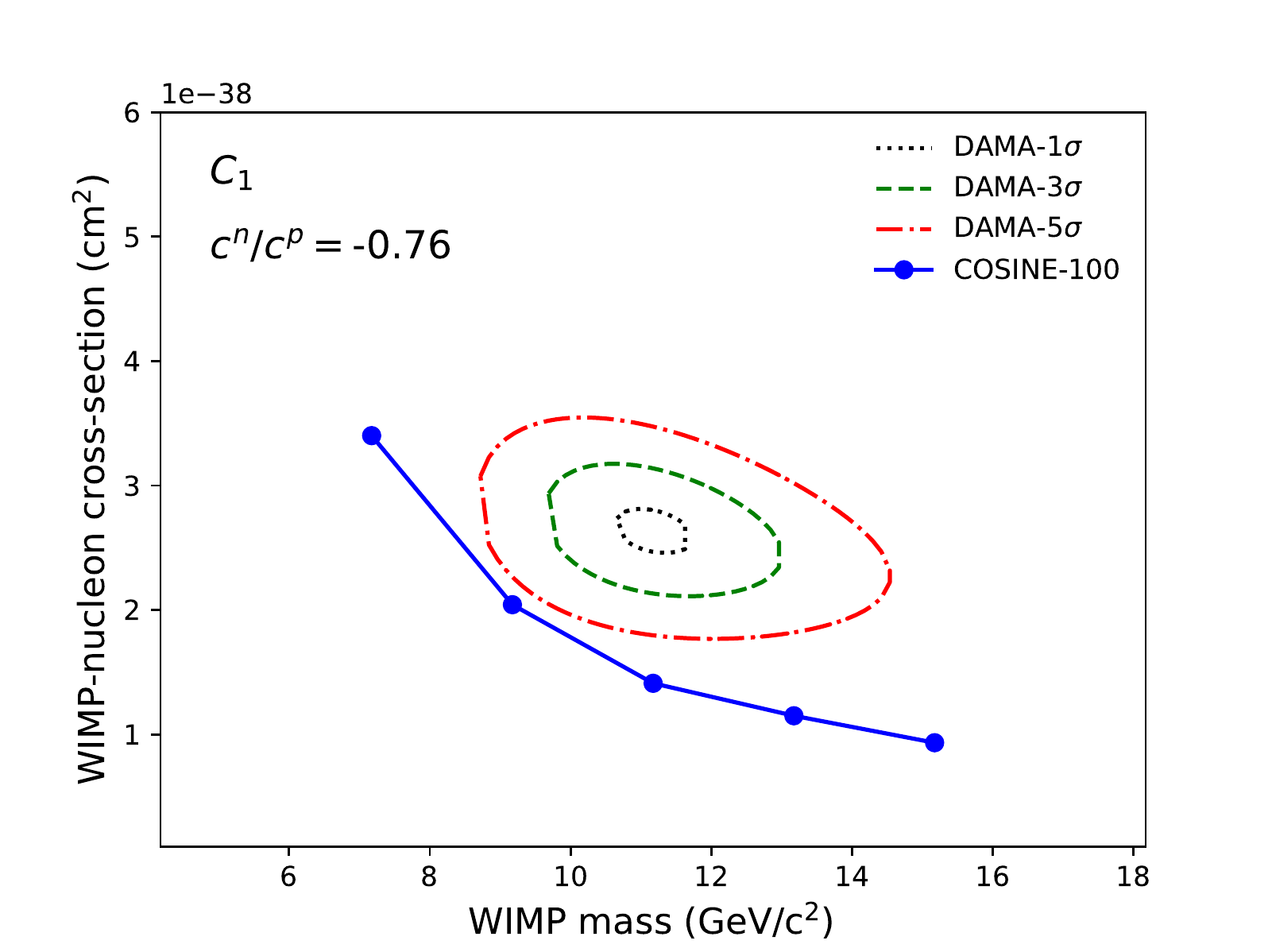}
  \includegraphics[width=0.32\columnwidth]{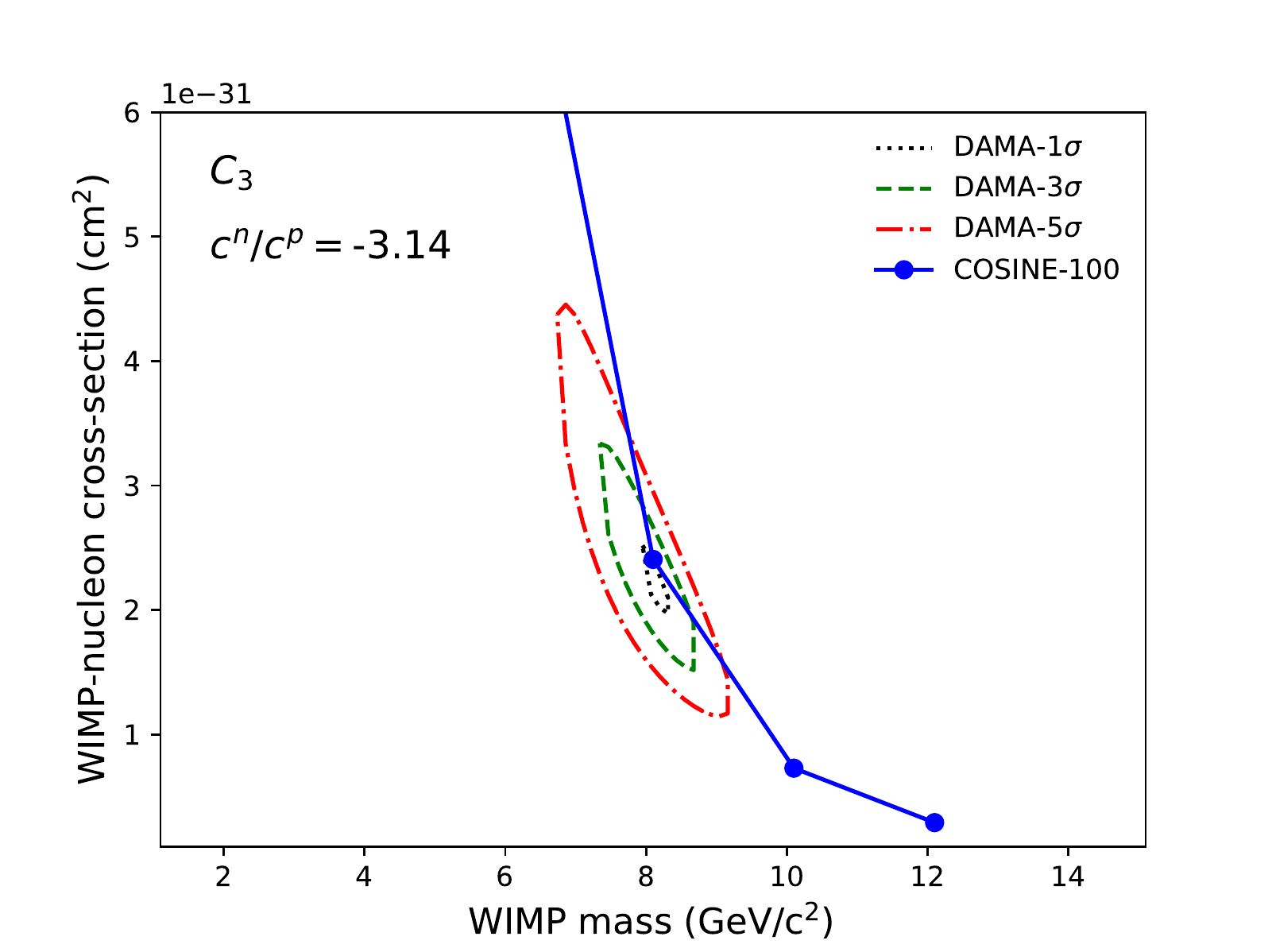}
  \includegraphics[width=0.32\columnwidth]{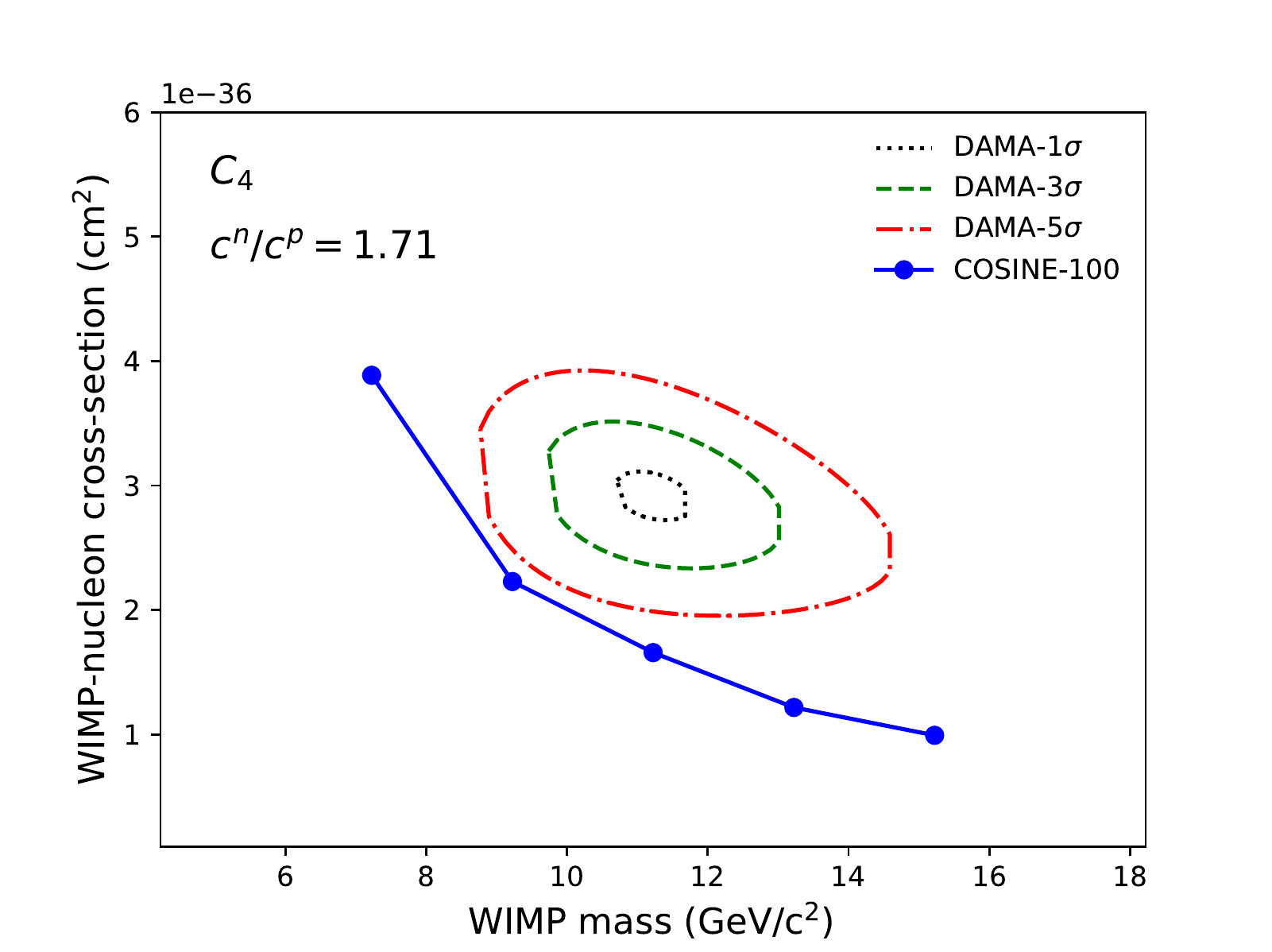}
  \includegraphics[width=0.32\columnwidth]{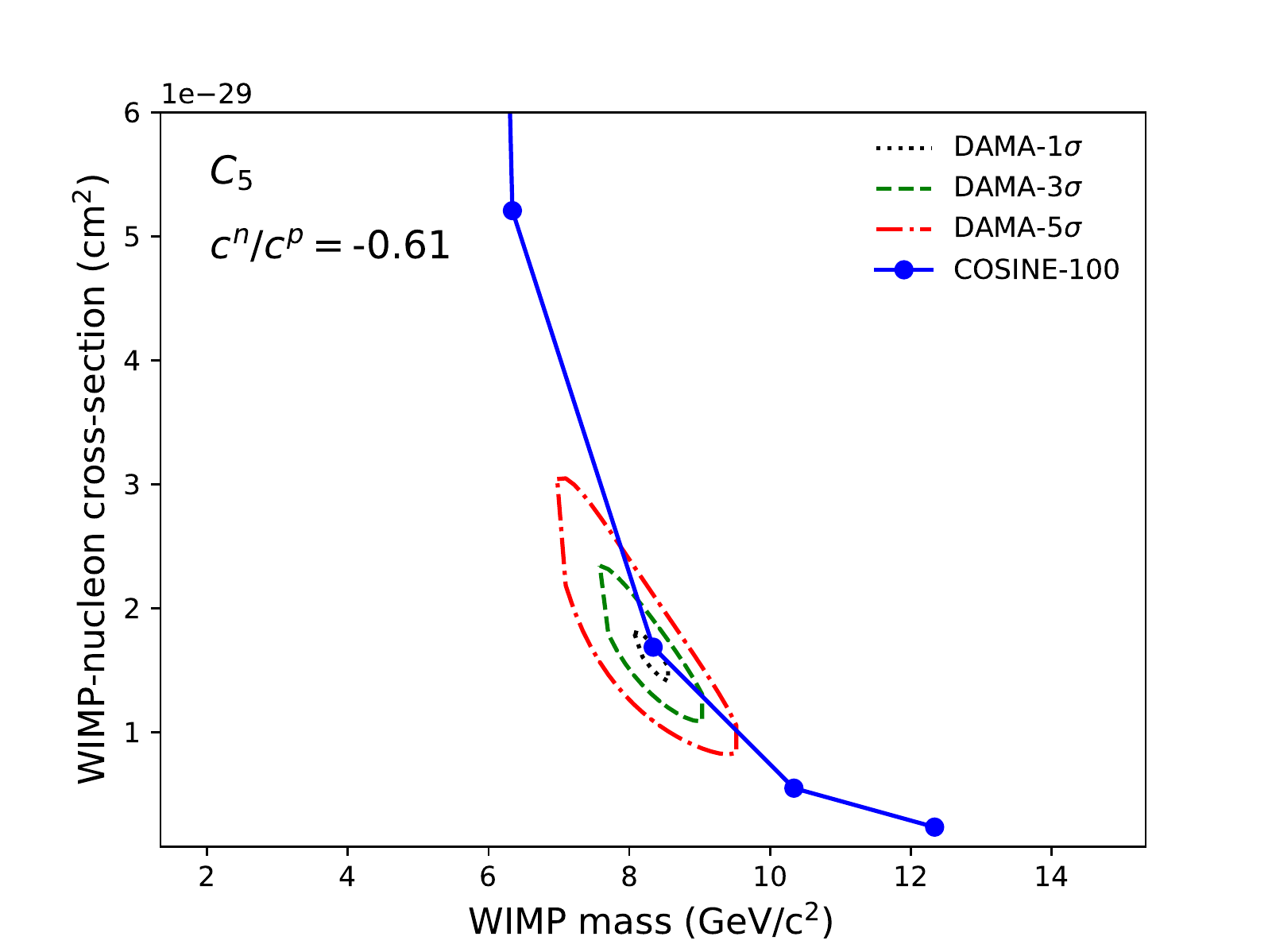}
  \includegraphics[width=0.32\columnwidth]{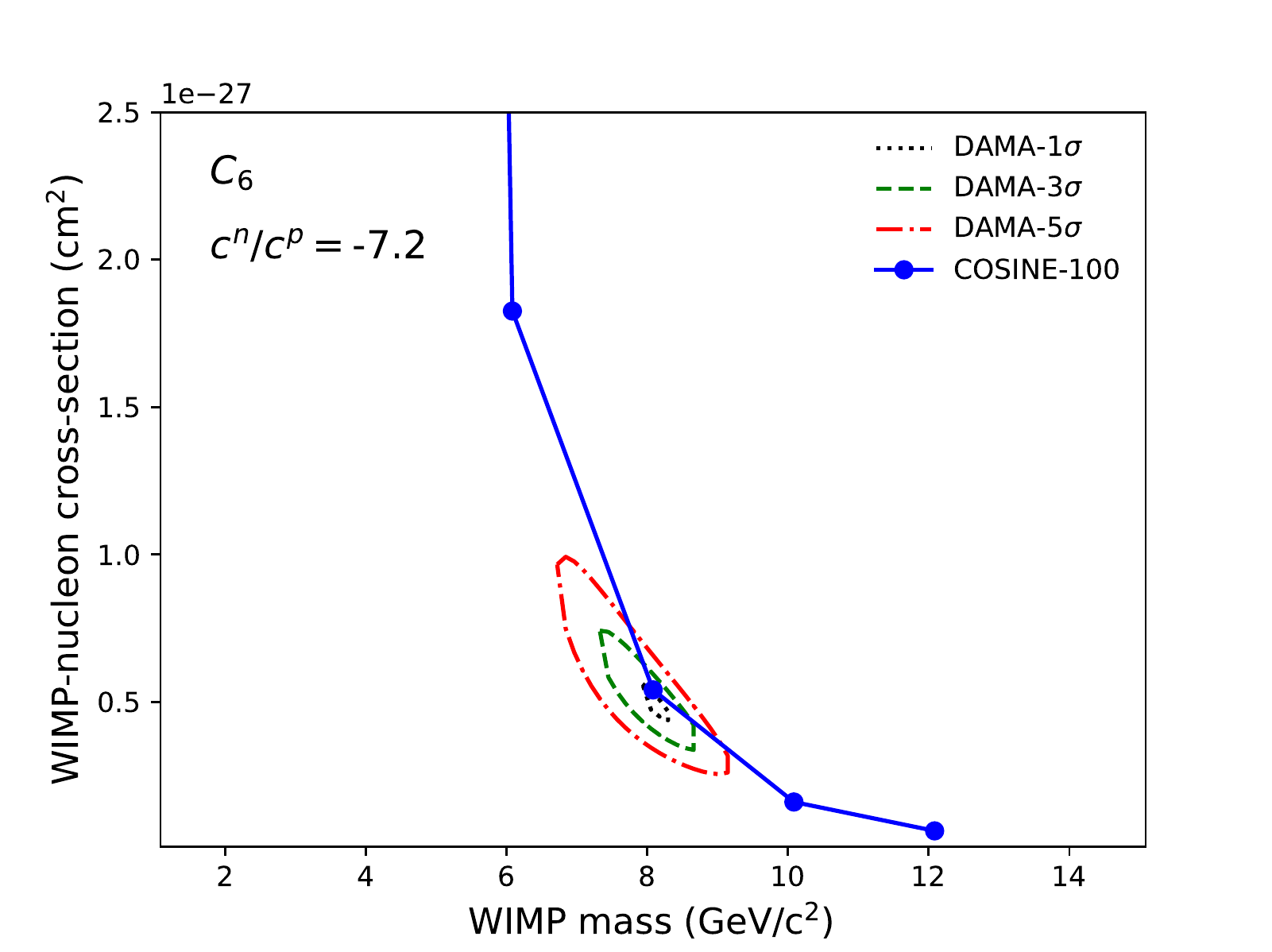}
  \includegraphics[width=0.32\columnwidth]{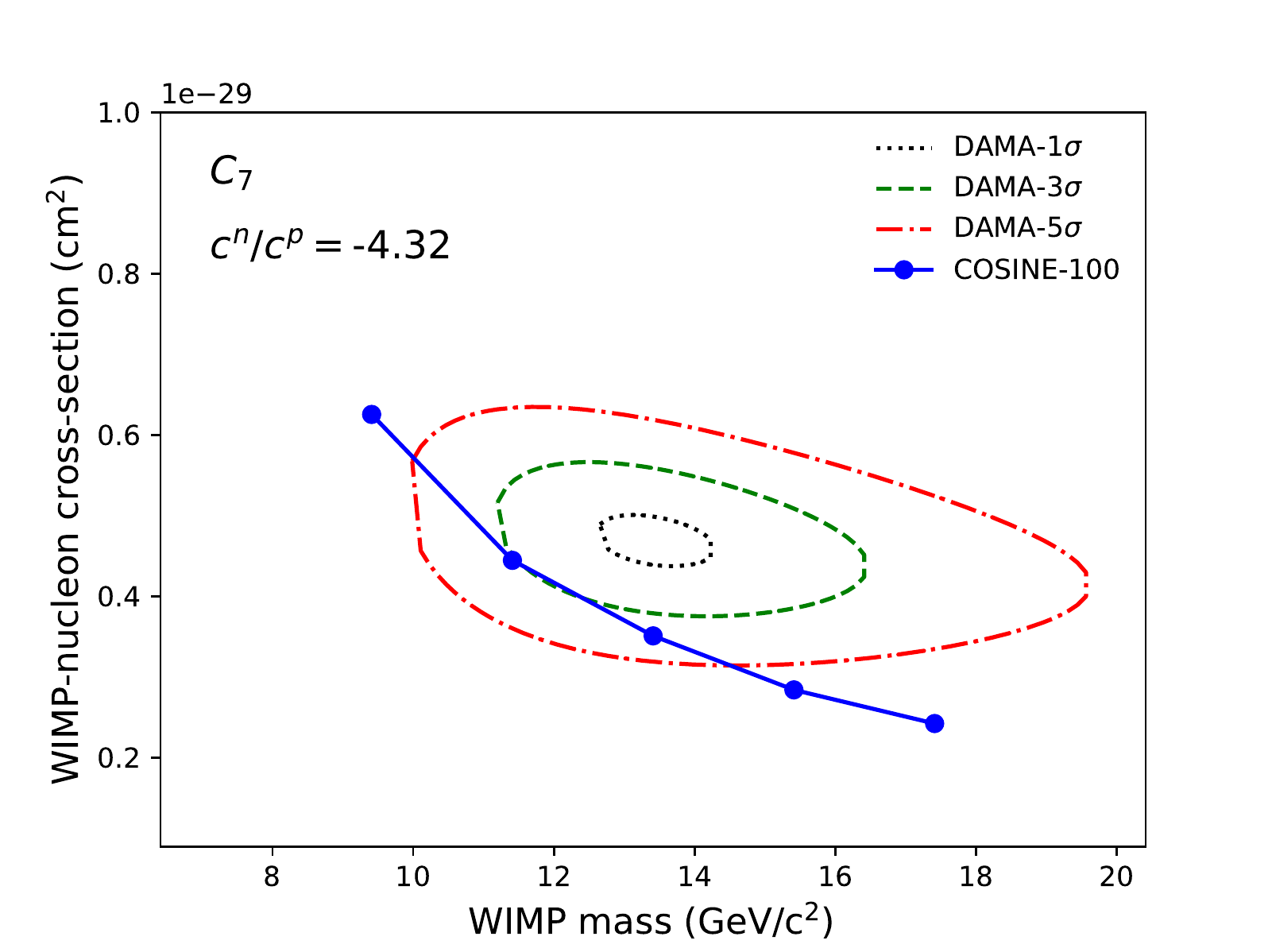}
  \includegraphics[width=0.32\columnwidth]{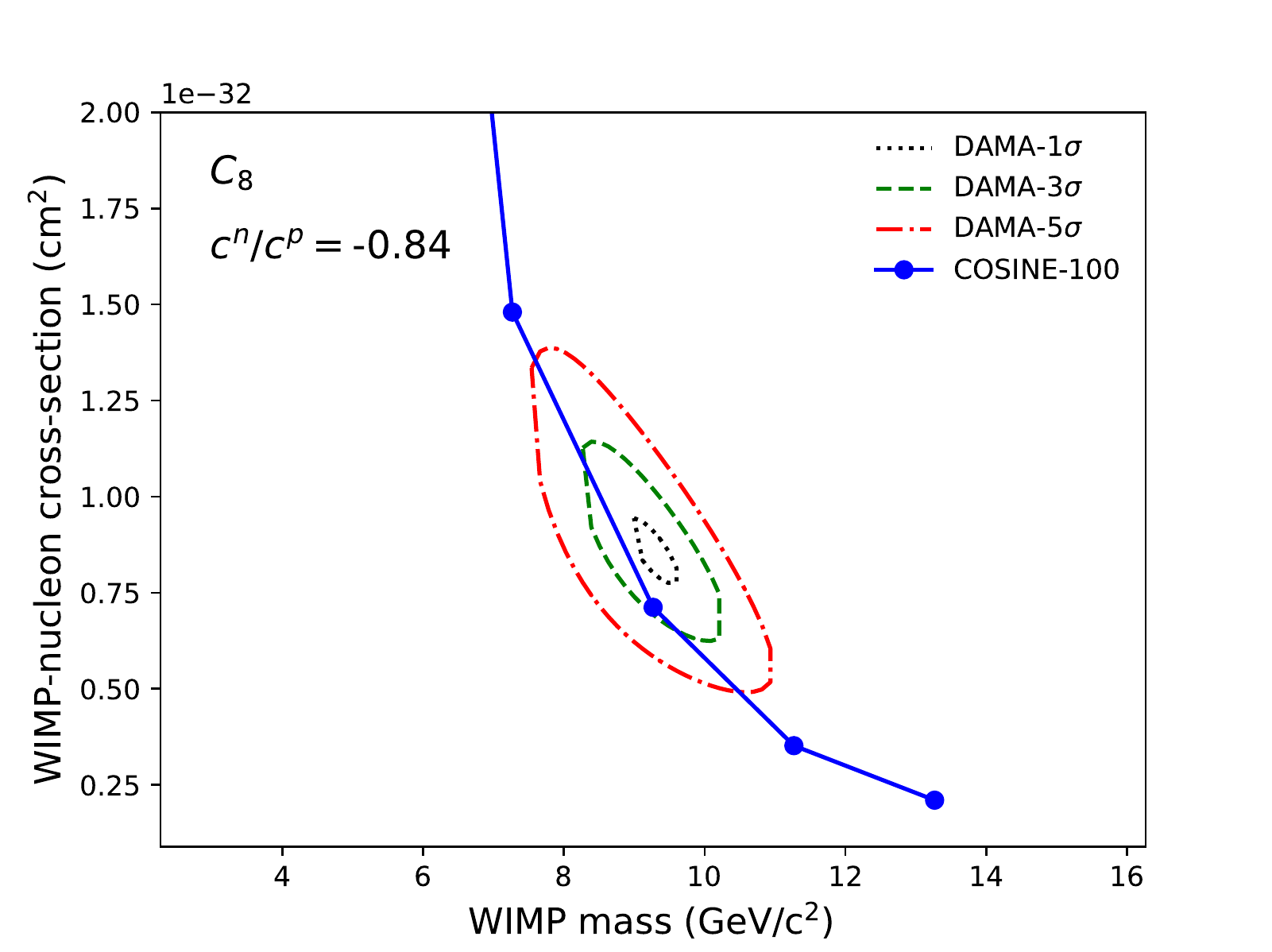}
  \includegraphics[width=0.32\columnwidth]{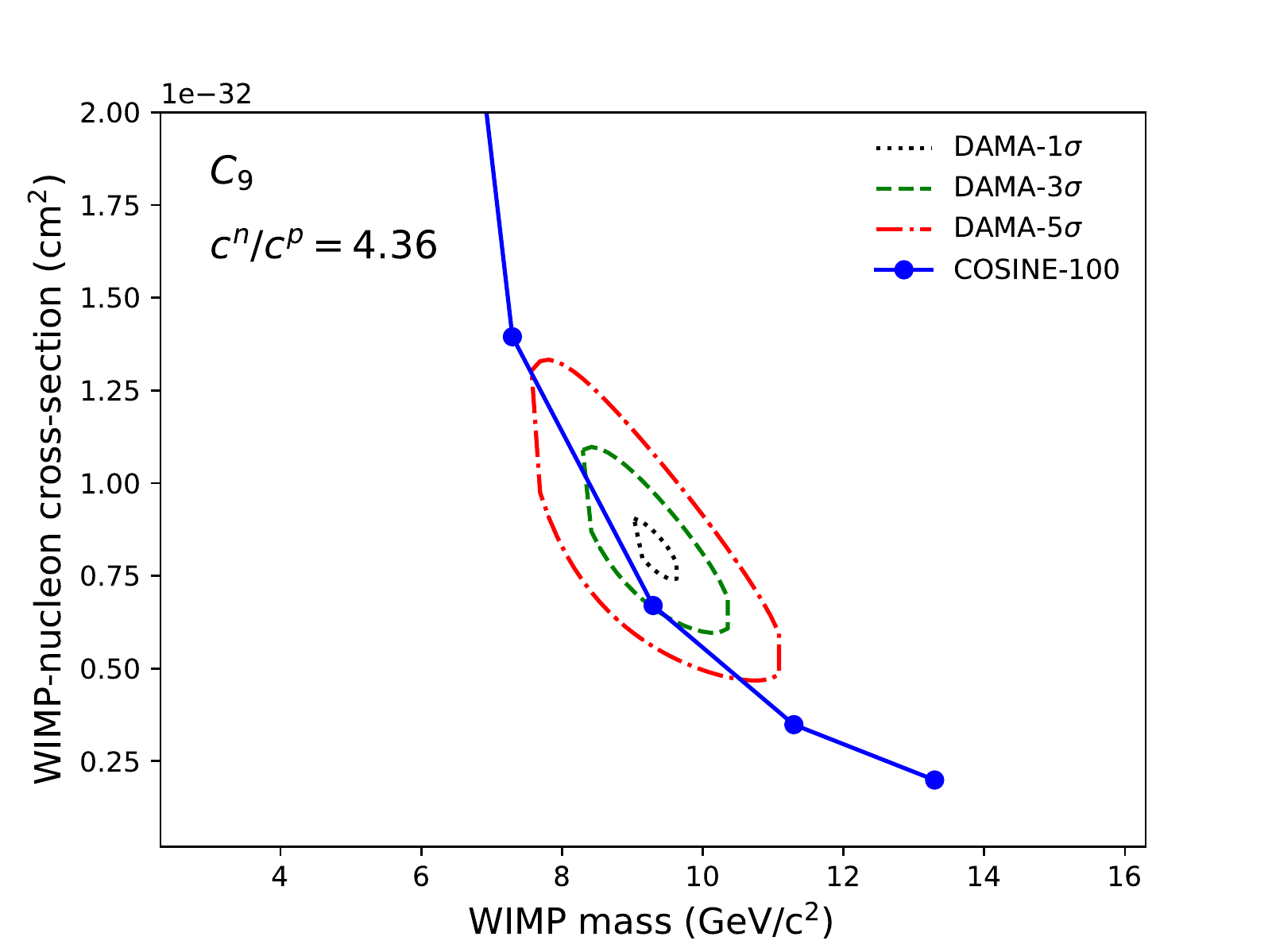}
  \includegraphics[width=0.32\columnwidth]{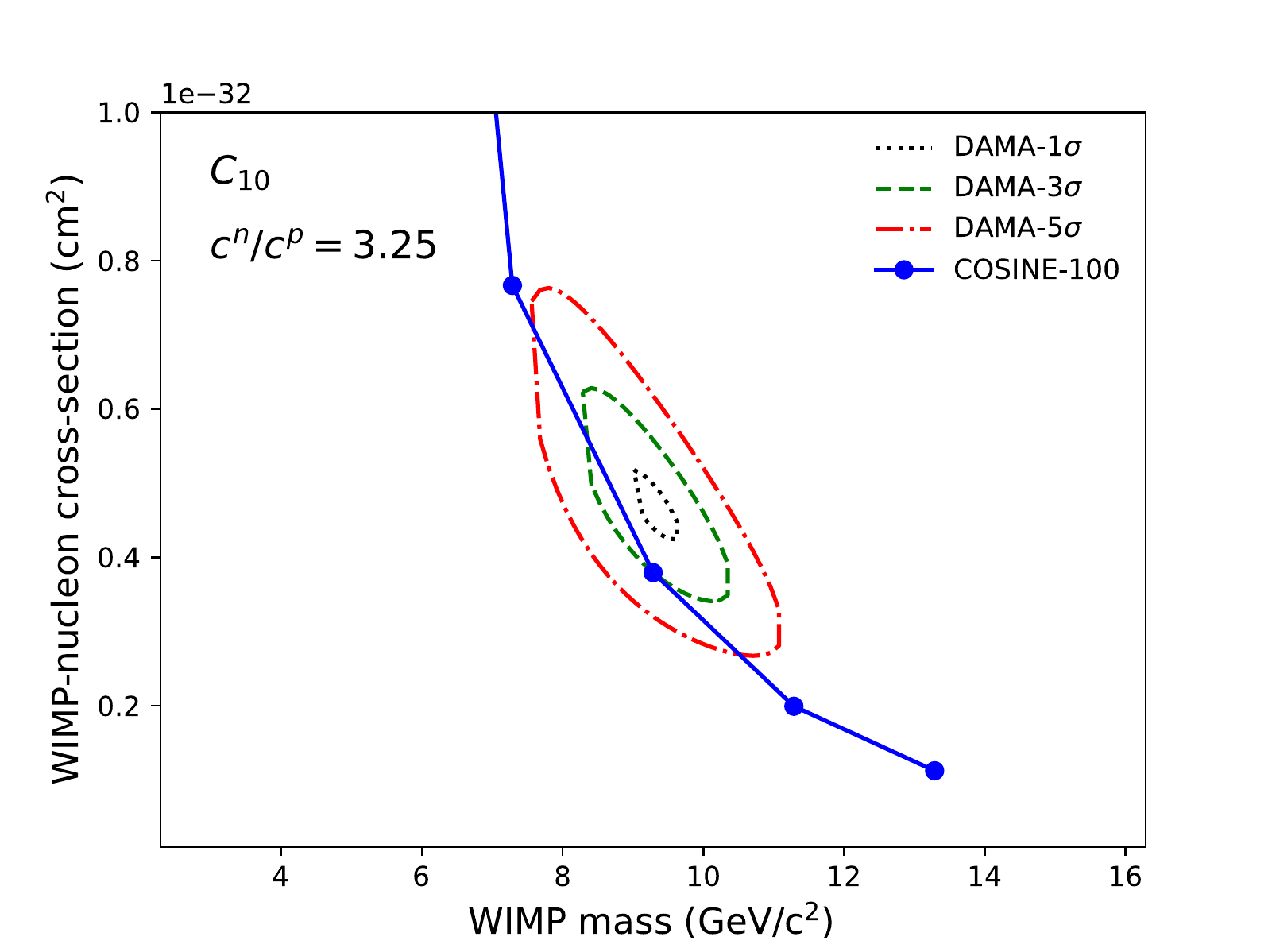}
  \includegraphics[width=0.32\columnwidth]{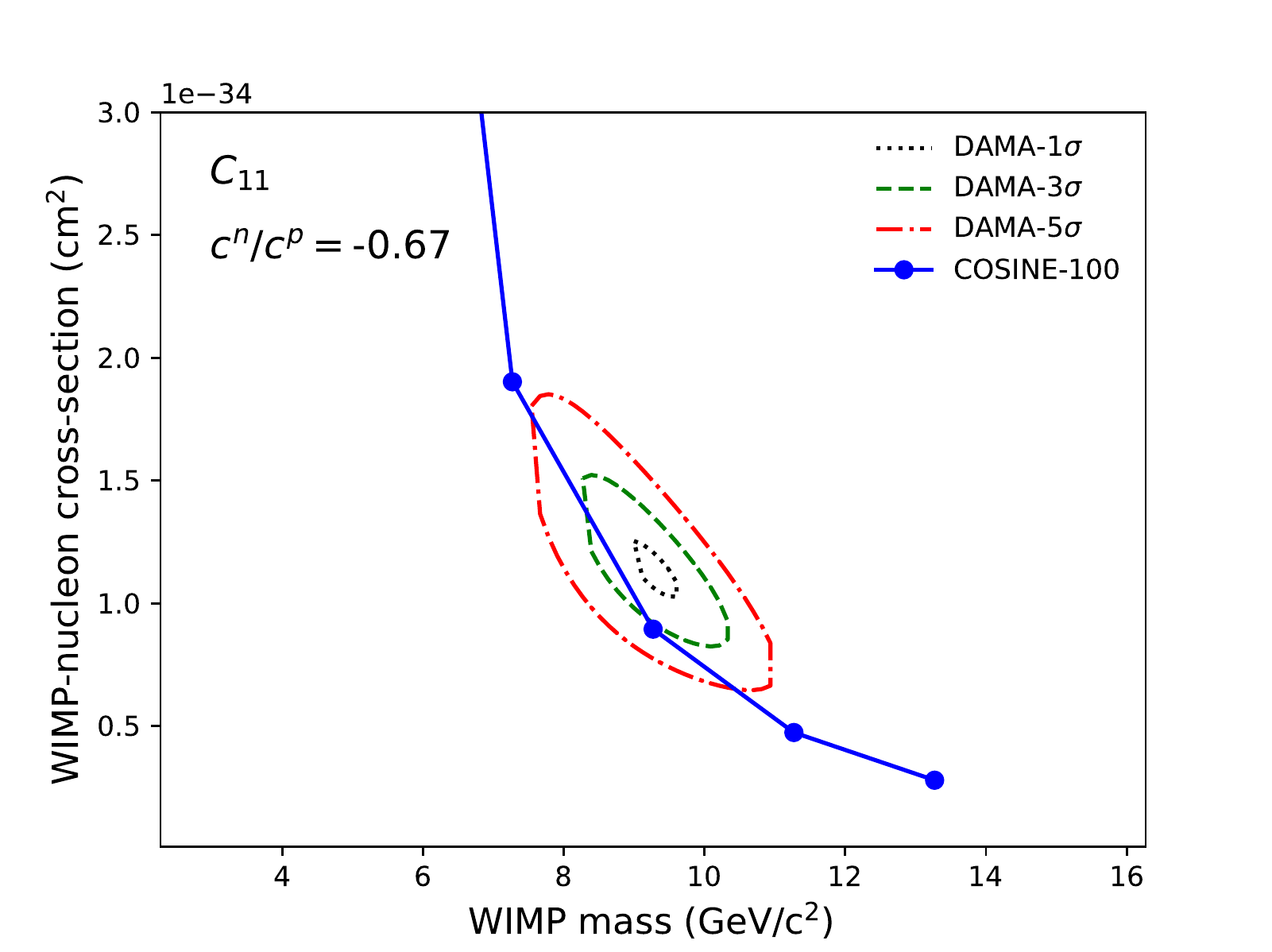}
  \includegraphics[width=0.32\columnwidth]{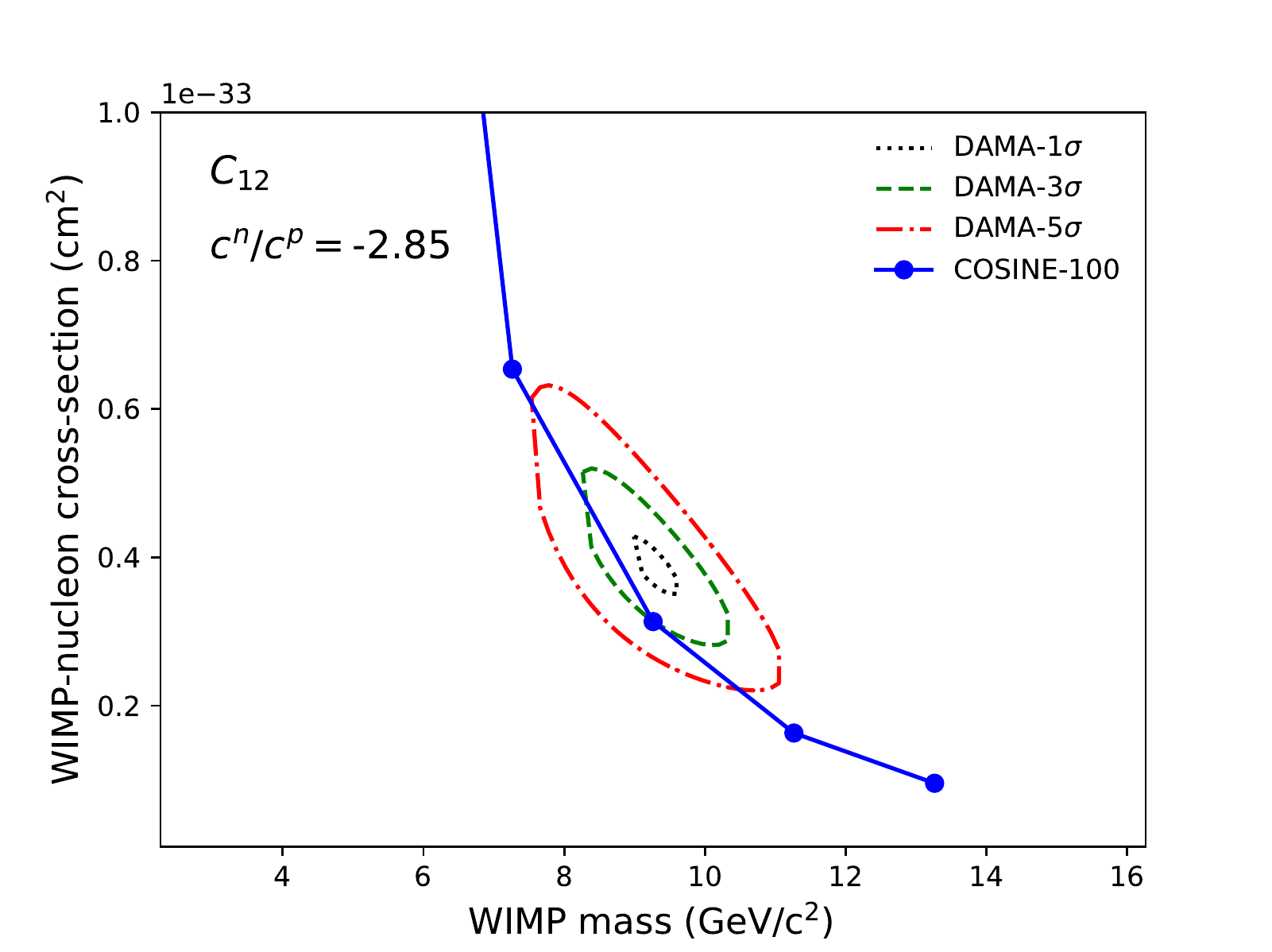}
  \includegraphics[width=0.32\columnwidth]{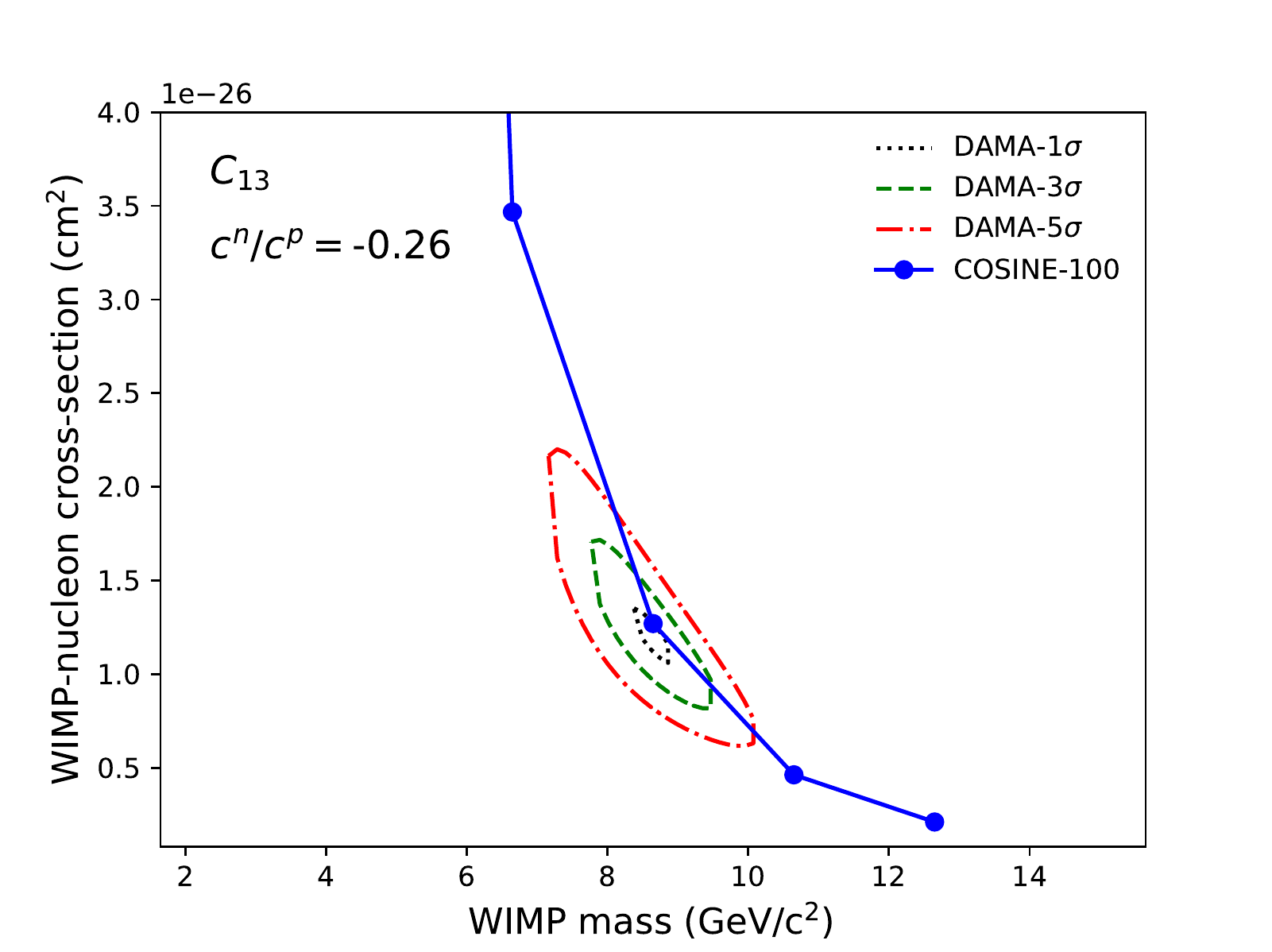}
  \includegraphics[width=0.32\columnwidth]{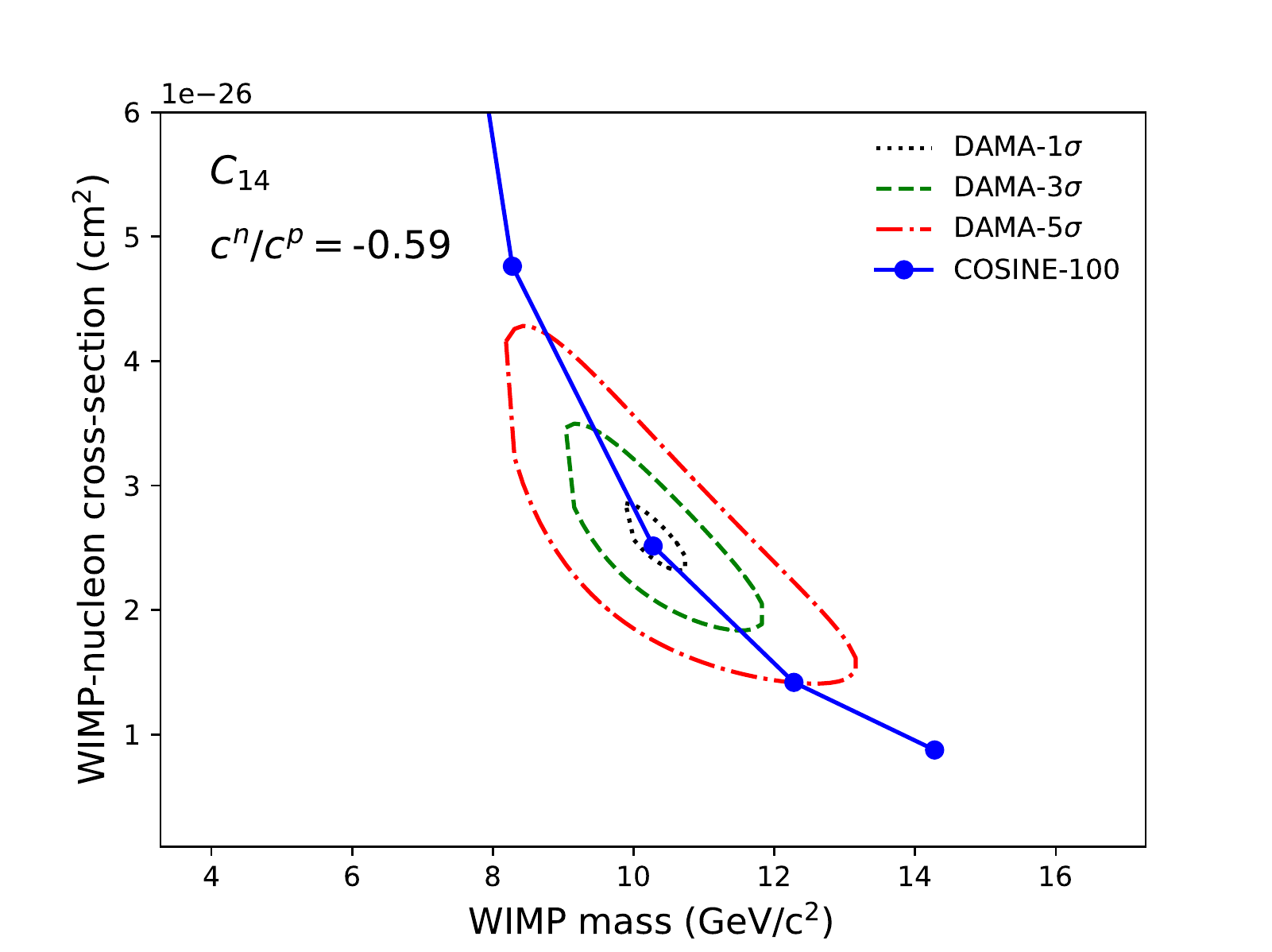}
  \includegraphics[width=0.32\columnwidth]{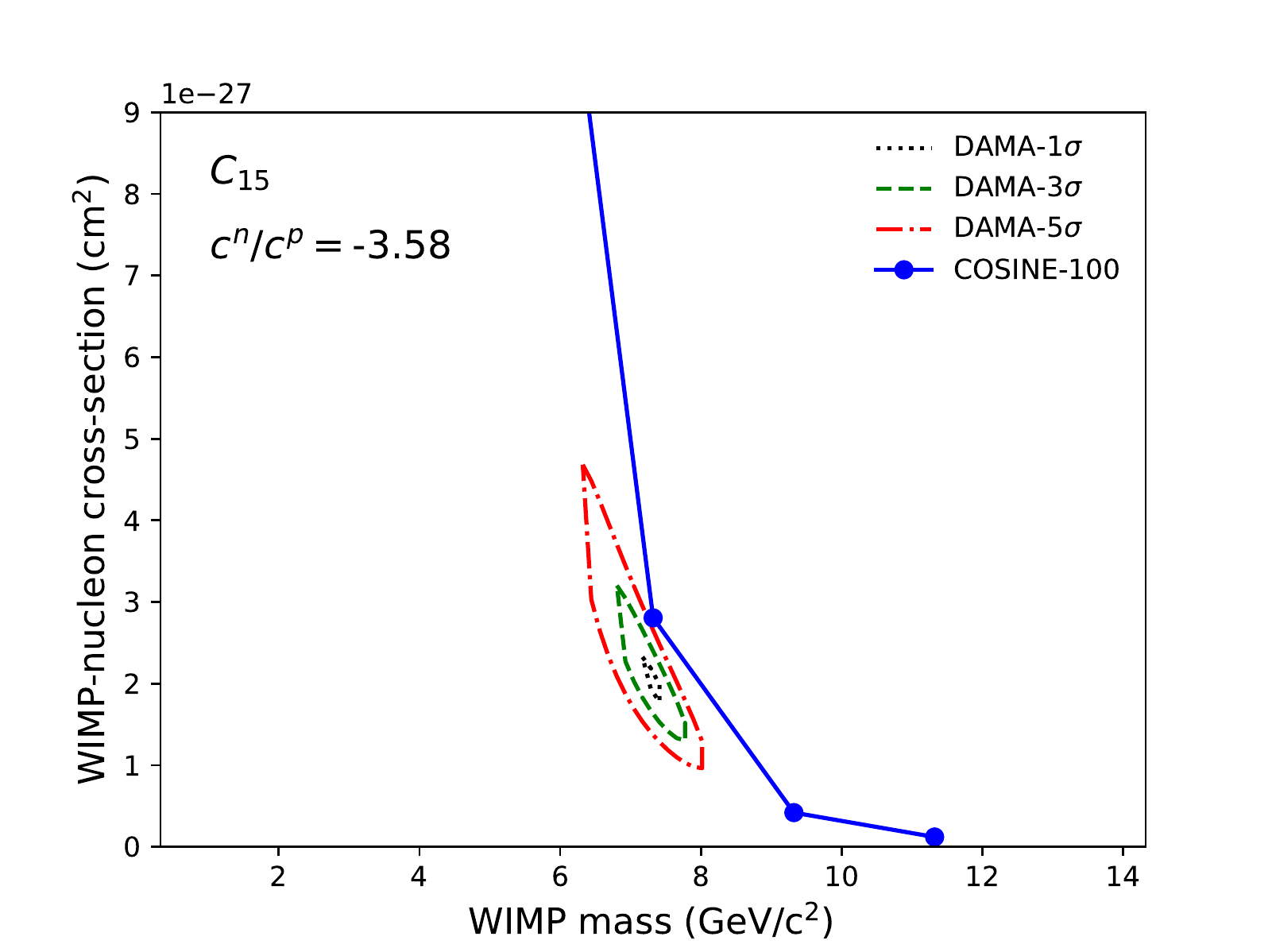}
\end{center}
\caption{Low WIMP mass DAMA modulation region (1--$\sigma$,
  3--$\sigma$ and 5--$\sigma$) and COSINE--100 90\% C.L. exclusion
  plot to the effective WIMP--proton cross section $\sigma_p$ of
  Eq. (\ref{eq:conventional_sigma}) for all the 14 NR effective
  operators of Table~\ref{tab:operators}. For each operator the
  $r$=$c^n/c^p$ neutron--over--proton ratio is fixed to the
  corresponding low--mass best fit value in
  Table~\ref{tab:best_fit_values}.}
\label{fig:low_mass}
\end{figure}

\begin{figure}
\begin{center}
  \includegraphics[width=0.32\columnwidth]{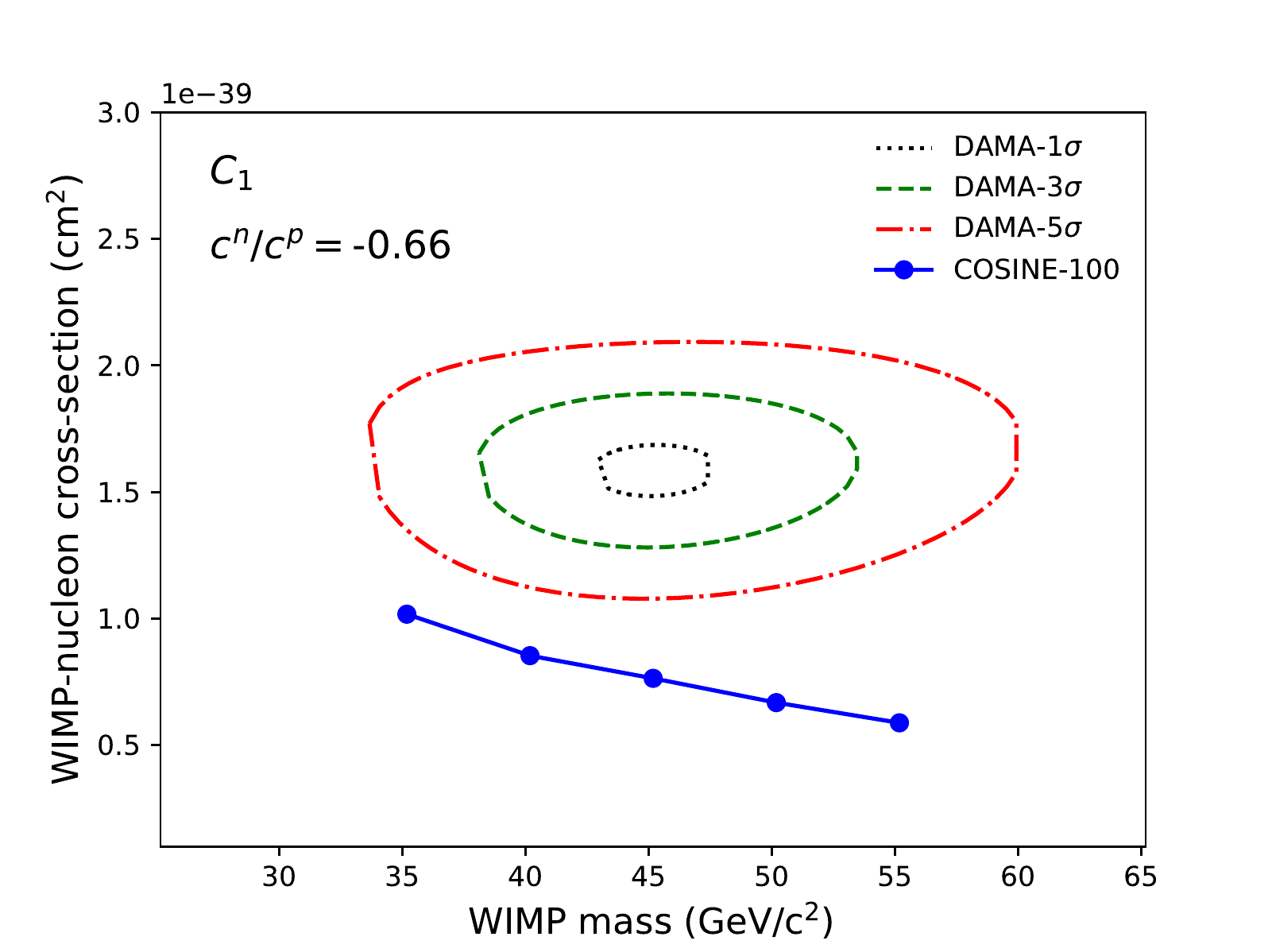}
  \includegraphics[width=0.32\columnwidth]{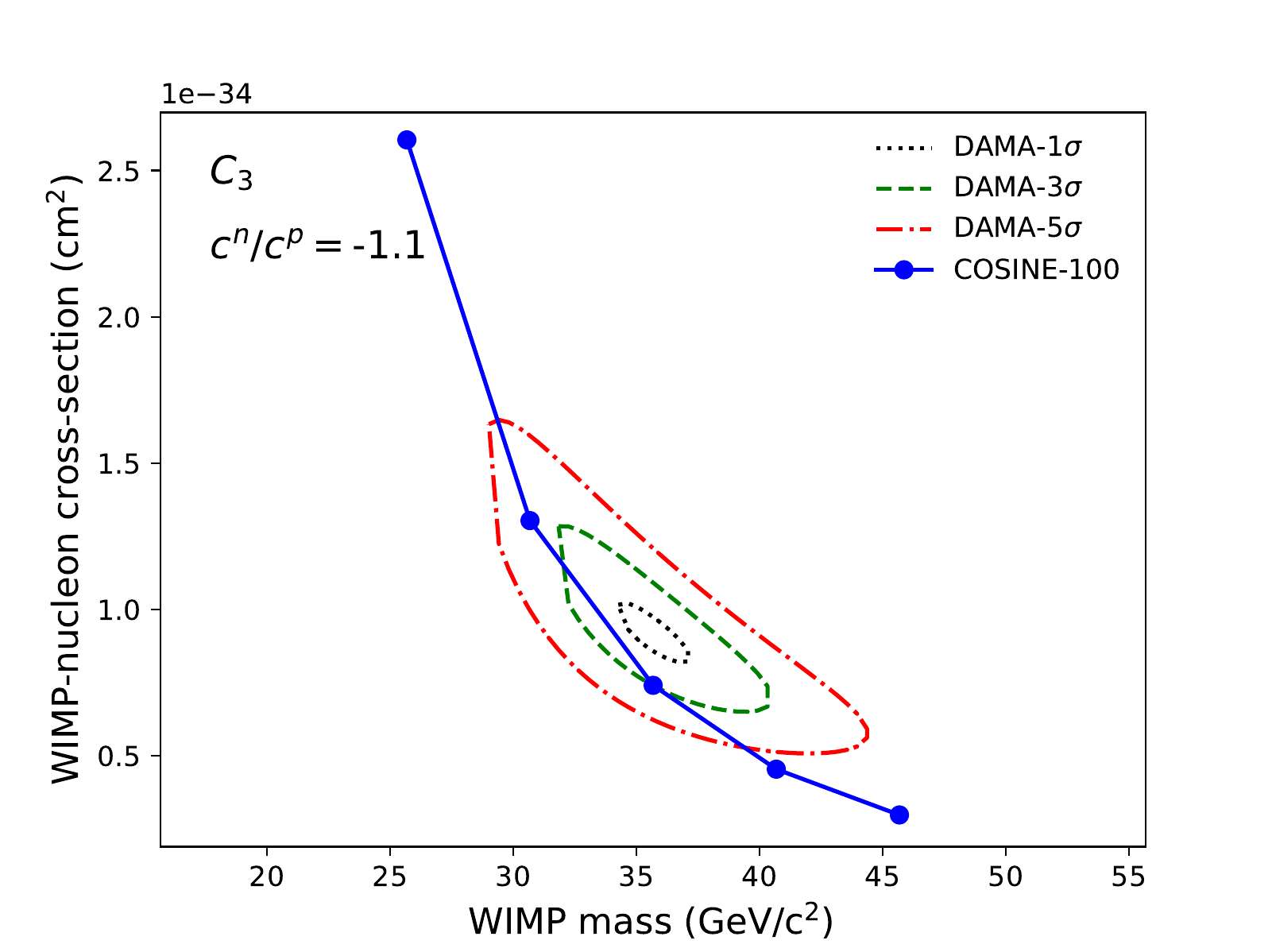}
  \includegraphics[width=0.32\columnwidth]{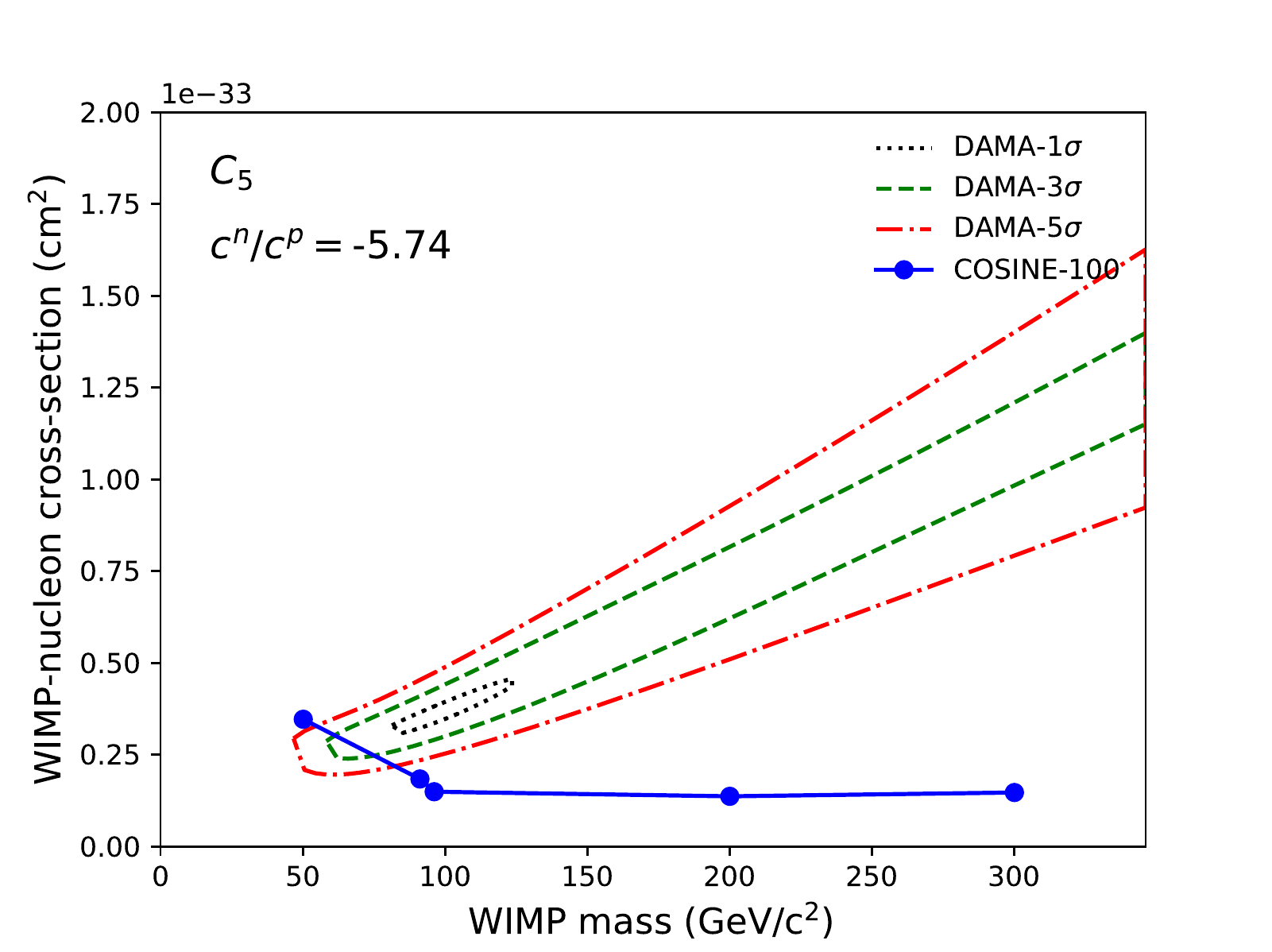}
  \includegraphics[width=0.32\columnwidth]{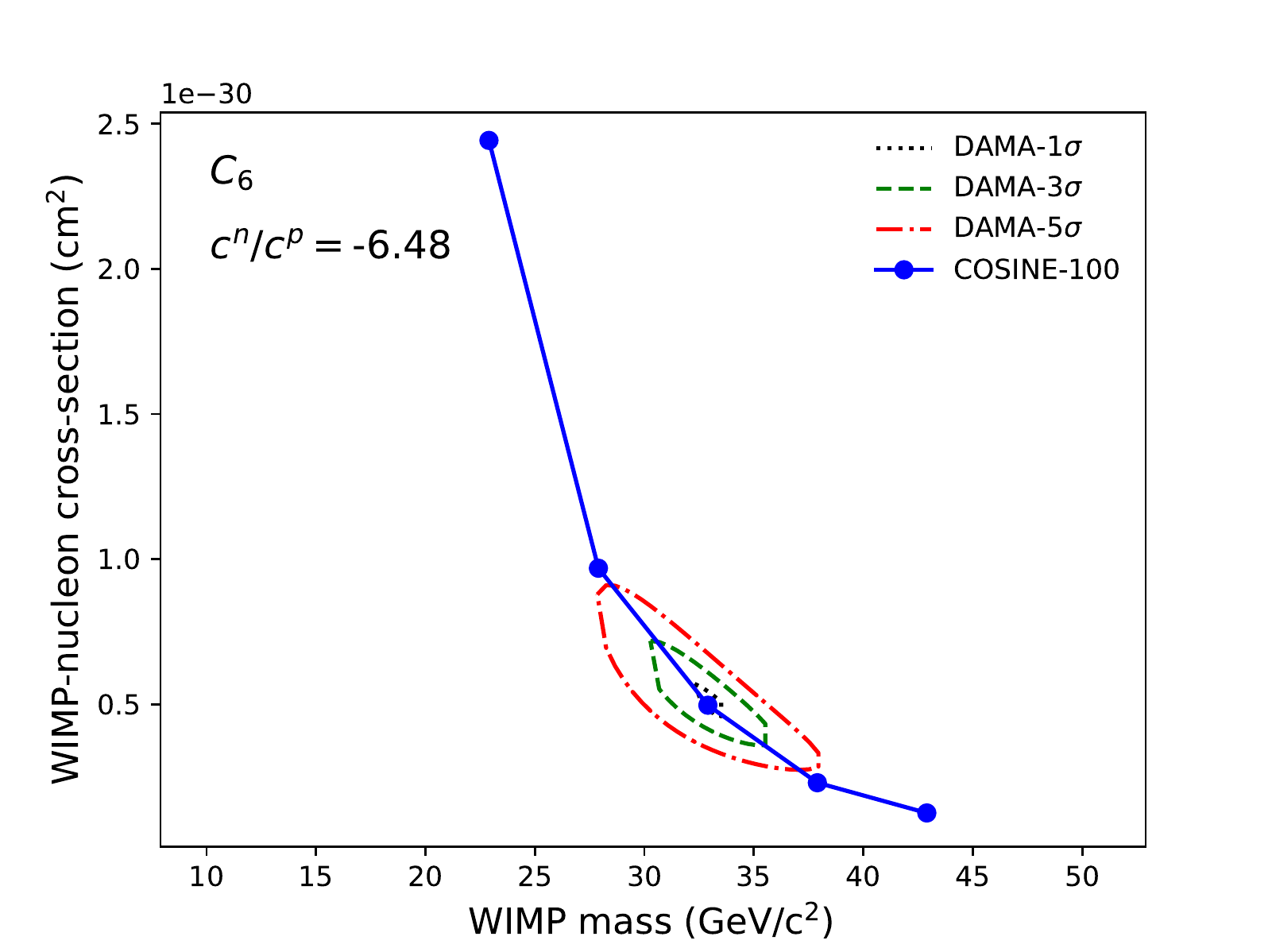}
  \includegraphics[width=0.32\columnwidth]{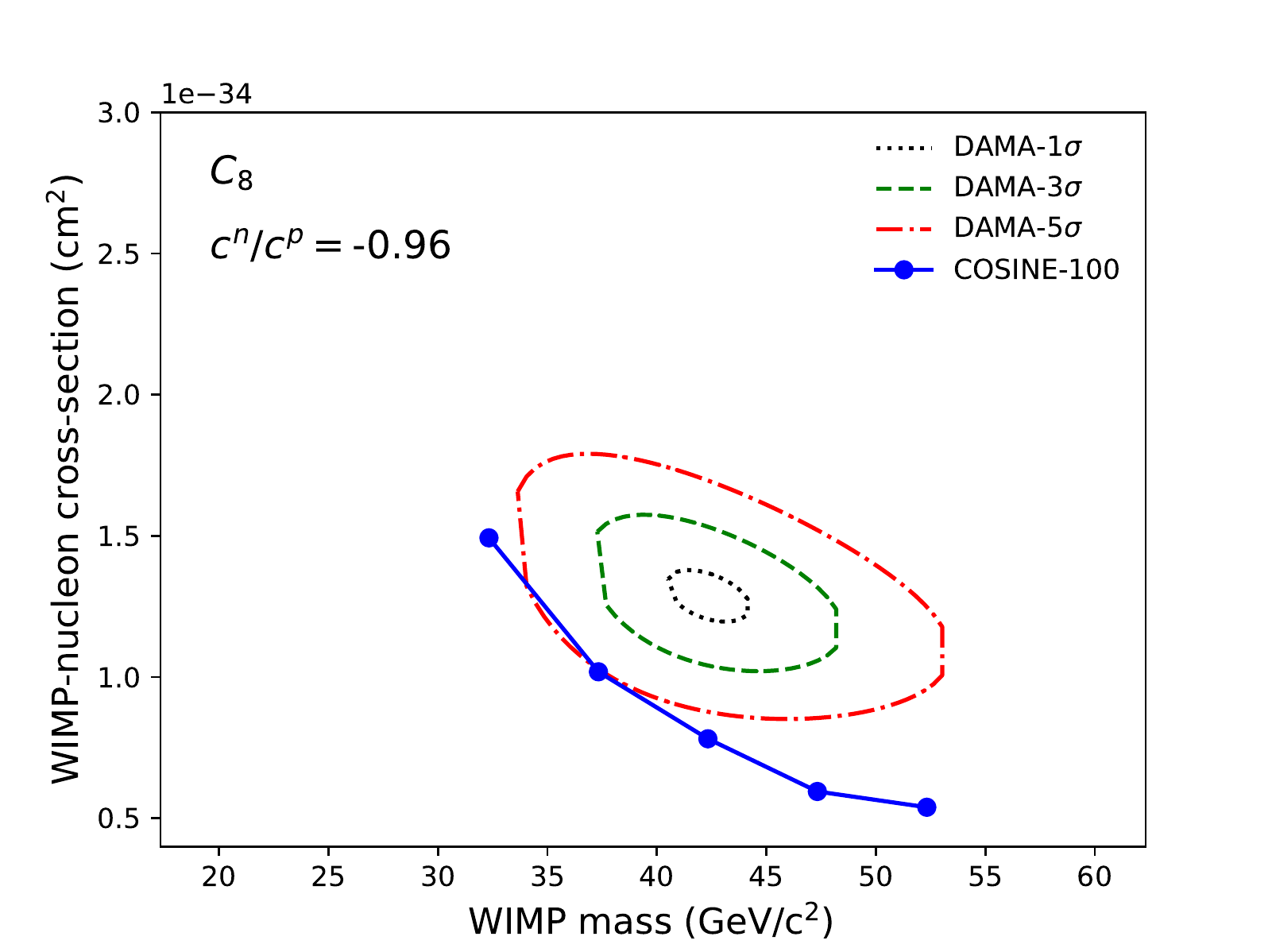}
  \includegraphics[width=0.32\columnwidth]{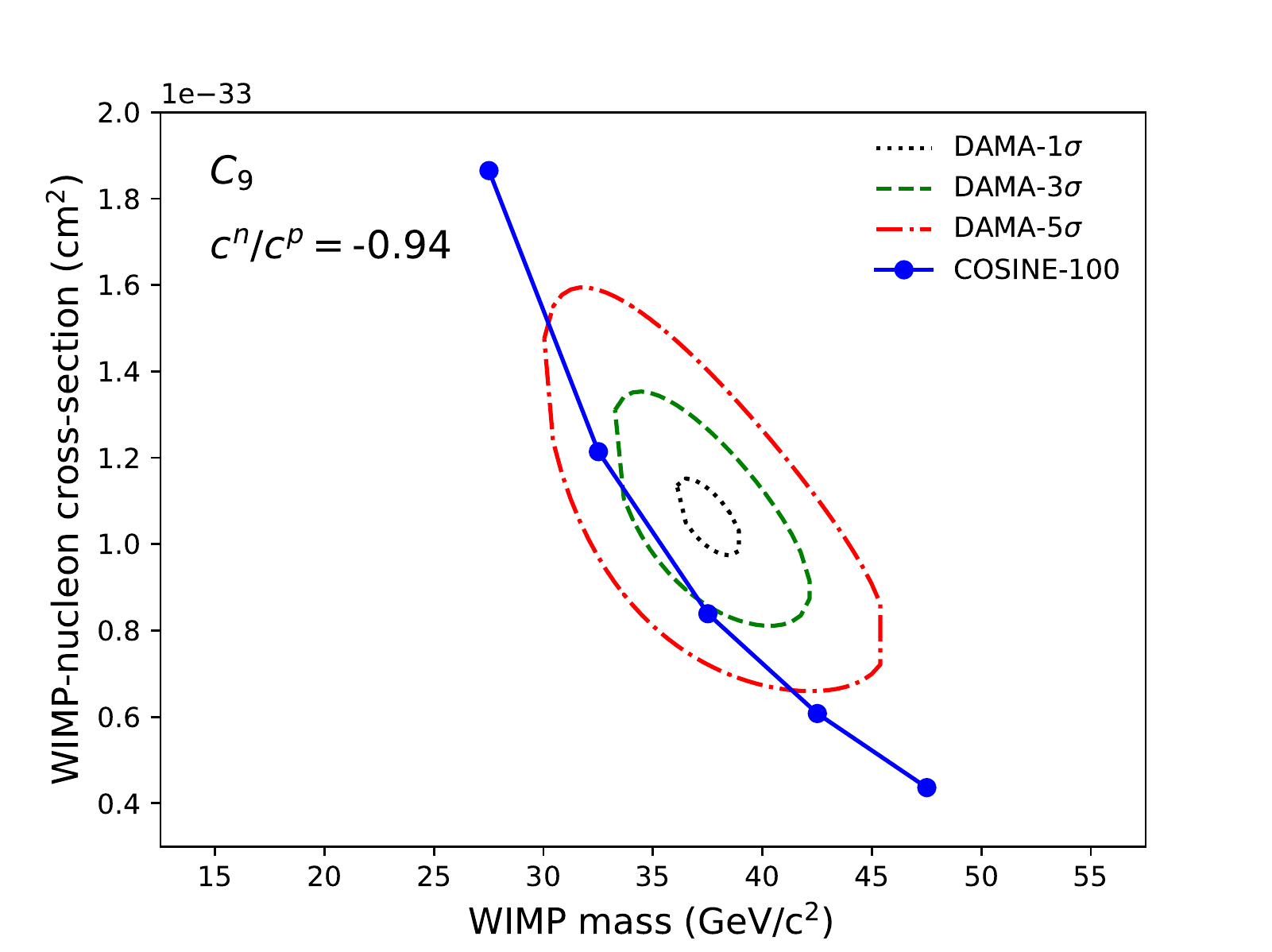}
  \includegraphics[width=0.32\columnwidth]{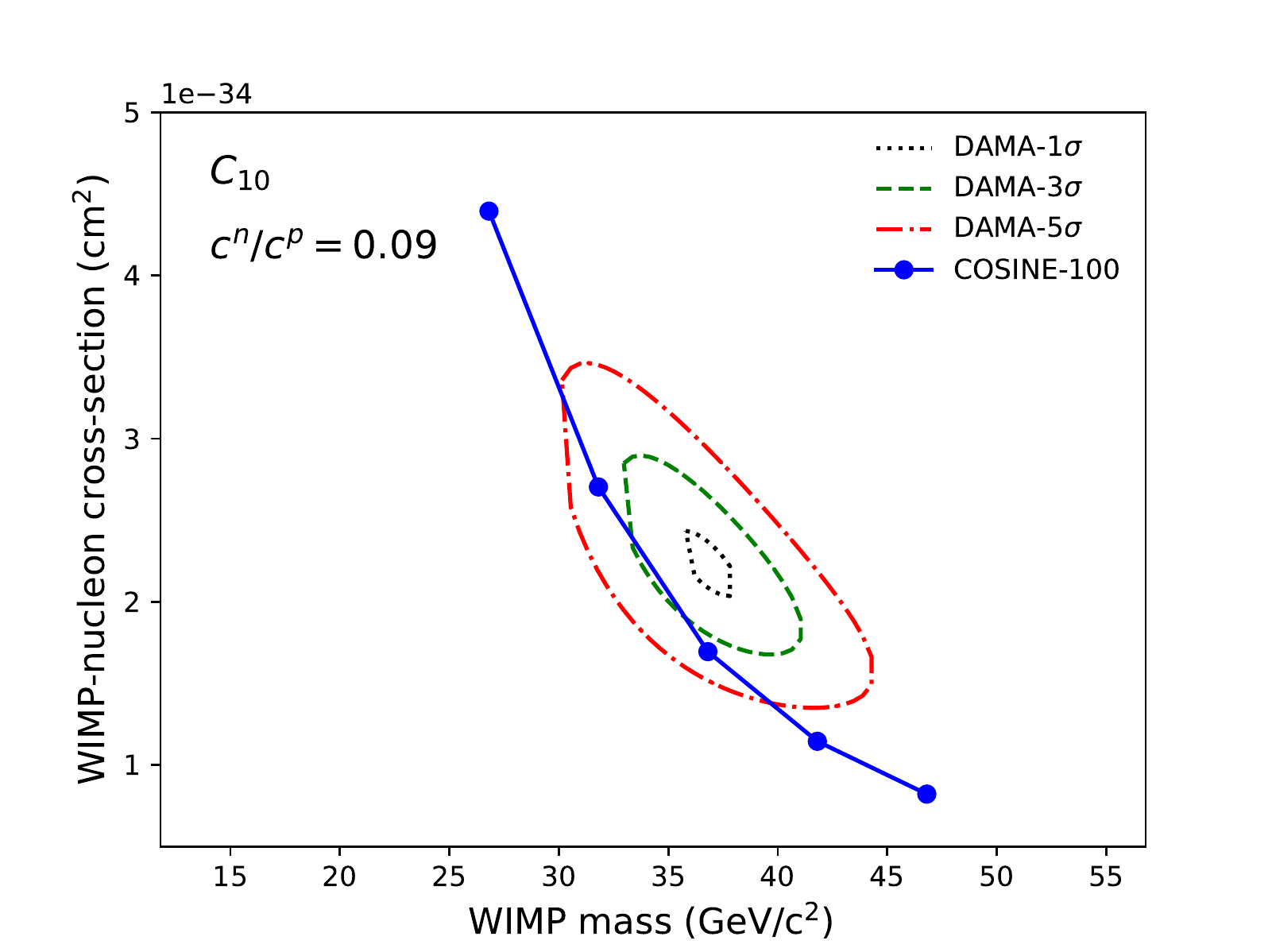}
  \includegraphics[width=0.32\columnwidth]{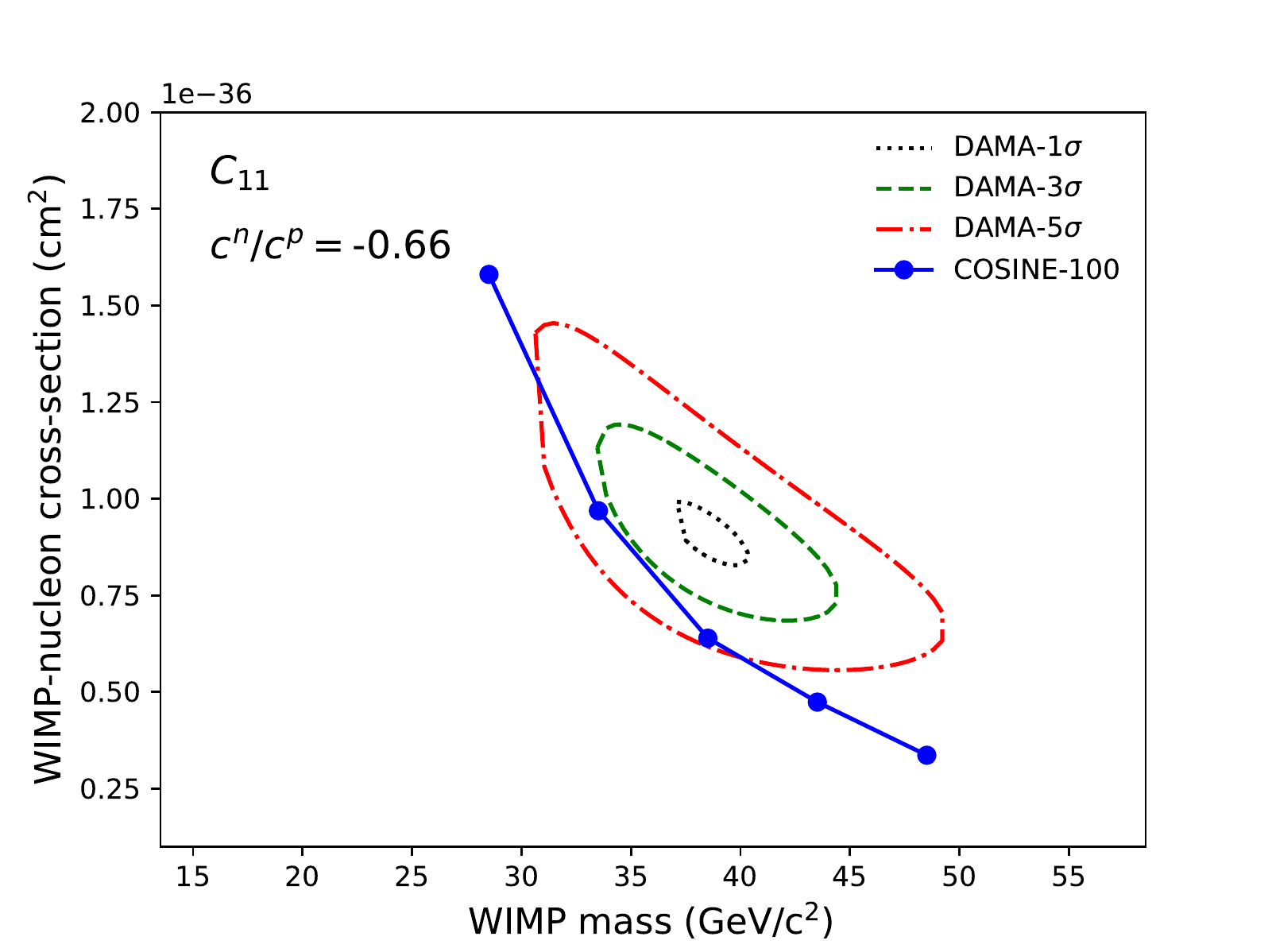}
  \includegraphics[width=0.32\columnwidth]{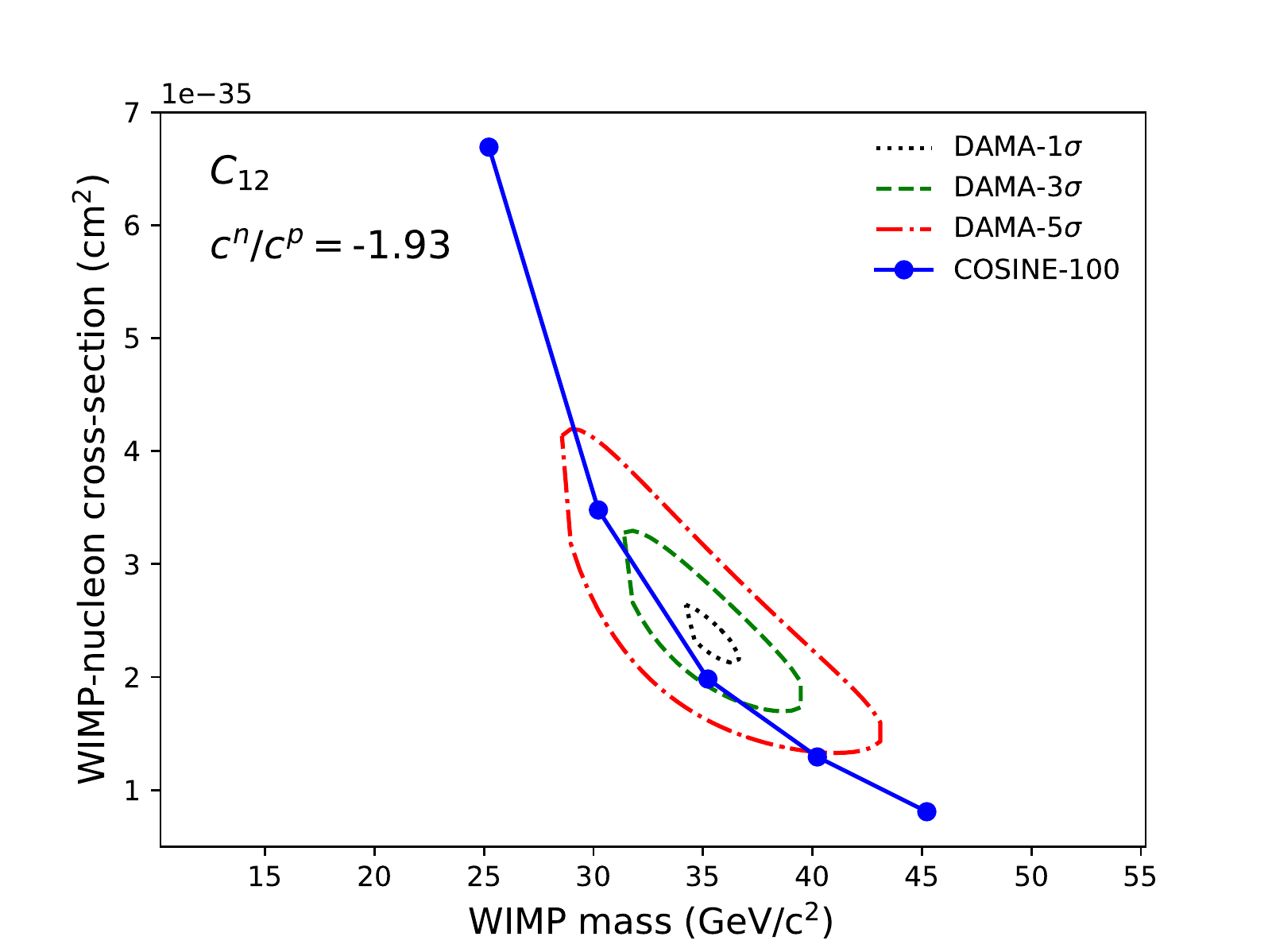}
  \includegraphics[width=0.32\columnwidth]{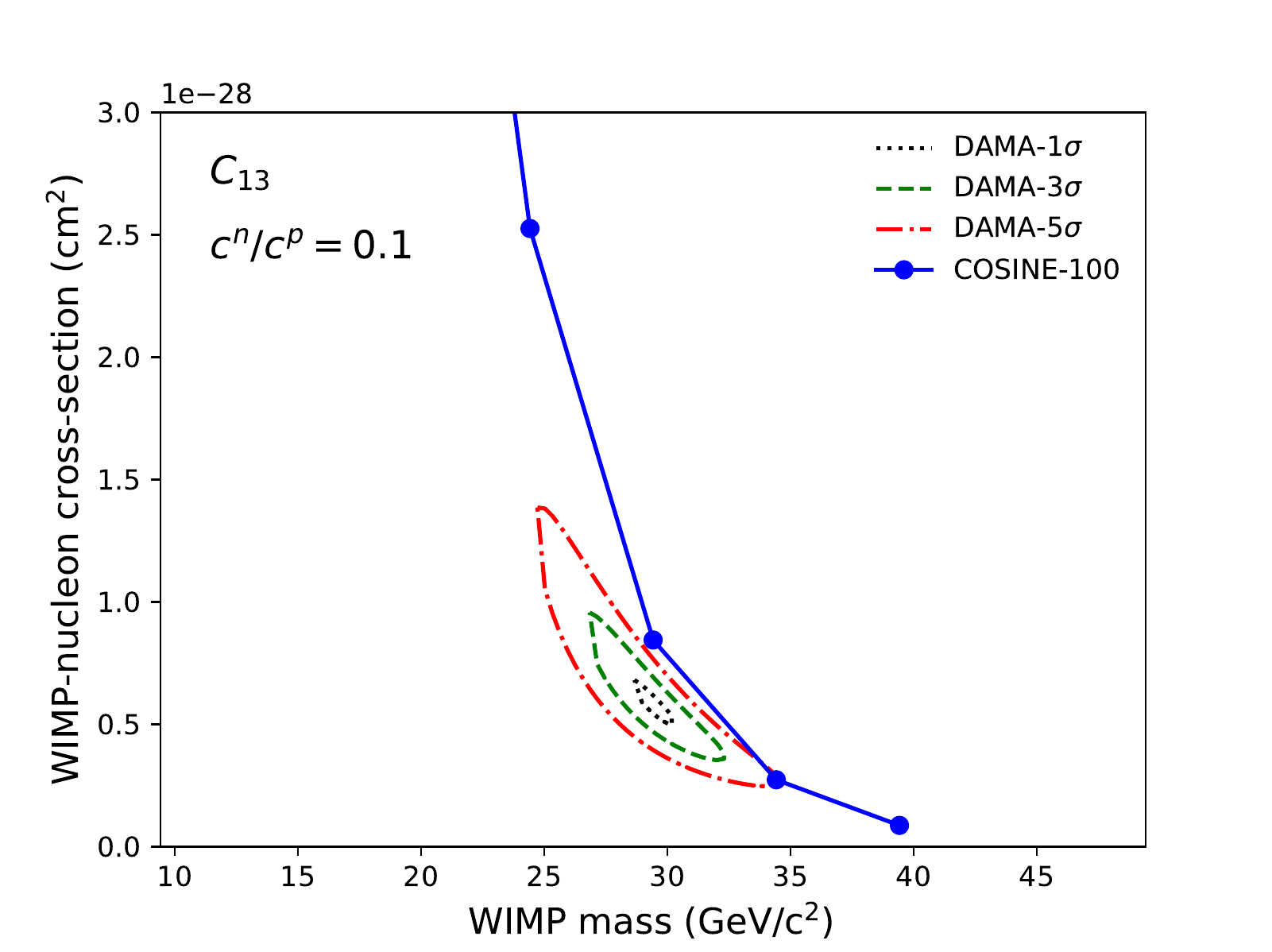}
  \includegraphics[width=0.32\columnwidth]{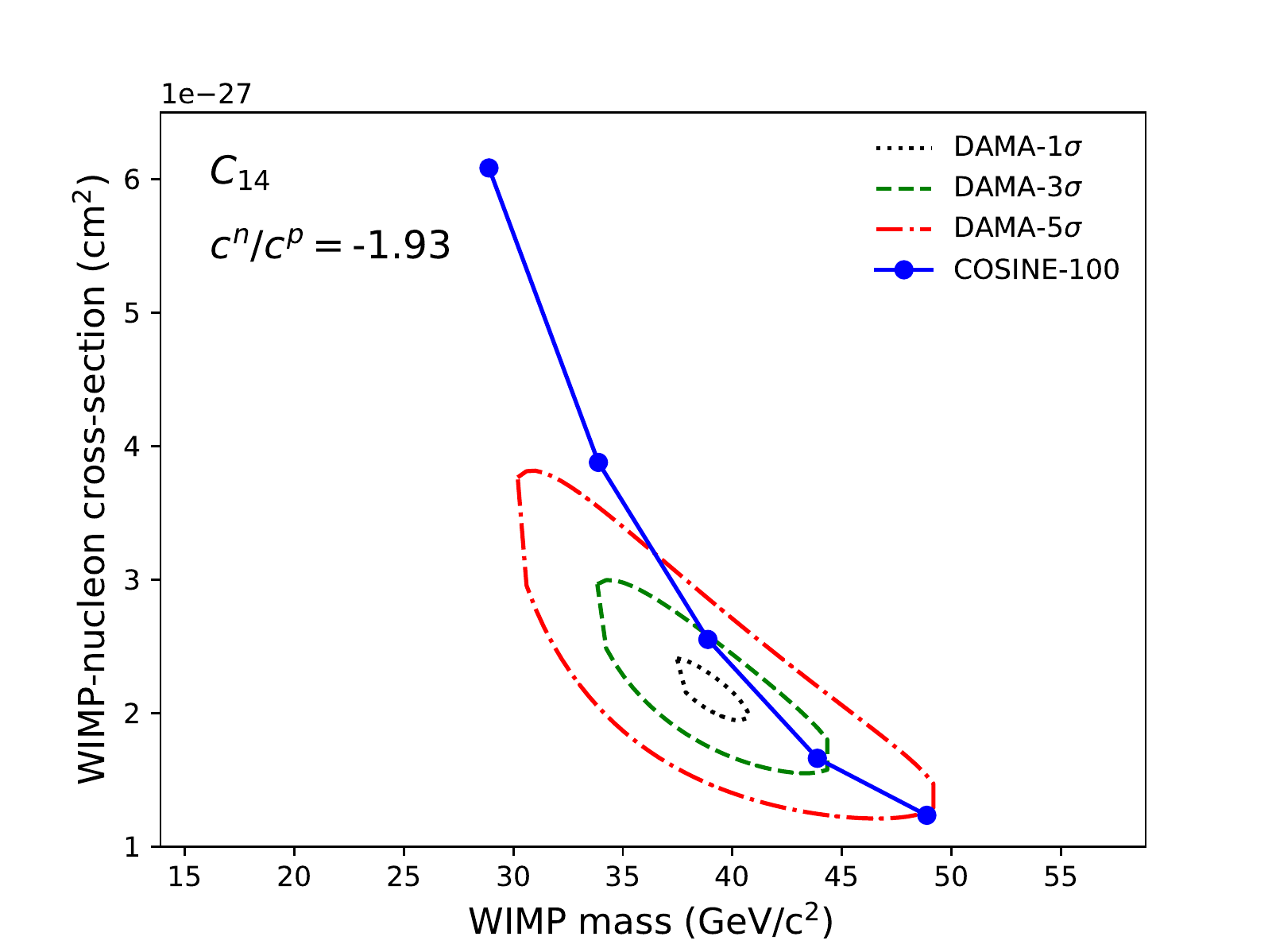}
  \includegraphics[width=0.32\columnwidth]{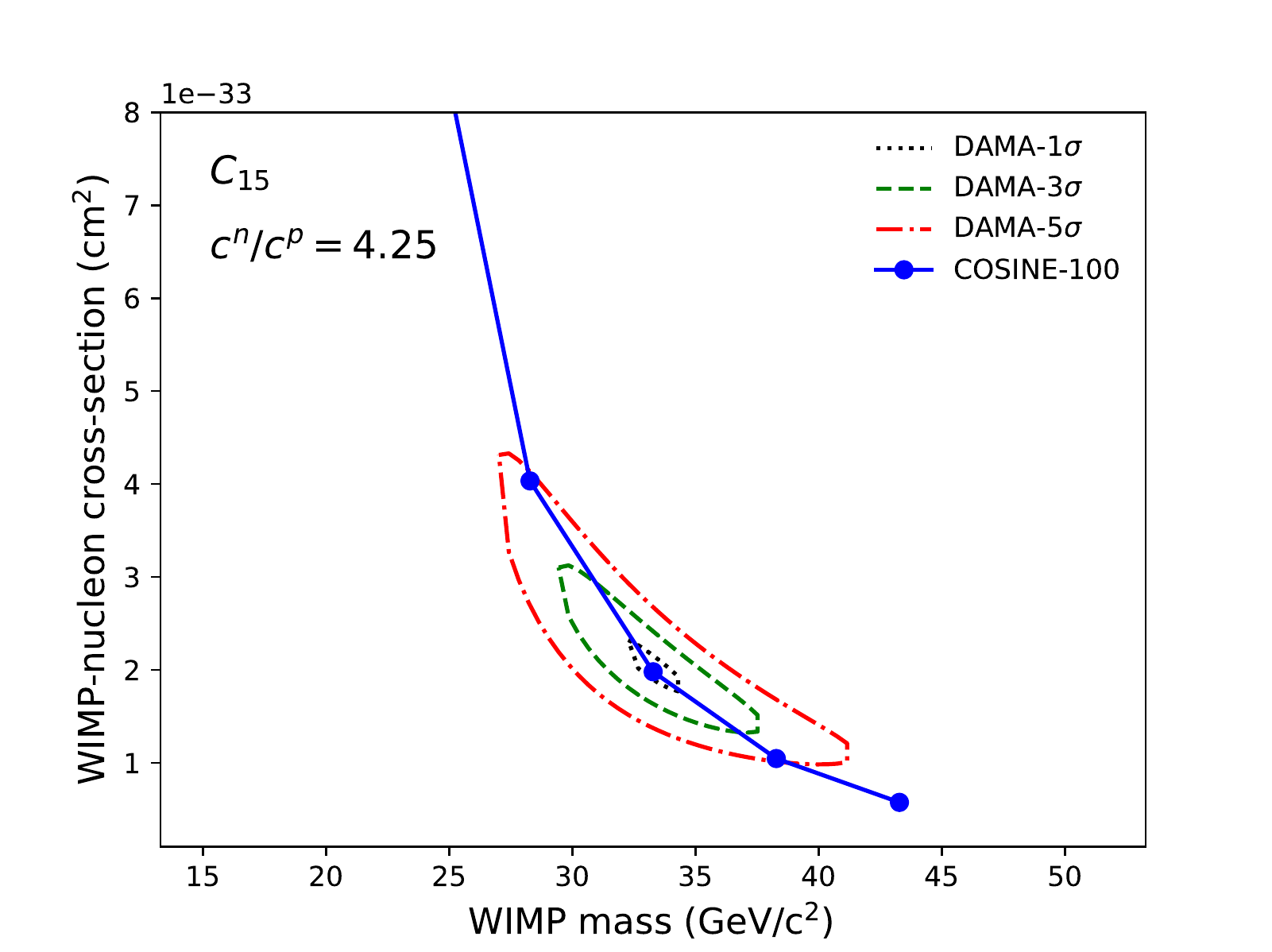}
\end{center}
\caption{High WIMP mass DAMA modulation region (1--$\sigma$,
  3--$\sigma$ and 5--$\sigma$) and COSINE--100 90\% C.L. exclusion
  plot to the effective WIMP--proton cross section $\sigma_p$ of
  Eq. (\ref{eq:conventional_sigma}) for all the 14 NR effective of
  Table~\ref{tab:operators}. For each operator the $r$=$c^n/c^p$
  neutron--over--proton ratio is fixed to the corresponding high--mass
  best fit value in Table~\ref{tab:best_fit_values}.}
\label{fig:high_mass}
\end{figure}

The result of our analysis is summarized in Figs.~\ref{fig:low_mass}
and \ref{fig:high_mass} where, for each NR effective coupling, the
DAMA modulation regions at 1--$\sigma$, 3--$\sigma$ and 5--$\sigma$ is
compared to the COSINE--100 90\% C.L. exclusion limit in the
$m_{\chi}$--$\sigma_p$ plane.  The present analysis is focused on the
comparison between DAMA and COSINE-100. However, it is worth reminding
here that limits from detectors using target materials different from
$NaI$ exclude effective cross sections ranging from one to three
orders of magnitude below the DAMA region~\cite{dama_eft_2018}. 

In each plot of Fig.~\ref{fig:low_mass}
the neutron--over--proton coupling ratio $r=c^n/c^p$ is fixed to the
corresponding low--mass best--fit value of
Table~\ref{tab:best_fit_values} and the WIMP mass interval is centered
accordingly. In Fig.~\ref{fig:high_mass} the same is done for the
high--mass best fit values of Table~\ref{tab:best_fit_values} (in the
latter figure the cases of couplings $c_4$ and $c_7$ are not included
since they do not provide a good fit to the DAMA modulation
amplitudes).

Figs.~\ref{fig:low_mass}-\ref{fig:high_mass} show that the exclusion
plot on $\sigma_p$ from COSINE--100 has a different impact on the DAMA
best fit modulation region depending on the specific non--relativistic
model. Namely, as far as the DAMA low--mass minima of
Fig.~\ref{fig:low_mass} are concerned, the tension between DAMA/LIBRA
and COSINE--100 is maximal for the couplings $c_1$ and $c_4$, while
for the DAMA high--mass minima of Fig.~\ref{fig:high_mass} this
happens for the couplings $c_1$ and $c_8$: in all such cases the 90\%
C.L. bound from COSINE--100, represented by the (blue) solid line,
rules out all the 5--sigma DAMA region shown as the (red) dot--dashed
contour. On the other hand, in all other cases the 5--sigma DAMA
region is not completely excluded by the corresponding 90\%
C.L. COSINE--100 upper bound, with two instances ($c_{15}$ at low WIMP mass
$c_{13}$ at high WIMP mass) for which all the DAMA modulation region
is allowed by the COSINE--100 constraint.

The main motivation of probing the modulation effect claimed by
DAMA/LIBRA using a sodium--iodide target with COSINE--100 is to obtain
results that depend as little as possible on the unknown properties of
the WIMP particle. On the other hand, the model dependence observed in
Figs.~\ref{fig:low_mass}--\ref{fig:high_mass} is due to two main
reasons: i) the change in the signal spectral shape; ii) the expected modulation fractions
$S_{m,[E_1^{\prime},E_2^{\prime}]}/S_{0,[E_1^{\prime},E_2^{\prime}]}$
in DAMA.

As far as the spectral shape of the expected WIMP signal is concerned,
each of the effective models listed in Table~\ref{tab:operators} is
characterized by a different dependence on the exchanged momentum $q$
(and so on the recoil energy $E_R=q^2/(2 m_T)$), both through the
nuclear response functions $W_{Tk}(q)$ (with $k$=$M$,
$\Phi^{\prime\prime}$, $\tilde{\Phi}^{\prime}$,
$\Sigma^{\prime\prime}$, $\Sigma^{\prime}$, $\Delta$) and through
additional powers of $q$ in the scattering amplitude (as summarized in
Table~\ref{table:eft_summary}). The raw energy spectra in COSINE--100
calculated using Eq.~(\ref{eq:dr_de}) are shown with a red solid line
in Fig.~\ref{fig:recoil} for each NR operator. Indeed, while in the
standard spin--independent and spin--dependent cases (corresponding to
$c_1$ and $c_4$) the expected differential rate is the featureless
superposition of two exponentially decaying spectra due to
WIMP--iodine and WIMP--sodium scattering events, in the case of other
NR operators the WIMP recoil spectrum can show a maximum at low energy
that may mimic one of the observed radiation peaks (such as the one
due to $^{40}$K) potentially affecting the sensitivity to the
signal. However, as shown in Fig.~\ref{fig:recoil} with the blue
dotted lines, when the quenching factors for Na and I and the
crystal resolution are applied to the raw spectra all the expected
rates are compressed to lower visible energies, so that this effect is
strongly reduced.  Moreover the expected rates in COSINE--100 become
almost insensitive to WIMP--iodine scattering events, that are driven
below the 2 keVee threshold.  An example of this effect is provided by
the $c_{13}$ coupling for which a slight 0.9\,$\sigma$ positive
fluctuation from zero is observed in the posterior probability,
weakening the bound as show in Fig.~\ref{fig:high_mass}.  This may by
partially ascribed to the shape of the signal spectrum, as shown in
Fig.~\ref{fig:recoil}.

However, a much more important source of model dependence in Figs.
\ref{fig:low_mass} and \ref{fig:high_mass} is due to the modulation
fractions. In fact, the data used in the present paper (and in the
result of Ref.~\cite{Adhikari:2018ljm}) are sensitive to the
time--averaged count rate, so that they are used to put upper bounds
on the quantity $S_{0,[E_1^{\prime},E_2^{\prime}]}$ defined in
Eq.~(\ref{eq:s0}). On the other hand, the DAMA effect WIMP
interpretation is in terms of the $S_{m,[E_1^{\prime},E_2^{\prime}]}$
quantities of Eq.~(\ref{eq:sm}). Crucially, the ratio of the two
quantities depends both on the WIMP velocity distribution (for which
we assume here a standard Maxwellian as given in
Eq.~(\ref{eq:maxwellian})) and on the specific operator assumed to
dominate in the Hamiltonian of Eq.~(\ref{eq:H}) among those in
Table~\ref{tab:operators}. The variation of the modulation fraction
with the NR model is shown in Table~\ref{tab:modulation}, where for
each NR operator we provide the minimum value of the modulation
fraction $(S^{DAMA}_m/S^{DAMA}_0)_{E^{\prime}<3.5~\rm keVee}$ in the
three DAMA energy bins for 2 keVee$\le E^{\prime}\le$3.5 keVee, where
the bulk of the DAMA modulation effect above the COSINE-100 threshold
is concentrated.  Such variations are an effect of the same modified
spectral features discussed in Fig.~\ref{fig:recoil}. In particular,
larger modulation fractions appear for WIMP scattering events off
sodium targets, which are sensitive to the high--speed tail of the
velocity distribution, and when the scattering amplitude is multiplied
by powers of the transferred momentum $q$ (as summarized in
Table~\ref{table:eft_summary}) due to the enhanced dependence of the
expected rate on the $v_{min}$ parameter.

One can notice that in the same energy intervals of
Table~\ref{tab:modulation} the time--averaged spectrum in DAMA is
above $\simeq$ 0.8 events/day/kg/keVee (see Fig. 1
in~\cite{Bernabei:2018yyw}). An analysis of the DAMA unmodulated data
similar to the one that we perform in the present paper for
COSINE--100 is not available and beyond our capability (since it would
require a detailed understanding of the background, possibly from
Monte Carlo simulations, and access to the systematics of the
experiment).  If, instead, a vanishing background is conservatively
assumed in DAMA, no constraint on the modulation effect can be
obtained. In fact the minimal modulation fraction of
Table~\ref{tab:modulation} ($\simeq$ 0.065 events/kg/day/keVee)
implies an upper bound $S_m\lsim$ 0.05 events/kg/day/keVee, which is
above the corresponding measured values of the modulated amplitudes
(see Table~\ref{tab:dama_data}).

As far as COSINE--100 is concerned, a given upper bound on the
time--averaged rate $S_0$ is converted into a bound on the yearly
modulated component $S_m$ in DAMA whose strength is inversely
proportional to the expected modulated fraction $S_m/S_0$.  In
Table~\ref{tab:modulation} one can see that the smallest values for
$(S^{DAMA}_m/S^{DAMA}_0)_{E^{\prime}<3.5~\rm keVee}$ (and so the
strongest limits on the modulated amplitudes) correspond to the
standard SI and SD couplings ($c_1$ and $c_4$) at the level of about
7\%. Indeed, as shown in Figs.~\ref{fig:low_mass} and
\ref{fig:high_mass} such values are small enough for the corresponding
COSINE--100 exclusion plots to exclude all the DAMA 5--$\sigma$
region.  However, for several models ($c_3$, $c_5$, $c_6$, $c_{13}$
and $c_{14}$ at low WIMP mass and $c_{14}$ at high WIMP mass) the
modulation fraction $(S^{DAMA}_m/S^{DAMA}_0)_{E^{\prime}<3.5~\rm
  keVee}$ is as high as $\simeq$ 12\%, and about half of the
corresponding 5 sigma C.L. DAMA regions are allowed. For most other
models the modulation fraction is at an intermediate value
$\simeq$0.10 so that most of the DAMA 5-sigma region is
excluded. Notice that the two highest numerical values in
Table~\ref{tab:modulation} ($\simeq$ 0.14 for the high--mass DAMA
best--fit of $c_{13}$ and $\simeq$ 0.15 for the low--mass DAMA
best--fit of $c_{15}$) correspond to the two less--constraining cases
in Figs.~\ref{fig:low_mass} and \ref{fig:high_mass} for which all the
DAMA modulation region is allowed by the COSINE--100 bound.
Therefore, the lower energy threshold of COSINE-100 would improve the
bound at low WIMP masses because WIMP--iodine scattering events in the
energy range 1 keVee$\le E^{\prime}\le$2 keVee drive $S_m/S_0$ to
lower values. The case of $c_5$ in Fig.~\ref{fig:high_mass} is a
peculiar one: as discussed in~\cite{dama_eft_2018} the corresponding
effective operator leads to a velocity--dependent cross section for
which the $\chi$--square can saturate to a constant, acceptable value
at large WIMP masses. This explains the peculiar elongated shape for
$c_5$ in Fig.~\ref{fig:high_mass}.

\section{Conclusions}
\label{sec:conclusions}

Assuming a standard Maxwellian for the WIMP velocity distribution, in
the present paper we have discussed the bounds from the null WIMP
search result of the COSINE-100 experiment on the DAMA/LIBRA--phase2
modulation effect within the context of the non--relativistic
effective theory of WIMP--nucleus scattering.  To this aim we have
systematically assumed that one of the effective operators allowed by
Galilean invariance dominates in the effective Hamiltonian of a
spin--1/2 DM particle.

We find that, although DAMA/LIBRA and COSINE--100 use the same
sodium--iodide target, the comparison of the two results still depends
on the particle--physics model.  This is mainly due to two reasons: i)
the WIMP signal spectral shape;
ii) the expected modulation fractions, when the upper
bound on the time--averaged rate in COSINE--100 is converted into a
constraint on the yearly modulated component in DAMA/LIBRA. We find
that the latter effect is the dominant one.  In particular, for
several effective operators we find that the expected modulation
fractions are larger than in the standard spin--independent or
spin--dependent interaction cases.  As a consequence, for such
operators compatibility between the modulation effect observed in
DAMA/LIBRA and the null result from COSINE--100 is still possible.
COSINE-100 has been taking stable data for more than 2.5 years
and 1\,keVee threshold analysis
is forthcoming. This would improve the bound
at low WIMP masses because WIMP--iodine scattering events in the
energy range 1 keVee$\le E^{\prime}\le$2 keVee drive $S_m/S_0$ to
lower values.

\acknowledgments The Sogang group's research was supported by the Basic
Science Research Program through the National Research Foundation of
Korea(NRF) funded by the Ministry of Education, grant number
2016R1D1A1A09917964. Jong--Hyun Yoon acknowledges support from the Magnus Ehrnrooth Foundation. 
We thank the Korea Hydro and Nuclear Power
(KHNP) Company for providing underground laboratory space at Yangyang.
The work of COSINE-100 is supported by: the Institute for Basic Science (IBS) under
project code IBS-R016-A1, Republic of Korea; UIUC campus research
board, the Alfred P. Sloan Foundation Fellowship, NSF Grants
No. PHY-1151795, PHY-1457995, DGE-1122492 and DGE-1256259, WIPAC, the
Wisconsin Alumni Research Foundation, Yale University and DOE/NNSA
Grant No. DE-FC52-08NA28752, United States; STFC Grant ST/N000277/1
and ST/K001337/1, United Kingdom; and Grant No. 2017/02952-0 FAPESP and CAPES Finance Code 001, Brazil.  

\providecommand{\href}[2]{#2}\begingroup\raggedright\endgroup

\end{document}